\documentclass[preprint,aps,nofootinbib]{revtex4}
\usepackage{graphicx}
\usepackage{color}

\def\bold#1{\setbox0=\hbox{$#1$}%
     \kern-.025em\copy0\kern- -\wd0
     \kern.05em\copy0\kern-\wd0
     \kern-.025em\raise.0433em\box0 }

\newcommand{\zpr}{\mbox{$Z'$}}

\newcommand{\mzp}{\mbox{$M_{Z'}$}}

\newcommand{\upr}{\mbox{$U(1)'$}}

\newcommand{\gev}{\,\mbox{GeV}}
\newcommand{\tev}{\,\mbox{TeV}}

\def\ga{\mathrel{\raise.3ex\hbox{$>$\kern-.75em\lower1ex\hbox{$\sim$}}}}
\def\la{\mathrel{\raise.3ex\hbox{$<$\kern-.75em\lower1ex\hbox{$\sim$}}}}

\def\m12{m_{1\!/2}}

\def\lsim{\mathrel{\raise.3ex\hbox{$<$\kern-.75em\lower1ex\hbox{$\sim$}}}}
\def\gsim{\mathrel{\raise.3ex\hbox{$>$\kern-.75em\lower1ex\hbox{$\sim$}}}}

\def\vs{v_s^{}}
\def\vsi{v_{si}^{}}

\begin{document}
\begin{titlepage}
\pagestyle{empty}
\baselineskip=21pt
\rightline{hep-ph/0405244}
\rightline{MADPH--04--1376}
\rightline{UPR--1043T}
\rightline{UCD--2004--20}
\vskip 1in
\begin{center}
{\large{\bf  \boldmath
The Higgs Sector in a \upr \ Extension of the MSSM
}}
\end{center}
\begin{center}
\vskip 0.2in 
{\bf Tao Han$^{1,4}$, Paul Langacker$^2$, Bob McElrath$^3$}
\vskip 0.1in 
$^1${\it Department of Physics, University of Wisconsin,
    Madison, WI~53706} \\ 
$^2${\it Department of Physics and Astronomy, University of Pennsylvania,
Philadelphia, PA 19104}\\
$^3${\it Department of Physics, University of California,
Davis, CA 95616} \\
$^4${\it Institute of Theoretical Physics, Academia Sinica,
           Beijing 100080, China} 

\vskip 0.2in {\bf Abstract}
\end{center}
\baselineskip=18pt
\noindent 
We consider the Higgs sector in an extension of the MSSM with extra
SM singlets,  involving an extra \upr \ gauge symmetry, 
in which the domain-wall problem is avoided and the effective $\mu$
parameter is decoupled from the new gauge boson
\zpr \ mass. The model involves
a rich Higgs structure  very different from that of the MSSM.
In particular, there are large mixings between Higgs doublets and the SM
singlets, significantly affecting the Higgs spectrum,
production cross sections, decay modes, existing exclusion limits, and
allowed parameter range. Scalars considerably lighter than the LEP2
bound (114 GeV) are allowed, and the range $\tan \beta \sim 1$ is both
allowed and theoretically favored. 
Phenomenologically, we concentrate our study on the lighter
(least model-dependent, yet characteristic) Higgs particles 
with significant $SU(2)$-doublet
components to their wave functions, for the case of no
explicit $CP$ violation in the Higgs sector.
We consider their spectra, including the dominant radiative
corrections to their masses from the top/stop loop.
We computed their  production cross sections and reexamine the existing
exclusion limits at LEP2.
We outline the searching strategy for some representative scenarios
at a future linear collider.
We emphasize that gaugino, Higgsino, and singlino decay modes are
indicative of extended models and have been given little attention.
We present a comprehensive list of model scenarios in the Appendices.

\end{titlepage}
\baselineskip=18pt
%%%%%%%%%%%%%%%%%%%%%%%%%%%%%%%%%%%%%%%%%%%%%%%%%%%%%%%%%%%%%%%%%%%%%
\setcounter{footnote}{0}

\section{Introduction}

Supersymmetry (SUSY) is probably the leading candidate for physics beyond the 
Standard Model (SM).  By adding partners of opposite statistics to the SM particles,
it is able to cancel the quadratically divergent contribution to the
Higgs mass.  The leading phenomenological model for SUSY is the Minimal
Supersymmetric Standard Model (MSSM), which incorporates two Higgs doublets
rather than one as in the Standard Model.  Two are required to give
masses to both up-type and down-type fermions, and to prevent anomalies
coming from triangle diagrams involving the Higgs superpartner, the
Higgsino.

The MSSM suffers from  the  ``$\mu$ problem''~\cite{muprob}.  
The superpotential for the MSSM contains the
supersymmetric mass term $\mu H_2 H_1$.  The minimization condition for
the MSSM scalar potential relates $\mu$ to $M_Z$ and soft SUSY breaking
parameters.  One expects all these quantities to be the same order of
magnitude to avoid the need for miraculous cancellations.  However, $\mu$ is an
input-scale parameter and therefore should have mass ${\mathcal O}(M_{Pl})$ 
or ${\mathcal O}(M_{GUT})$.  This has led to a widespread
belief that the MSSM must be extended at high energies to include a
mechanism which relates $\mu$ to the SUSY breaking mechanism.

One possibility is the next-to-minimal Supersymmetric Standard Model
(NMSSM)~\cite{nmssm}, which has been studied
extensively~\cite{nmssm, nmssmpheno}. 
The NMSSM solves the
$\mu$ problem by relating $\mu$ to the vacuum expectation value of a
Standard Model singlet.  The model contains a single extra gauge singlet
superfield, $S$, and a superpotential of the form:
\begin{equation}
W_{NMSSM}={1\over6}kS^3+{1\over2}\mu_SS^2+h SH_2H_1 + W_{MSSM},
\end{equation}
where $W_{MSSM}$ is the superpotential of the MSSM without an elementary
$\mu$ term~\footnote{Some treatments include an elementary $\mu$ term
in addition to the effective one in order to avoid the cosmological domain
wall problem, but this reintroduces the $\mu$ problem~\cite{extended}.}.
If the scalar component of $S$ has a vacuum expectation value, an effective $\mu$ term
$h \langle S \rangle$ is induced. 
However, to have the appropriate hierarchy of
mass scales, the term $\mu_S$ should be disallowed.  Otherwise,
 it should also naturally have a mass near $M_{Pl}$, which would make it
unnatural for $S$ to obtain a weak-scale vacuum expectation value.  
Similar statements apply to the addition of a term  linear
in $S$ to $W_{NMSSM}$.
These can be removed by invoking a $Z_3$ discrete symmetry in the superpotential.
However this would lead to domain walls in the early universe, a
situation which is strongly disfavored cosmologically~\cite{domains}. 
Attempts to remedy this either reintroduce some form of the $\mu$ problem
or lead to other difficulties~\cite{domains2}. 

In string motivated models,  bare
mass terms in the superpotential are generically of order the string scale $M_{\rm
string}$.  This means that any field impacting low energy physics must
not have a mass term like $\mu$ or  $\mu_S$.  Additionally, in 
many constructions (e.g.,  heterotic and intersecting brane) superpotential terms are
typically off-diagonal in the fields, which disallows the NMSSM $k S^3$ term.
Thus, we are led to consider larger models that may be derived from
string constructions.

All of these difficulties can be solved in extended models involving 
an additional (non-anomalous) \upr\  gauge symmetry, which is
very well-motivated as an extension to the MSSM.
$U(1)$ gauge groups arise
naturally out of many Grand Unified Theories (GUT's) and string
constructions~\cite{string}, as do the SM singlets needed
to break the \upr. Experimental limits on the \zpr \ mass and mixings 
are model-dependent, 
but typically one requires~\cite{zprimeev,CDFzprime} $\mzp > 500-800$ GeV and
a $Z-Z'$ mixing smaller than a few $\times 10 ^{-3}$. 

\upr \  models are similar to the NMSSM in that they 
involve a SM singlet field $S$
which yields an effective $\mu$ parameter $h \langle S \rangle$,
where the superpotential includes the term $h SH_2H_1$,
solving the $\mu$ problem~\cite{mugauge,mugauge2}.
 $S$ will in general be charged under the
\upr \  gauge symmetry, so that its expectation value
also gives mass to the new \zpr \ gauge boson. The extended gauge
symmetry forbids an elementary $\mu$ term
as well as terms like $S^n$ in the superpotential (the role of the $S^3$
term in generating quartic terms in the potential is played by
$D$-terms and possibly off-diagonal superpotential terms
involving additional SM singlets). 
Such models do not need to invoke discrete symmetries
(the $Z_3$ of the NMSSM is embedded in the \upr) so there
are no domain wall problems. Such constructions may
also solve other problems, such as naturally forbidding
$R$-parity violating terms which could lead
to rapid proton decay~\cite{chiral}. Other implications include
the presence of exotic chiral supermultiplets~\cite{chiral};
non-standard sparticle spectra~\cite{alternative};
possible flavor changing neutral current effects~\cite{FCNC}, with
implications for rare $B$ decays~\cite{bdecay};
new sources of CP violation~\cite{CP};
new dark matter candidates~\cite{darkmatter};
and enhanced possibilities for electroweak baryogenesis~\cite{elbary}.

As mentioned, string constructions frequently lead to the
prediction of one or more additional \upr \ symmetries at low
energies and to the existence of exotic chiral supermultiplets,
including SM singlets which can break the extra symmetries.
However, no fully realistic model has emerged. 
We therefore take the bottom-up approach and add the minimal
supersymmetric matter content necessary to solve the $\mu$ problem
without introducing extra undesirable global or discrete symmetries.

In this paper we explore the extended Higgs sector in a particular
\upr \ model involving several SM singlet fields~\cite{ell}.
This has the advantage of decoupling the
effective $\mu$ parameter from the \zpr \ mass, and leads
naturally to a sufficiently  heavy \zpr. It will be seen
that the Higgs physics is very rich and quite different from
that of the MSSM. We expect that the generic features will
be representative of a wider class of constructions, and that they
can be tested by
the next generation of high energy experiments and thus provide further
guidance toward constructing the correct SUSY theory.  In the event that
the data from the LHC deviates significantly from MSSM
expectations, it will be especially important 
to consider non-minimal models.  It is useful in planning for
future experimental analysis programs 
to have a variety of well-motivated alternatives in  mind.

In Section \ref{model} we describe the model.
We first outline the general structure of the model in Sec.~\ref{modelmore},
and discuss the electroweak symmetry breaking and the radiative
corrections to the light Higgs mass in Sec.~\ref{EWSB}. We then explore
the phenomenological constraints on the model parameters in
Sec.~\ref{constr}. In particular, we find that the MSSM upper bound on
the light Higgs mass and the lower
bound (direct search from LEP2) can both be relaxed. In order to carry
out further quantitative studies, we perform a comprehensive examination
of the mass spectrum for the Higgs bosons
in Sec.~\ref{secmass}. We classify the Higgs
bosons according to their similarity to the MSSM spectrum and experimental 
signatures. 
The decay modes of the Higgs bosons and their production cross sections at 
$e^+e^-$ colliders are studied in Sec.~\ref{decay}, including
phenomenological implications and search strategies.
We summarize our results in Sec.~\ref{concl}.

\section{The Model}
\label{model}

\subsection{General structure}
\label{modelmore}

The model we consider, first introduced in~\cite{ell}, has the superpotential:
\begin{equation}
  W = h S H_2 H_1 + \lambda S_1 S_2 S_3 + W_{\rm MSSM} \label{wmodel}
\end{equation}
$S$, $S_1$, $S_2$, and $S_3$ are standard model singlets, but 
 are charged under an
extra $U(1)^\prime$ gauge symmetry. The off-diagonal nature of the second
term is inspired by string constructions, and the model is such that the
potential has an $F$ and $D$-flat direction in the limit $\lambda \rightarrow 0$,
allowing a large (TeV scale ) \zpr \ mass for small $\lambda$.
The use of an $S$ field different from the $S_i$ in the first term allows
a decoupling of \mzp \ from the effective $\mu$.  $W$ leads to the $F$-term
scalar potential:
\begin{eqnarray}
  \label{VF}
  \nonumber    
    V_F = & h^2
      \left(|H_2|^2 |H_1|^2 + |S|^2 |H_2|^2 + |S|^2 |H_1|^2\right) \\
    + & \lambda^2 
      \left( |S_1|^2 |S_2|^2 + |S_2|^2 |S_3|^2 + |S_3|^2 |S_1|^2 \right)
\end{eqnarray}
The $D$-term potential is:
\begin{eqnarray}
  \label{VD}
  V_D &=& {{G^2}\over 8} \left(|H_2|^2 - |H_1|^2\right)^2 \nonumber\\&
      + & {1\over 2} g_{Z'}^2\left(Q_S |S|^2 + Q_{H_1} 
      |H_1|^2 + Q_{H_2} |H_2|^2 + \sum_{i=1}^3 Q_{S_i}
      |S_i|^2\right)^2 ~,~\,
\end{eqnarray}  
where $G^2=g_1^{2} +g_2^2=g_2^2/\cos^2 \theta_W$. $g_1, g_2$,  and $g_{Z'}$ are the coupling
constants for $U(1), SU(2)$ and $U(1)^{\prime}$, respectively, and $\theta_W$ is the weak
angle.
$Q_{\phi}$ is the $U(1)^{\prime}$ charge of the field $\phi$. We will take $g_{Z'}
\sim \sqrt{5/3} g_1$ (motivated by gauge unification) for definiteness.

We do not specify a SUSY breaking
mechanism but rather parameterize the breaking with the soft terms
%\footnote{One
%could add an additional soft term of the form
%$m^2 _{S_1 S_2}S_1^\dagger S_2  + {\rm H. C.}$ to $V_{soft} $.
%This could lead to Higgs sector CP violation and therefore mixing
%between scalars and pseudoscalars. Although such a term is useful
%for electroweak baryogenesis~\cite{elbary}, it is beyond the scope of the
%present investigation.}
\begin{eqnarray}
\label{vsoft}
    V_{\rm soft} &=& m_{H_1}^2 |H_1|^2 + m_{H_2}^2 |H_2|^2 + m_S^2 |S|^2 +
       \sum_{i=1}^3 m_{S_i}^2 |S_i|^2
       \nonumber\\&
        - & (A_h h S H_1 H_2 + A_{\lambda} \lambda S_1 S_2 S_3 + {\rm H. C.})
    \nonumber \\
&+& (m_{SS_1}^2 S S_1 + m_{SS_2}^2 S S_2 + {\rm H. C.})
\end{eqnarray} 
The last two terms are necessary to break two unwanted global $U(1)$ 
symmetries, and require $Q _{S_1}=Q _{S_2}=-Q_S$. 
The potential $V=V_F+V_D +V_{soft}$ was studied
in~\cite{ell}, where it was shown that for appropriate parameter
ranges it is free of unwanted runaway directions and has an
appropriate minimum.
We denote the vacuum expectation values of $H_i,  S,$ and $S_i$
by $v_i,  \vs,$ and $\vsi$, respectively, i.e., without a factor of
$1/\sqrt{2}$.
Without loss of generality we can choose $A_h h > 0$, $A_\lambda \lambda > 0$ and
$m^2 _{SS_i}<0$ in which case the minimum occurs for the expectation
values all real and positive.

So far we have only specified the Higgs sector, which is the focus of
this study.  Fermions must also be charged under the $U(1)^\prime$
symmetry in order for the fermion superpotential Yukawa terms
$W_{fermion} = \bar{u} {\bf y_u} Q H_2 - \bar{d} {\bf y_d} Q H_1 -
\bar{e} {\bf y_e} L H_1$ to be gauge invariant.
%\footnote{The signs in
%the superpotential are chosen so that when $H_2$ and $H_1$ get VEV's,
%fermion masses are positive.  An extra minus sign comes from the
%antisymmetric $\epsilon^{\alpha \beta}$ which is used to tie up the
%$SU(2)$ indices.}  
The $U(1)^\prime$ charges for fermions do not
contribute significantly to Higgs production or decay, if sfermions and 
the $Z^\prime$ superpartner are heavy.  We therefore
ignore them this study.

Anomaly cancellation in \upr \ models generally requires the introduction
of additional chiral supermultiplets with exotic SM 
quantum numbers~\cite{mugauge2,chiral,elbary,wang}.
These can be consistent with gauge unification, but do introduce
additional model-dependence.  The exotics can
be given masses by the
same scalars that give rise to the heavy $Z^\prime$ mass.  The exotic sector is not the
focus of this study.  We therefore consider the scenario in which the $Z^\prime$ and
other matter necessary to cancel anomalies is too heavy to significantly affect
the production and decays of the lighter Higgs particles.

\subsection{Higgs sector and electroweak symmetry breaking}
\label{EWSB}

The Higgs sector for this model contains 6 CP-even scalars and 4 
physical CP-odd
scalars, which we label $H_1 ... H_6$ and $A_1 ... A_4$, respectively, in
order of increasing mass.

We compute the six CP-even scalar masses including the dominant 1-loop
contribution coming from the top/stop loop.  Using the effective
potential approach~\cite{effpot}, one writes down the radiatively
corrected effective potential including leading order~(0) and 1-loop
corrections~(1)
\begin{equation}
    V_{\rm eff} = V^{(0)} + V^{(1)} + ...
\end{equation}
and then requires that the effective potential be minimized to obtain
the vacuum expectation values for the fields.  In practice we find it
simpler to eliminate the soft $(\rm mass)^2$ parameters using the
minimization conditions rather than solve for vacuum expectation values.

The full scalar potential is then
\begin{equation}
    V_{\rm eff} = V_D + V_F + V_{\rm soft} + 
    \frac{3}{16 \pi^2}\frac{m_t^4}{v^4}(v_2^2 + v_1^2)^2
    \left(\frac{\Re H_2}{\sqrt{2}v_2} + 1\right)^4\left\{
    \frac{3}{2} - \ln{\left[\frac{m_t^2}{m_{\tilde{t}}^2}\left(\frac{\Re H_2}{\sqrt{2}v_2}+1\right)^2\right]}
    \right\}
\label{full}
\end{equation}
At one loop only the $H_2$ gauge eigenstate gets a correction.  At two
loops both $H_1$ and $H_2$ get corrections from the top and stop.
We have written $v$, $v_1$ and $v_2$ in such a way so that $v_1$
and $v_2$ can be considered rescaled quantities, while $v,  m_t,$ and
$m_{\tilde t}$ (which only occur in ratios) are fixed at their
physical values (as given in Table \ref{model1params}). 
To avoid computational
round-off error we treat $V_{\rm eff}$ as a dimensionless quantity with
all values ${\mathcal O}(1)$, and rescale dimensionful quantities by
$v/\sqrt{(v_2^2+v_1^2)}$ after a viable minimum is found.

In the $\Re{(H_2, H_1, S, S_1, S_2, S_3)}$ 
gauge basis, the resulting mass matrix can be
parameterized as $M_{tree}^2 + \delta M^2$, where
\begin{eqnarray}
\label{deltamhu}
\delta M^2 = \left( {\delta m^2_{H_2} \atop 0_{5 \times 1}} 
                  {0_{1 \times 5} \atop 0_{5 \times 5}} \right), 
                  \hspace{2cm}
\delta m^2_{H_2} = \frac{3}{4\pi^2} \frac{m_t^4}{v_2^2} \ln{m_{\tilde{t}}^2\over m_t^2}
\end{eqnarray}
in the no-stop mixing limit $m_{\tilde{t_1}}=m_{\tilde{t_2}}$~\cite{effpot}.
Since in this model $\tan \beta  \simeq 1$ generically, the contribution from
the bottom loop is negligible so we do not include it.  We also neglect
renormalization scale-dependence and assume the renormalization scale
$Q^2=m_{\tilde{t}}^2$. The singlets
cannot couple directly to the top at tree level, so the large top loop
does not contribute to the masses of any of the new singlets except by
mixing.  The correction $\delta m^2_{H_2}$ has the value $(90 \gev)^2$  for
the MSSM with $m_{\tilde{t}}=1 \tev$ in the large $\tan{\beta}$ limit.  
In the MSSM this is then split among the $h$
and $H$ mass eigenstates.  All other quantities are evaluated at tree level,
using tree level relations.

We find viable electroweak symmetry breaking minima by scanning over the vacuum
expectation values of the six CP-even scalar fields.  We require that
the CP-even mass matrix be positive
definite numerically, which guarantees a local minima, while
simultaneously eliminating the soft mass squared for each field.  
The soft masses reported in the appendices are evaluated including the above
1-loop correction.
The CP-odd mass matrix is guaranteed to be positive semi-definite at tree level (and
thus, all VEV's are real) by appropriate redefinitions of the fields and
choices of parameters as described in Sec.~\ref{parameterspace}.  The
expressions for the first-derivative conditions to eliminate the soft
masses are given in~\cite{ell}.
%
%This potential can suffer from what is known as D-flat directions where
%the potential has a runaway direction.  Due to the vanishing of the
%D-term scalar potential (\ref{VD}).
%The mass matrix positive definite is the second-derivative condition
%ensuring a local minimum.  
The procedure outlined guarantees a local
minimum for each parameter point, but does not guarantee a global
minimum.  
%Thus D-flat directions in the potential do not present a
%problem as far as finding viable model points in parameter space, but
%like most analyses of this type, we cannot be sure that any given model
%point is in a global minimum.

The parameters $m_{SS_1}^2$ and $m_{SS_2}^2$ must be chosen to avoid
directions in the potential that are unbounded from below.  
We require
\begin{equation}
    m_S^2 + m_{S_i}^2 + 2 m_{S S_i}^2 > 0,
\end{equation}
to avoid unbounded directions with  $\vs=\vsi$ and the other VEV's vanishing.

We scan over vacuum expectation values such that the three singlets
$S_1$, $S_2$, and $S_3$ typically have larger VEV's than the other three fields.
We allow points in our Monte Carlo scan that fluctuate from all VEV's equal up to
$\langle S\rangle$ approximately 1 TeV and $\langle S_i\rangle$
approximately 10 TeV, as we specify in Table \ref{model1params}.
This generically results in a spectrum with 1-5
relatively light CP-even states, often with one of them lighter than the LEP2
mass bound, but having a relatively small overlap with the MSSM $H_2$ and
$H_1$.  It is necessary that at least one of the singlets have an
$\mathcal{O}({\rm TeV})$ vacuum expectation value, so that the mass of
the $Z^\prime$ gauge boson is sufficently heavy that it evades current
experimental bounds, and any extra matter needed to cancel anomalies is
heavy enough to not significantly affect light Higgs production or
decay.

A bound exists on the mass of the lightest Higgs particle in any
perturbatively valid supersymmetric theory~\cite{Kane:1992kq,Quiros:1998bz}. 
The limit on the lightest MSSM-like CP-even Higgs mass in this model is:
\begin{eqnarray}
        \nonumber
    M^2_h \le h^2 v^2 &+& 
    (M_Z^2 - h^2 v^2) \cos^2{2 \beta}
        +2 g_{Z^\prime}^2 v^2 (Q_{H_2} \cos^2 \beta  + \sin^2 \beta  Q_{H_1})^2 \\
        &+&{3 \over 4} \frac{m_t^4}{\pi^2 v^2}
        \ln {m_{\tilde{t_1}} m_{\tilde{t_2}} \over m_t^2}.
\label{mhlimit}
\end{eqnarray}
This is obtained by taking the limit as the equivalent of the $B$-term 
in
the MSSM goes to infinity, $B = A_h h v_s \rightarrow \infty$, in the
$2\times 2$ submatrix containing $H_2$ and $H_1$.  In the MSSM this is
equivalent to taking $M_A \rightarrow \infty$,  the decoupling
limit.  
%If we require perturbativity at some high GUT scale, $h$ is
%allowed to be ${\mathcal O}(1)$ due to the new contributions to its
%renormalization group equations.  In the interest of exploring the low
%energy effective potential, we allow $h$ to be large enough to require
%another scale to enter before the Planck scale, if $h$ is to remain
%perturbative.
%Note that the second term of Eq. \ref{mhlimit} vanishes for 
%$\tan \beta = 1$, therefore since $tan \beta \simeq 1$ generically in
%these models, the lightest Higgs mass is determined mostly by the new
%contributions to the potential.  In this model as with any model with
%many Higgs particles, a situation can arise where the MSSM-like
%couplings are shared among many states, allowing unusually heavy states
%or unusually light states that evade current experimental bounds.
This expression is the same as in the NMSSM, except for the 
$g _{Z'}$  ($D$-term) contribution.
Perturbativity to a GUT or Planck scale places an upper limit
${\mathcal O}(0.8)$  on $h$~\cite{ell}, which is less stringent than the corresponding
limit in the NMSSM~\cite{NMSSMbnd} due to the \upr \ contributions to its
renormalization group equations.
Larger values would be
allowed if another scale entered before the Planck scale.
We will allow $h$ as large as 1 in the interest of exploring the low
energy effective potential.
The second term of Eq.~(\ref{mhlimit}) vanishes for 
$\tan \beta = 1$. Since $\tan \beta \simeq 1$ generically in
these models, the lightest Higgs mass is determined mostly by the new
$F$ and $D$-term contributions proportional to $h^2$ and $g_{Z^\prime}^2$.  
In this model, as with any model with
many Higgs particles, a situation can arise in which the MSSM-like
couplings are shared among many states, allowing unusually heavy states
or unusually light states that evade current experimental bounds.

The four CP-odd masses can in principle be found algebraically but the
results are complicated and not very illuminating.
Perhaps the most striking feature of the mass spectrum is that the 
$A_1$ is allowed to be very light, a feature shared with the NMSSM~\cite{nmssm}.  
This is caused by a 
combination of small $m_{SS_1}^2$ or $m_{SS_2}^2$ and
a small value of $\vs$ compared to the $\vsi$.  In the limit that $\vsi\ (i =$ 1 or
2)  is the largest scale in the problem, the lightest $A$ mass is
\begin{equation}
\label{ma1}
    m_{A_1}^2 = -m_{SS_i}^2\frac{\vs \vsi}{v_{si}^2 + v_{s3}^2} +
    \mathcal{O}\left(\frac{1}{v_{si}^4}\right).
\end{equation}
In the limit that $s_3$ is large we obtain
\begin{equation}
\label{ma2}
    m_{A_1}^2 = -4 m_{SS_i}^2\frac{\vs \vsi}{v_s^2+ v_{s1}^2+ v_{s2}^2} + 
    \mathcal{O}\left(\frac{1}{v_{s3}}\right).
\end{equation}
In our scans, $-m_{SS_i}^2$ is approximately in the range ($0-1000$  GeV)$^2$.
An example where this occurs is presented in Appendix~(\ref{a1lightest}).
However,  this requires a hierarchy between the off-diagonal soft
masses $m_{S S_i}$ and the other soft masses $m_S$ and $m_{S_i}$.  This
might be difficult to achieve depending on the SUSY breaking mechanism.
A similar analysis holds for $H_1$, but an algebraic expression cannot be
derived since the eigenvalues of a $6\times 6$ matrix cannot be expressed algebraically.
Examples of spectra with a light $H_1$ are given in 
Appendices~(\ref{h1lightest},\ref{h1smdominant}).

To make comparisons to the MSSM, we define the ``MSSM fraction''
$\xi_{\rm MSSM}$.
%  We define this to
% help guide the reader's intuition, $\xi_{\rm MSSM}$ is not fundamental
% in any way.  
For a given Higgs state $H_i\ (A_i)$ in the mass basis,
\begin{equation}
    \xi^i_{\rm MSSM} = \sum_{j=1}^2 (R^{ji})^2,
\end{equation}
where $R$ is the matrix that rotates the interaction fields 
to the mass basis, and the index $j$ runs over MSSM states.  
In the case of the CP-even Higgs 
this corresponds to adding in quadrature the
eigenvector components of a state in the $H_2$ and $H_1$ directions.
When a state is MSSM-like, $\xi_{\rm MSSM}=1$ and it has no mixing 
with singlet Higgs bosons.
%For CP-odd states  $A_i$, $N=1$.
If the two lightest CP-even states and the lightest CP-odd state all have
$\xi_{\rm MSSM} \simeq 1$, the theory is MSSM-like and the extra scalars
are decoupled.  An example of such a spectrum is presented in 
Appendix~(\ref{mssmlike}).  
There is always a massless CP-odd scalar with $\xi_{\rm
MSSM} \simeq 1$, and another with $\xi_{\rm MSSM} \simeq 0$.  These are
the Goldstone bosons corresponding to the $Z$ and $Z^\prime$ gauge
bosons, respectively.  

A similar quantity is defined for the neutralinos by summing in
quadrature over the eigenvector components corresponding to $\tilde{B}$,
$\tilde{Z}$, $\tilde{H_2}$ and $\tilde{H_1}$.  The ``Singlino fraction''
$\xi_{\tilde{s}}$ and ``Zino-prime fraction'' $\xi_{\tilde{Z^\prime}}$ are
defined in an analogous manner.

\section{Phenomenological Constraints}
\label{constr}

Due to the introduction of the Higgs singlets, there are 
several more parameters than  in the MSSM Higgs sector.
We follow the global symmetry breaking structure of Model I of Ref.~\cite{ell}.
Existing experimental measurements already constrain any new model.  In
our parameter space scans, we apply the constraints in the following
subsections.

\subsection{Parameter Space}
\label{parameterspace}

\begin{table}[tb]
\begin{tabular}{lll}
\hline 
$g_1=0.36$          & $g_2=0.65          $ & $g_{Z^\prime}=\sqrt{5/ 3} g_1$  \\
$m_t = 174.3 \gev $    & $v = 174 \gev          $ &  
$m_{\tilde{t_1}} = m_{\tilde{t_2}}=1 \tev $ \\
\hline 
$Q_{H_2}={1}/{4}$  & $Q_{H_1}={1}/{4}   $ & $Q_S=-{1}/{2}$  \\
$Q_{S_1}={1}/{2}$  & $Q_{S_2}={1}/{2}   $ & $Q_{S_3}=-1$ \\
\hline 
$h=-1\ ...\ 1$           & $\lambda=-0.2\ ... \ 0.2           $ & \\ 
\hline 
$A_h=0.0\ ...\ 50$       & $A_{\lambda}=$0.0\ ...\ 50   &  \\
$|m_{SS_1}^2|=0\ ...\ 100~~~$ & $|m_{SS_2}^2|=$0\ ...\ 100~~~  & $m_{S_1S_2}=0$  \\
%$g_1$&=&$0.357419$         & $g_2$&=&$0.651215          
%$ & $g_{Z^\prime}$&=&$\sqrt{5 \over 3} g_1$  
$M_2 =-10\ ...\ 10$         & $M_1 = -5\ ...\ 5   $ & $M_1^\prime = -20\ ...\  20$ \\ 
\hline 
$v_{1,2} =1\ ...\  e$         & $\vs = 1\ ...\  e^2   $ & $v_{s1,2,3} = 1\ ...\  e^4$ \\ 
\hline 
\end{tabular}
\caption{Input values and the ranges for model parameters.}
\label{model1params}
\end{table}
We list the model parameters in Table \ref{model1params}.
Besides the SM gauge couplings, $g_{Z^\prime}$ is chosen as
 $\sqrt{5/3}g_1$ that unifies with $g_2$ and $g_3$ in simple GUT
models.  However, we do not require unification of the gaugino masses.
We have fixed the \upr \ charges for definiteness.
The parameters $A_h$, $A_\lambda$, $m_{SS_1}$, and $m_{S_1S_2}$, and
$M_2$ are of course dimensionful, as are the expectation values
$v_i, v_s,$ and $v_{si}$.  For our computation we choose arbitrary units to start,
and use the analytical minimization conditions for the VEV's as
given in Ref.~\cite{ell}, eliminating the soft mass square parameters.  We check
that the VEV's obtained from scanning are a minimum by explicitly verifying that the
matrix of second derivatives is positive definite.
After finding a viable minimum, we rescale all dimensionful parameters by
$v/\sqrt{v_2^2+v_1^2}$, where $v$ is fixed at 174 GeV, which
shifts the Higgs vacuum expectation value to its measured value.  
The other dimensionful parameters $m_t$, $v$, $m_{SUSY}$, $m_{\tilde{t_1}}$ and
$m_{\tilde{t_2}}$  enter the Higgs potential at
one loop as given by Eq.~(\ref{full}). However, they enter in
ratios so that the units cancel out.  
$\tan\beta = v_2/v_1$ is therefore an output.

In the MSSM  the sign of $\mu$ is a free parameter.
In this model $\mu=hv_s$, and $v_s$ is taken to be positive at the
minimum, meaning that the sign of $\mu$ is really the sign of $h$.  We
can absorb any overall phase of $A_h h$ by a redefinition of the fields,
in exactly the same way a phase of $B$ can be absorbed in the MSSM, and
$B=A_h h$ taken to be positive.  
%This leaves the relative sign between
%$A_h$ and $h$ (either both positive or both negative) which becomes the
%``sign of $\mu$''.  We do the same with $\lambda$ and $A_\lambda$ by
Any phase appearing in the soft-masses $-m_{S
S_1}^2$ and $-m_{S S_2}^2$ can be absorbed by a field redefinition on
$S_1$ or $S_2$, so that  $m^2_{SS_i}$ are negative.
$\lambda A_\lambda$ can be taken to be positive by
redefining  $S_3$.
%, giving another sign degree of freedom between them.  
 This uses up our freedom to redefine our fields,
leaving  a true phase in $h$ and $\lambda$ that cannot be rotated
away.  With these field redefinitions $A_h h$, $A_\lambda
\lambda$, $-m_{S S_1}^2$, and $-m_{S S_2}^2$
are all positive
\footnote{These two parameters are chosen negative to conform to the 
convention of Ref.~\cite{ell}.}, and all of the VEV's will be real and
positive at the minimum.
% \footnote{The
%negative sign appearing in the $m_{S S_1}^2$ and $m_{S S_2}^2$ values of
%Ref. \cite{ell} has been absorbed by this field redefinition.  These two
%parameters can be taken real and positive.}
%
There is not enough freedom left to rotate away a phase appearing in
a possible additional term $m_{S_1 S_2}^2 S_1^\dagger S_2$.  
We have thus taken this parameter to be zero.
A phase in this parameter would provide for CP
violation in the Higgs sector, and therefore lead to mixing between
scalars and pseudoscalars.  Although such a term is useful for
electroweak baryogenesis~\cite{elbary}, it is beyond the scope of the
present investigation.

With these conventions the gaugino masses can be either positive or
negative.  The scalar potential and therefore vacuum expectation values
are insensitive to the signs of $h$ and $\lambda$ at tree level, since it
always appears as $A_h h$ and $A_\lambda \lambda$ whose phases can be
rotated away, or $|h|^2$ and $|\lambda|^2$.  Only the charginos and
neutralinos are sensitive to the signs of $h$ and $\lambda$.  
The neutralino and chargino mass matrices are given in Ref.~\cite{ell}. 
We allow both positive and negative values for $h$, $\lambda$, 
and the gaugino masses.

\subsection{$Z^\prime$ Mass and $Z-Z^\prime$ Mixing}

Limits on the \zpr \ mass and $Z-Z'$ mixing angle are model-dependent.
However, for typical models
 the $Z$ pole data indicate that the $Z-Z^\prime$
mixing must be less than  a few $\times 10^{-3}$~\cite{zprimeev},
while direct searches at the Tevatron
limit the mass of the $Z^\prime$ to be greater than $\sim 500-800$
GeV~\cite{CDFzprime}.  Therefore, we require of the $Z-Z^\prime$ mixing
angle:
\begin{equation}
\label{alphazzp}
\alpha_{Z-Z^\prime} = {1 \over 2}\arctan \left(
    \frac{2 M_{ZZ^\prime}^2}{M_{Z^\prime}^2-M_Z^2}
    \right) \lsim 5\times 10^{-3}
\end{equation}
where $M^2_{ZZ^\prime}$ is the off-diagonal entry in the mass-squared matrix,
and $M_Z^2$, $M_{Z^\prime}^2$ are the diagonal entries.  We require for the
mass:
\begin{equation}
M_{Z^\prime} \ge 500 \gev.
\label{mzp}
\end{equation}
The $Z^\prime$ would be produced at tree level at the Tevatron, since if
we require that fermions receive mass through the usual Higgs mechanism,
they must be charged under $U(1)^\prime$ to keep the superpotential
Yukawa terms gauge invariant.  We do not calculate the $Z^\prime$
production cross section to avoid the
necessity of having to specify the fermion $U(1)^\prime$ charges.  
This model does not
particularly care how heavy the $Z^\prime$ is.  Since there are four
singlets contributing to its mass, it is not difficult to give some of
them large vacuum expectation values,  resulting
in a heavy $Z^\prime$.  This occurs naturally for small $\lambda$.

A lighter $Z^\prime$ near the experimental limit is also not a problem.
The singlets must have smaller vacuum expectation values, and therefore
smaller masses, but since they do not couple directly to the Standard
Model except by mixing with the MSSM Higgs bosons, they can be
 light.  
 The typical scale for exotics introduced to cancel anomalies is near the
 \zpr \ mass.
 
%For the particular choice of parameters (Sec.\ref{model1params}), a light
%$Z^\prime$ near experimental limits is difficult to achieve if we
%constrain $\alpha_{Z-Z^\prime}$ tightly.  This is due to the fact that
%in order to get the $Z-Z^\prime$ mixing (Eq. ~\ref{alphazzp}) within
%experimental limits $M_{Z^\prime}$ must be taken rather heavy.  The
%way to reduce the mixing is to reduce the off-diagonal entries in the
%matrix, $M_{ZZ^\prime}=g_{Z^\prime}\sqrt{g^2+g^{\prime
%2}}(Q_{H_2}v_2^2-Q_{H_1}v_1^2)$.  However in the model as written
%originally, $M_{ZZ^\prime}=0$ would require $\tan \beta   =-2$ which is
%not possible.  Other choices of $Q_{H_2}$ and $Q_{H_1}$ can have a
%naturally light $Z^\prime$ near the experimental limit with naturally
%small mixing.  \FIXME{redo with original parameters as well?  Lightest
%Z' was 2.2 TeV.  My charge choice still uses Model I global symmetry
%breaking structure}  We choose $Q_{H_2}=Q_{H_1}$ and normalize all
%charges so that the largest is 1, as given in Sec.\ref{model1params}.

The mixing constraint in (\ref{alphazzp}) is most easily satisfied for
\mzp \ in the TeV range. Smaller values of \mzp, such as we allow,
generally require a suppressed value for $M^2_{ZZ^\prime}=g_{Z^\prime}\sqrt{g^2+g^{\prime
2}}(Q_{H_2}v_2^2-Q_{H_1}v_1^2)$. Since $\tan \beta$ is typically close to unity,
this is achieved for the choice $Q_{H_2}=Q_{H_1}$, which we have assumed\footnote{Small
mixing was achieved in~\cite{ell} with $Q_{H_2}\ne Q_{H_1}$ because of rather
large \zpr \ masses.}.

\subsection{Chargino and Neutralino Mass Bounds}

The chargino in this model is
essentially identical to the MSSM chargino.  There are no new tree-level
modifications to the chargino couplings or mass.
As reported by the LEP2 experiments, we require that the chargino mass be
\begin{equation}
M_{\chi^\pm} \ge 103 \gev.
\label{mchi}
\end{equation}

We place no constraints on the neutralino.  Current constraints are very
model-dependent.  Even with the MSSM it has
been demonstrated that the LSP can be as light as $6 \gev$~\cite{Hooper:2002nq}.  
With the additional assumption that the LSP is
mostly singlet, it can be lighter still 
(including massless~\cite{Gogoladze:2002xp}). 
Such light singlinos may provide a very
interesting candidate for the dark matter component of the 
universe~\cite{mySinglinoDM}.

\subsection{LEP2 Bounds on the Higgs Masses}
\label{hzha}

The pair creation of the charged Higgs boson $H^+H^-$ in $e^+e^-$
collisions provides a model-independent channel for the Higgs search. 
The non-observation of this signal at LEP2  requires 
\begin{equation}
M_{H^\pm} \ge 71.5 \gev.
\label{mhch}
\end{equation}

Limits were placed on cross sections for $e^+e^- \rightarrow Zh$ and
$e^+ e^- \rightarrow Ah,\ AH$ at LEP2 up to energies of 209 GeV.  We impose
on our model that it not violate these bounds by directly computing
these cross sections. 
For a theory with two Higgs doublets plus any number of  singlets, 
the cross section for a Higgs boson radiation off a $Z$  is simply
related to the SM cross section by
\begin{equation}
\sigma(e^+e^- \rightarrow Z H^i) 
    = (R^{i1}_H \sin\beta - R^{i2}_H \cos\beta)^2 
        \sigma^{}_{SM}(e^+e^- \rightarrow Zh) ,
\label{hzxsect}
\end{equation}
where $R_H$ is the matrix that diagonalizes the CP-even Higgs mass matrix.
Similarly, the cross section for  Higgs pair production via $Z$ exchange is
obtained by scaling the MSSM result
\begin{equation}
\sigma(e^+e^- \rightarrow H^i A^j) 
    = (R^{i1}_H R^{j1}_A - R^{i2}_H R^{j2}_A)^2
    \frac{ \lambda(M_{A^j}, M_{H^i})^{3/2} }{(12M_Z^2/s+\lambda(M_Z, M_h))
        \lambda(M_Z,M_h)^{1/2}}
    \sigma_{SM}(e^+e^- \rightarrow hZ)
\end{equation}
%
%\begin{equation}
%\sigma(e^+e^- \rightarrow H^i A^j) 
%    = (R^{i1}_H R^{j1}_A - R^{i2}_H R^{j2}_A)^2
%\sigma^{}_{MSSM}(e^+e^- \rightarrow H^i A^j),
%\label{haxsect}
%\end{equation}
%
where $R_A$ is the matrix that diagonalizes the CP-odd Higgs mass matrix.
%$R_A$ is that for the CP-odd Higgs, 
%and $\lambda$ is the usual 2-point kinetic function:
$\lambda(m_1, m_2)=(1-(m_1+m_2)^2/s)(1-(m_1-m_2)^2/s)$.  The explicit
indices $1$ and $2$ correspond to the $H_1$ and $H_2$ columns,
respectively.  The cross section can be written in this simple form
because the $ZhA$ vertex comes entirely from the covariant derivatives of
$H_2$ and $H_1$.

We impose that these two cross sections be less than the LEP2 limits for 
all mass eigenstates with each production channel
separately\footnote{In some cases the bounds would be strengthened if
one summed over the production channels.}.
For the $ZH$ case we use the LEP2 limit $M_h \ge 114.1 \gev$, which
leads to  a cross section of  184 fb as an upper bound at $\sqrt{s}=209 \gev$.
We then require that the cross section in this model be less than that value. 
Since the cross section increases as the Higgs mass is decreased, 
this gives a conservative estimate of when a model is inconsistent
with the LEP2 bound and is thus ruled out in terms of the Higgs mass
and other coupling parameters. 
For $HA$ channels, using the LEP2 MSSM
mass limits $M_h \le 91.0 \gev$ and $M_A \le 91.9 \gev$,
we compute the $hA$ cross section to be $48.3$ fb  at $\sqrt{s}=209 \gev$.
This is the value one
obtains omitting factors of cos/sin $\alpha$ and $\beta$.  It is exactly
correct when either $\sin(\beta-\alpha)=1$ or $\cos(\beta-\alpha)=1$ for
one of the CP-even Higgs states.  

We also require that the light Higgs bosons do not increase the $Z$ width
beyond experimentally measured bounds.  The decays $Z \rightarrow H_i
e^+ e^-$ and $Z \rightarrow H_i A_j$ are each required to have a partial
width less than $2.3$ MeV.

We have not attempted to include acceptance effects, 
such as may be associated with the nonstandard
decay modes for the $H^i$ and $A^j$. These effects would
tend to weaken the constraints.

\begin{figure}[tb]
    \includegraphics[scale=0.33]{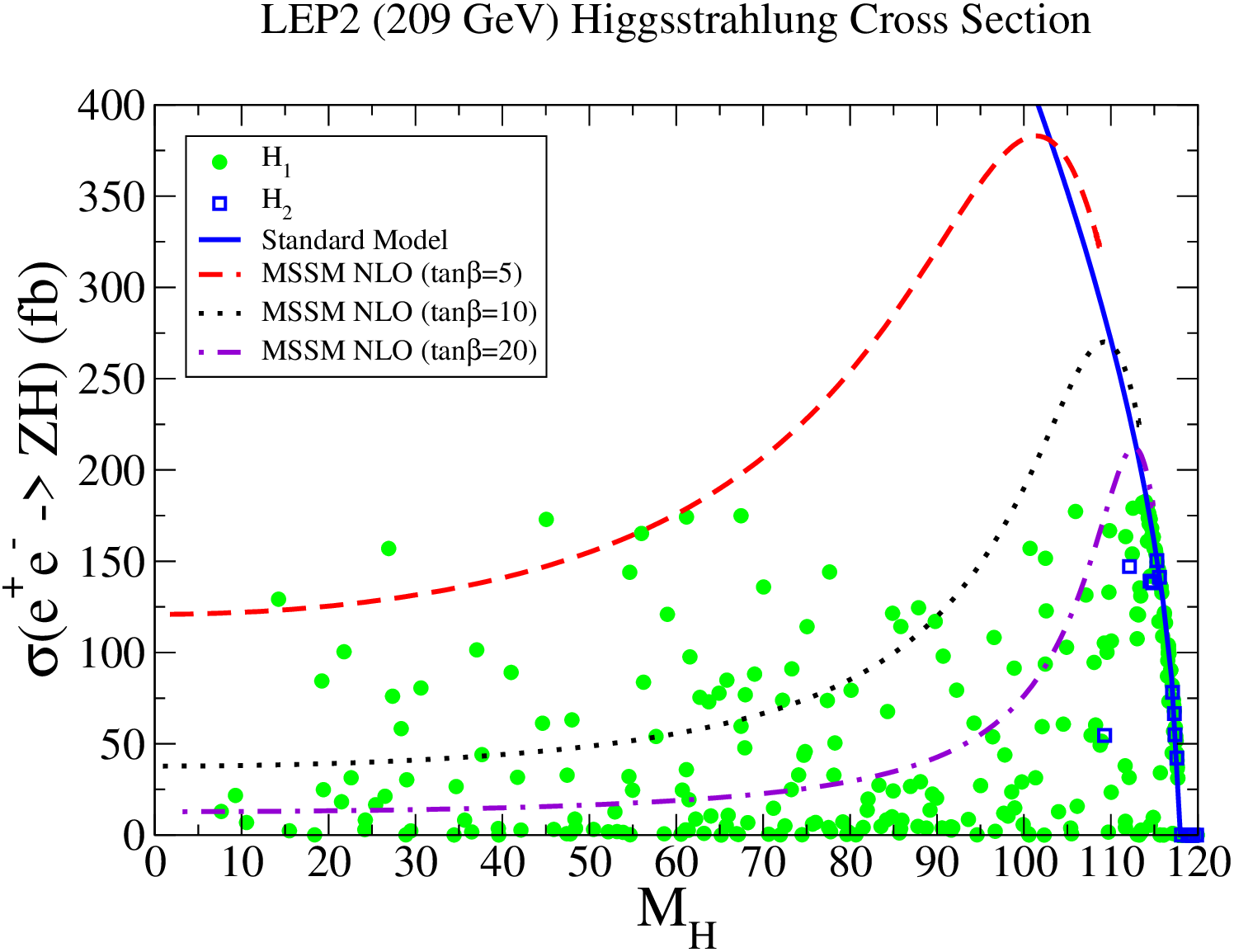}
    \includegraphics[scale=0.33]{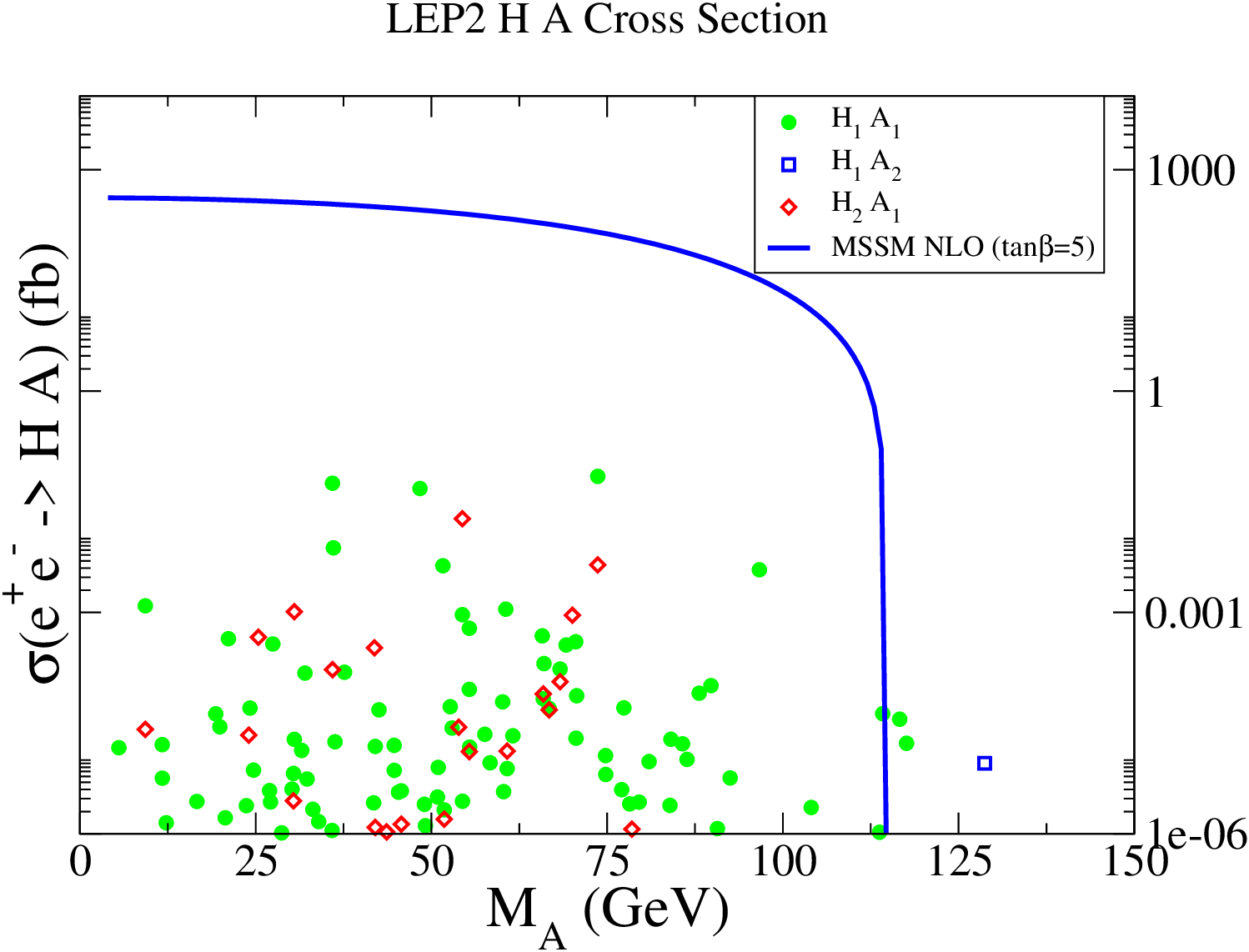}
    \caption {Cross sections at LEP2 (a) for $ZH_i$ production
and (b) for $A_iH_j$ production 
versus the relevant Higgs boson mass.  
In (a), the solid curve is the SM production, and the dashed, dotted 
and dash-dotted are for the MSSM with $\tan\beta=5,10,20$, respectively. 
In (b), the curves is with $\tan\beta=5$. 
}
    \label{xseclep}
\end{figure}

We show the production cross sections at LEP2 versus the relevant
Higgs boson mass parameter in Fig.~\ref{xseclep}
for (a) the $ZH_i$ channels and (b) $A_iH_j$ channels. 
Each symbol point indicates a solution satisfying all the constraints 
listed in this section. For comparison, the SM production rate is plotted 
by the solid curve in (a),  and the dashed, dotted, and dash-dotted, 
are for the MSSM with $\tan\beta=5,10,20$, respectively. 
In (b), the curve is with $\tan\beta=5$.  MSSM curves are at NLO using the 
software HPROD~\cite{HPROD}.
It is interesting that there are solutions
that have a CP-even Higgs as light as $M_{H_1}\approx 8$ GeV,
and a CP-odd state $M_{A_1}\approx 6$ GeV,
that satisfy the LEP2 bounds.  After removing the solutions incompatible
with the $ZH$ bound from LEP2, there are essentially no solutions
that lead to sizable cross sections in the $AH$ channel, 
as seen in Fig.~\ref{xseclep}(b). 
The size of these masses
reflects only the range of parameters we chose for scanning.  As long as
the light states are mostly singlet in composition, they can be
arbitrarily light.  As shown in Eq.~(\ref{ma1}) and Eq.~(\ref{ma2}), the $A_1$ can be tuned
to be very light. 

\section{Mass Spectrum and Couplings for Higgs Bosons}
\label{secmass}

\begin{figure}[t]
    \includegraphics[scale=0.33]{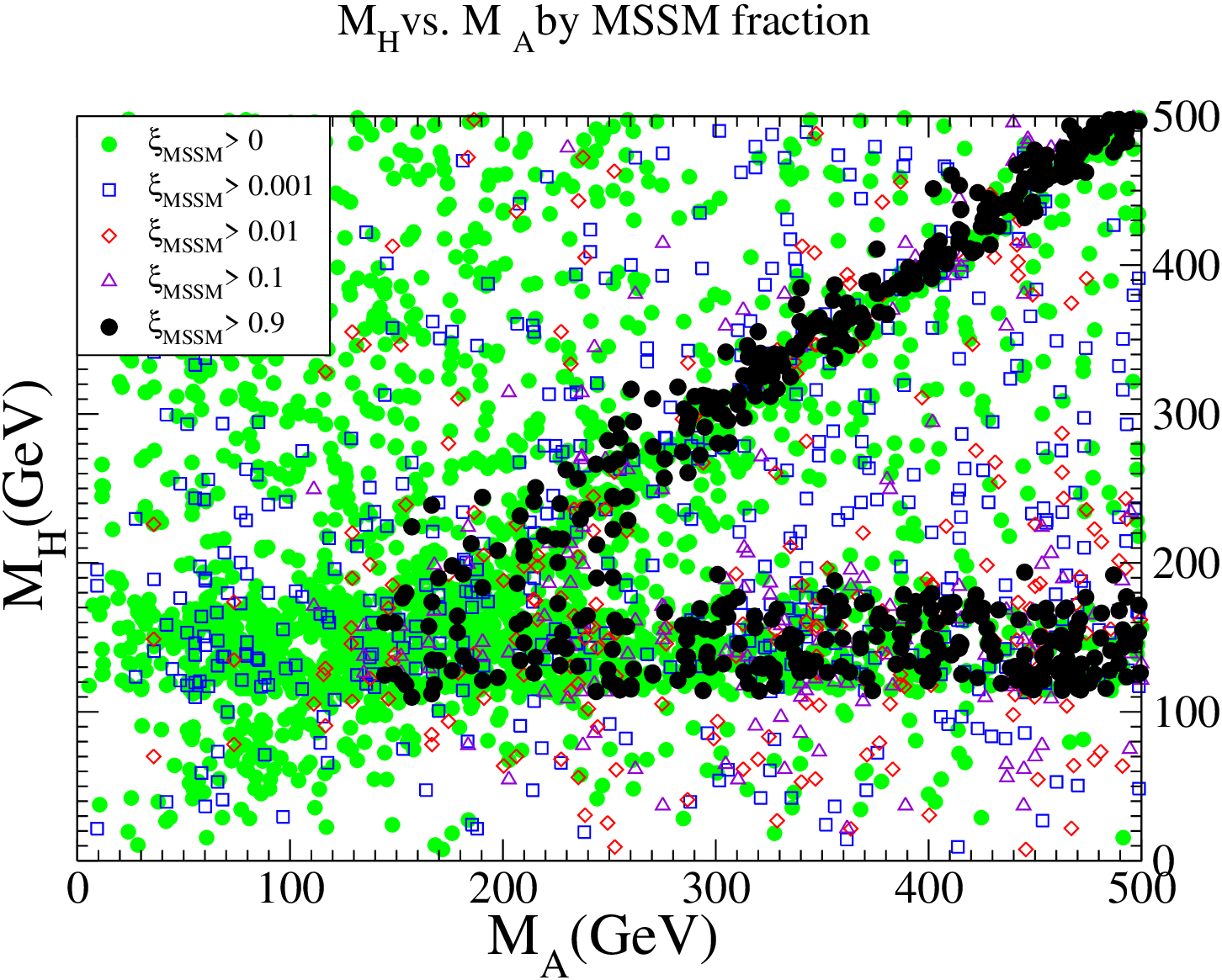}
    \includegraphics[scale=0.33]{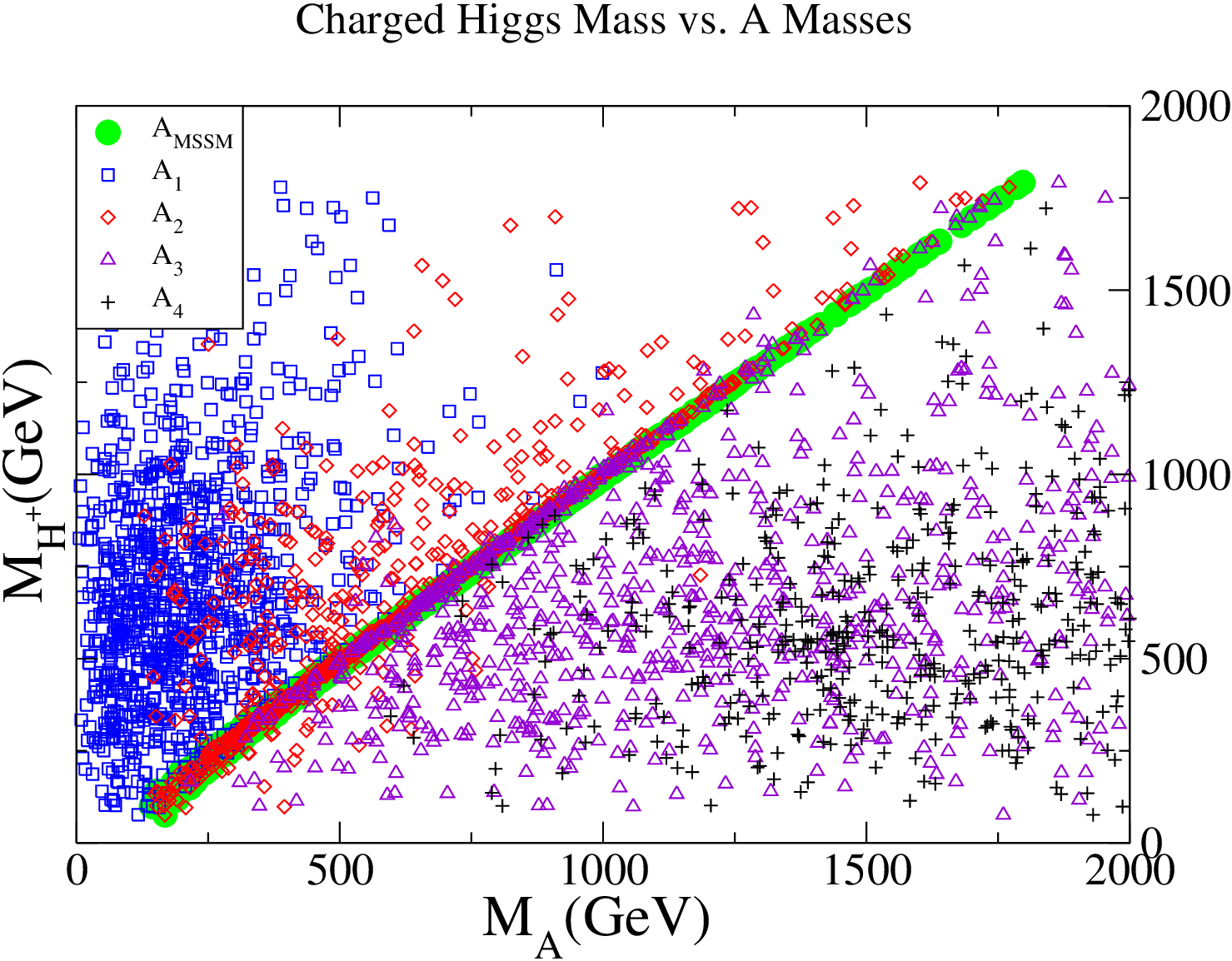}
    \caption{(a) $M_H-M_A$ mass plane, labeled according to MSSM fraction
    $\xi_{\rm MSSM}$.  For each point both $H_i$ and $A_i$ satisfy the
    condition $\xi_{\rm MSSM} > 0, 0.001, 0.01, 0.1$, or $0.9$.  All pairs
    $(M_{H_i}, M_{A_j})$ are plotted.  (b) $M_{H^+}-M_A$ mass plane with the
    MSSM $A^{\rm MSSM}$ mass $M^{\rm MSSM}_A = 2 A_h h v_s/\sin{2 \beta}$
    included for comparison.}
    \label{mhma}
\end{figure}

We first point out the relaxed upper bound on the mass of the lightest
CP-even Higgs boson. 
As given in  Eq.~(\ref{mhlimit}), the lightest CP-even Higgs boson mass
at tree level would vanish in the limit 
$h\to 0$, $g_{Z^\prime}\to 0$ and $\tan \beta \to 1$.
Using the parameters discussed in \ref{parameterspace}, 
the upper limit on the lightest Higgs boson mass at tree level
as given by the first two terms in Eq.~(\ref{mhlimit}) is $142\gev$.
Including the effects of Higgs mixing and the one-loop top correction, 
we find masses up to $\sim 168 \gev$.  
The mass could be made even larger  if we allowed
$h>1$, although the perturbativity requirement up to the GUT
scale at 1-loop level would imply that $h\le 0.8$.  We know that new
heavy exotic matter must enter this model to cancel anomalies, so it is
not necessarily justified to require $h$ to be perturbative to the
Planck scale by calculating its 1-loop running using only low energy
fields.

The masses of the various Higgs particles are a function of the mixing
parameters, and most of the simple MSSM relations among masses are
broken. It is quite common to have a light
singlet with sizable MSSM fraction that can still evade the LEP2 bounds.
Typical allowed light CP-even and odd masses
are shown  in Fig.~\ref{mhma}(a) for various ranges of MSSM fractions.
We see that it is possible to have light MSSM Higgs bosons below about
100 GeV without conflicting the LEP2  searches. This is because of the
reduced couplings to the $Z$ when the MSSM fraction becomes small.
One can clearly make out the usual MSSM structure when $\xi_{\rm
MSSM}$ is large, with the diagonal band for $\xi_{\rm MSSM} > 0.9$ being
$M^{\rm MSSM}_H \simeq M^{\rm MSSM}_A$, and the horizontal band being the
saturation of $M^{\rm MSSM}_h$ at its upper bound in the decoupling limit.
As $\xi_{MSSM}$ decreases, we can see points in the lower left that are
able to evade the LEP2 bounds on $M_{h,H}$ and  $M_A$.

The  mass range for the charged Higgs boson is demonstrated in 
Fig.~\ref{mhma}(b).   There is
 still a linear relationship between the charged Higgs mass and
the MSSM $A$ mass since the singlets do not affect the $H^+$ mass. 
However, after mixing there is not necessarily a state
with that mass, or the identity of the state is obscured.  
Most of the parameter space has a single state that can be identified as
MSSM-like, with $\xi_{\rm MSSM} \sim 1$; in such circumstances there 
is also generally an $H$ very close in mass to both the $A$ and $H^+$.
This is demonstrated in Appendix~(\ref{h1smdominant}) 
with $M_{A_3}=774$, $M_{H^+}=792$, and
$M_{H_4}=780$ GeV.  
%For $M_{H^+} < 200$ GeV,
However, the difference between $M_{H^+}$ and the $M_{A_i}$ can be 50
GeV or more due to mixing, especially when the MSSM-like state is not
clearly identifiable.  Such an example is presented in Appendix~(\ref{maxamix}).

\begin{figure}[tb]
    \includegraphics[scale=0.33]{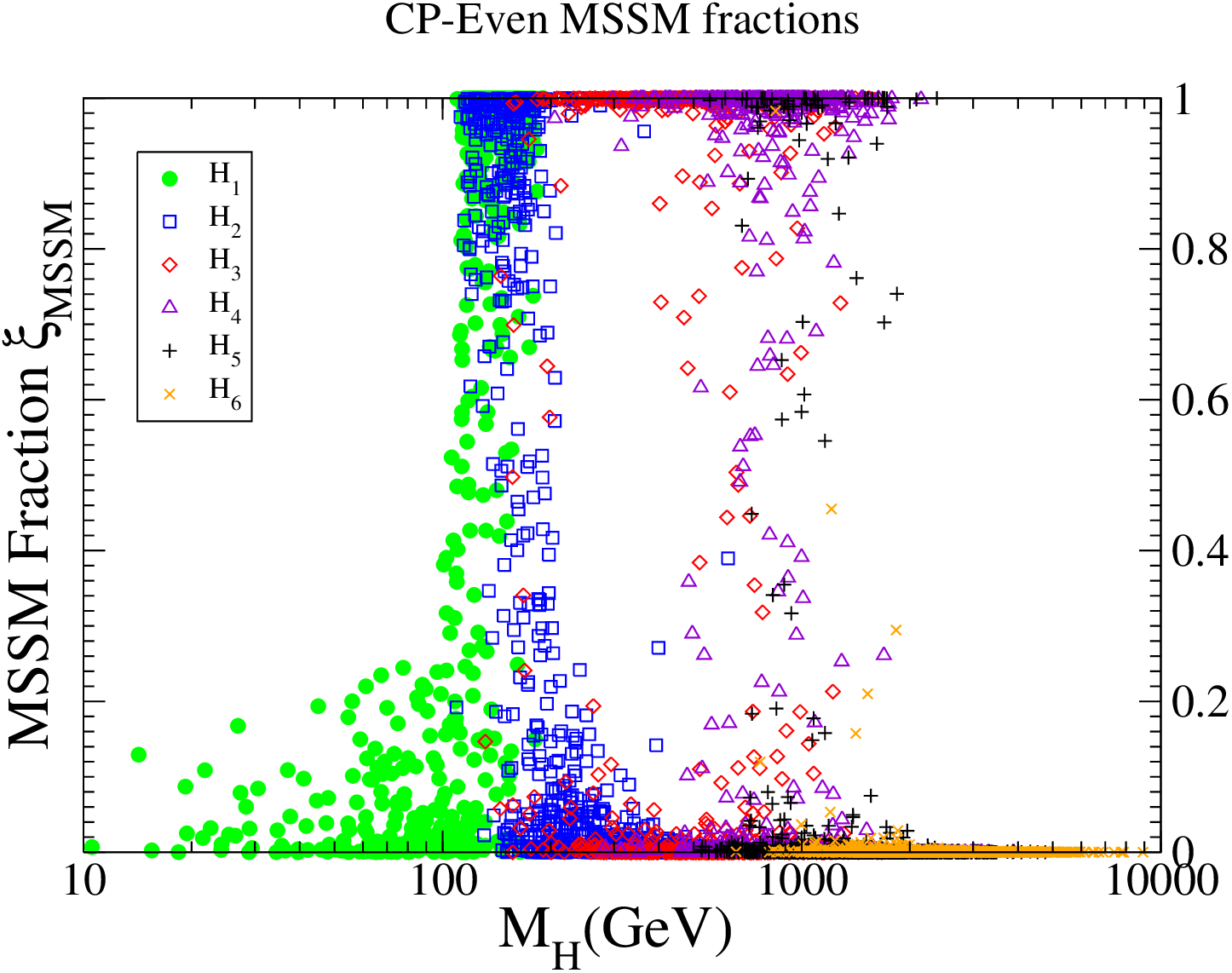}
    \includegraphics[scale=0.33]{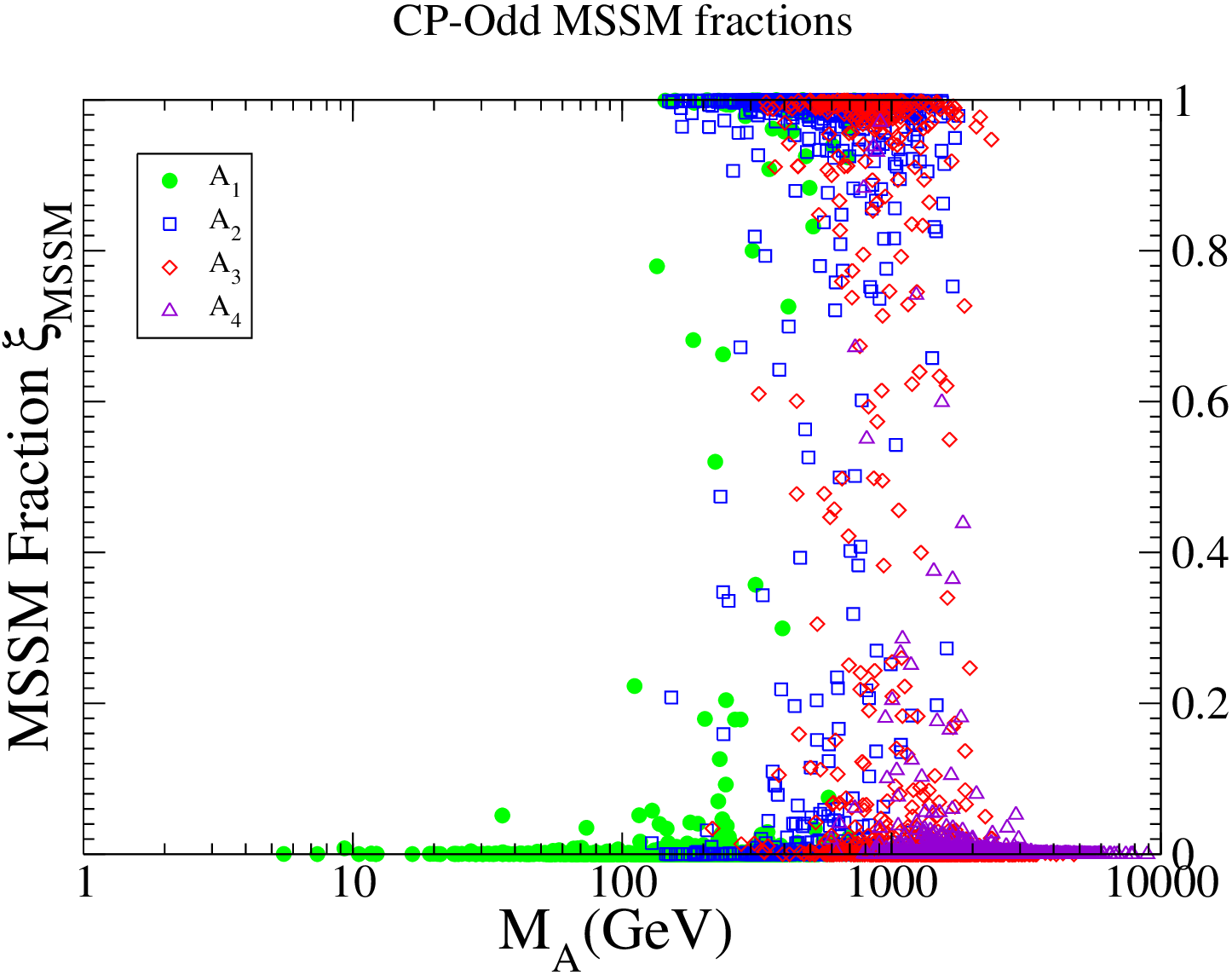}
    \caption{The MSSM fraction (a) for the CP-even and (b) for the CP-odd states.}
    \label{mssmfrac}
\end{figure}

\begin{figure}[tb]
    \includegraphics[scale=0.33]{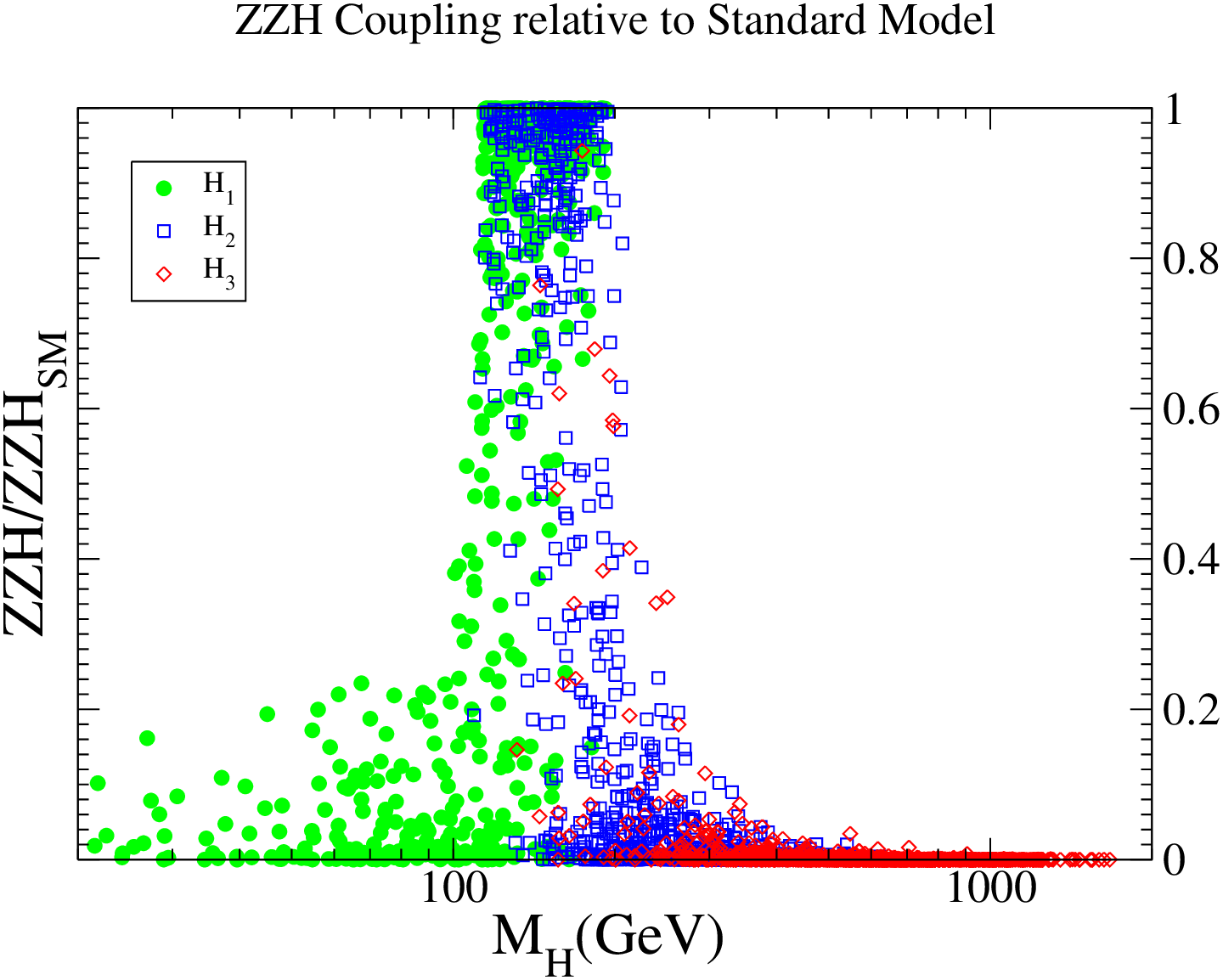}
    \includegraphics[scale=0.33]{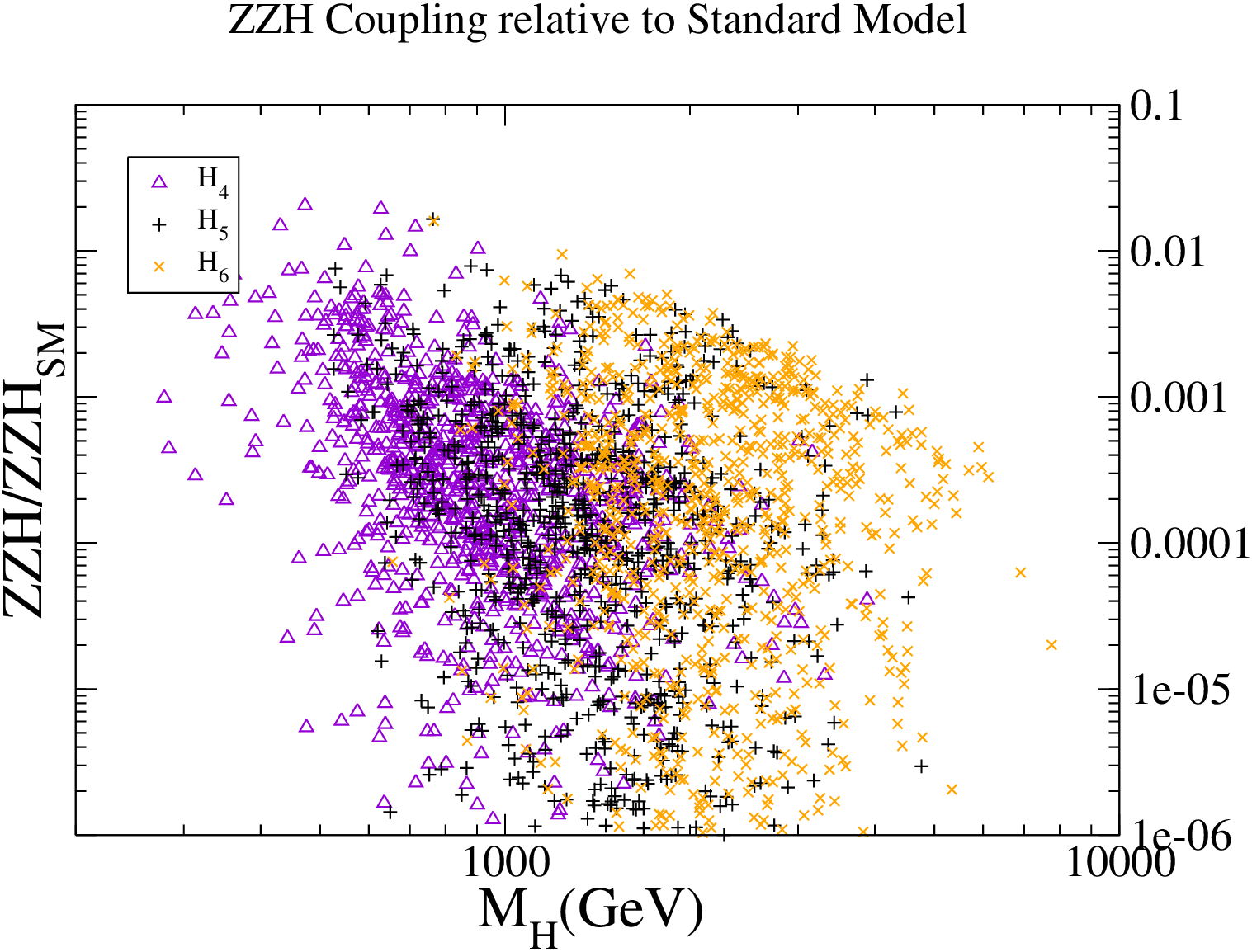}
    \caption {$ZZH$ coupling of the CP-even Higgs, relative to the
    Standard Model $ZZH$ coupling.}
    \label{ZZHcoupling}
\end{figure}

The MSSM fractions are shown versus the masses of  $H_i$ and $A_i$
in Fig.~\ref{mssmfrac}. It becomes more transparent that 
lighter Higgs bosons can be consistent with the
LEP bound as long as the MSSM fraction is less than about 0.2.
Another way to illustrate this is via the
$ZZH$ coupling relative to its SM value, as
shown in Fig.~\ref{ZZHcoupling}. We see that the
LEP2 bound for $M_H >114$ GeV is restored only for those $H_i$
states in which the couplings to $Z$ become substantial.
This figure
is remarkably similar to Fig. \ref{mssmfrac} because the $ZZH$ coupling is
$\sqrt{\xi_{\rm MSSM}} \cos(\alpha-\beta) \lambda_{\rm ZZH,SM}$ where
$\alpha$ is the angle that diagonalizes the CP-even mass matrix in the MSSM.

\begin{figure}[tb]
    \includegraphics[scale=0.33]{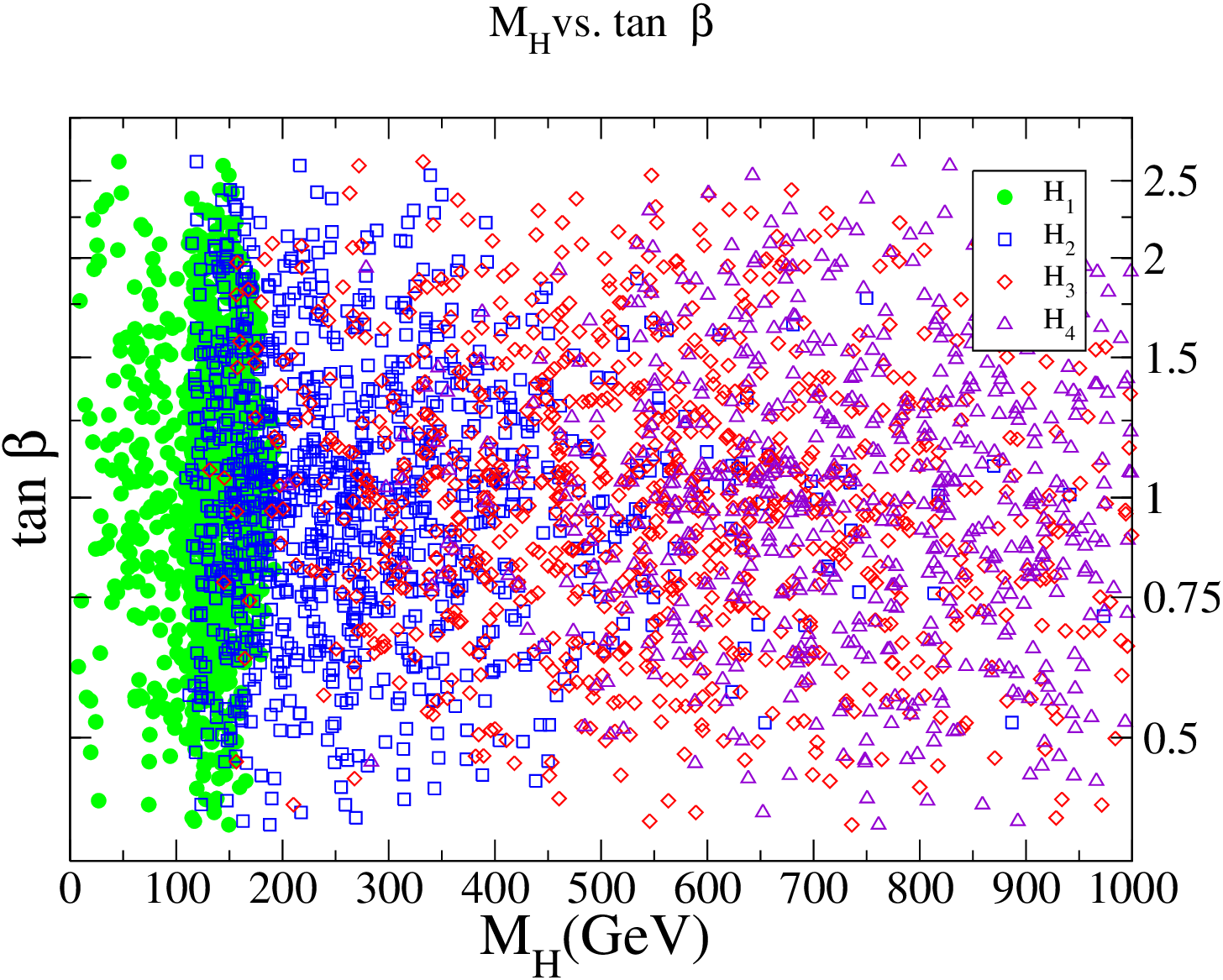}
    \includegraphics[scale=0.33]{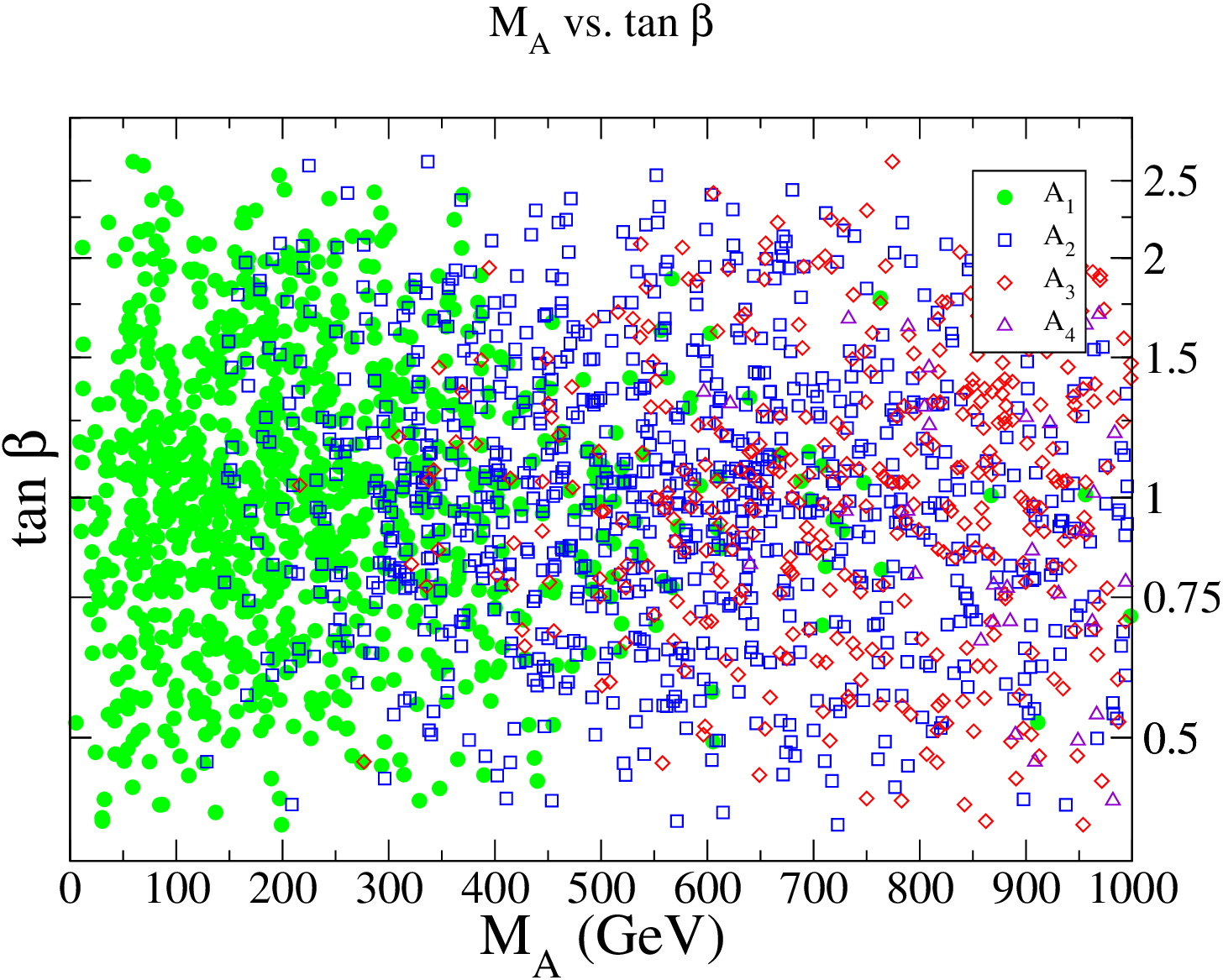}
    \caption{Range of $\tan\beta$ versus (a) the CP-even, and (b) CP-odd masses.}
    \label{tanbeta}
\end{figure}

One of the most important parameters in the SUSY Higgs sector
is $\tan\beta$. In the model under consideration, 
$\tan\beta \approx 1$ is favored (because $A_h$ must be large enough to 
ensure $SU(2)$ breaking). 
We show the value of $\tan\beta$ versus 
the allowed ranges of masses of  $H_i$ and $A_i$ in Fig.~\ref{tanbeta}.  
Though the model naturally favors $\tan\beta \approx 1$, there are
solutions deviating from this relation. The actual range reflects our
parameter scanning methodology shown in Table~\ref{model1params},
which results in $1/e < \tan\beta < e$.

%The remainder of the mass spectrum can be grouped roughly into several
%categories according to how many states share the MSSM couplings.
%    \begin{itemize}
%    \item{MSSM-like}
%    \item{MSSM-like with light singlet $H_1$}
%    \item{NMSSM-like (3 CP-even states share MSSM couplings)}
%    \item{NMSSM-like with light singlet $H_1$}
%    \item{4 Higgs' sharing the MSSM fraction}
%    \end{itemize}
%For this categorization we only consider the lightest states, since
%they modify the decay modes of the lightest Higgs with MSSM-like
%couplings (the one that presumably would be discovered first).  However
%the remaining singlets may be spaced either between the two MSSM-like
%states or above, or any combination thereof.  Analyzing the decays of
%the heavy MSSM-like Higgs introduces considerable model-dependence.
%Therefore we concentrate on the lowest-lying states only.

\section{Higgs Boson Decay and Production in $e^+e^-$ Collisions}
\label{decay}

Due to the rather distinctive features of the Higgs sector different
from the SM and MSSM, it is important
to study how the lightest Higgs bosons decay in order to explore
their possible observation at future collider experiments.
The lightest Higgs bosons can
decay to quite non-standard channels, leading to distinctive,
yet sometimes difficult experimental signatures.
For the Higgs boson production and signal observation, 
we concentrate on an $e^+e^-$  linear collider. It is known that
a linear collider can provide a clean experimental environment
to sensitively search for and accurately study new physics signatures.
If the Higgs bosons are discovered at the LHC, a linear collider would 
be needed to disentangle the complicated signals in this class of 
models. If, on the other hand, a Higgs boson is not observed at the LHC
due to the decay modes difficult to observe at the hadron
collider environment, a linear collider will serve as a discovery machine.

\subsection{Lightest CP-Even State $H_1$}

\begin{figure}[tb]
    \includegraphics[scale=0.65]{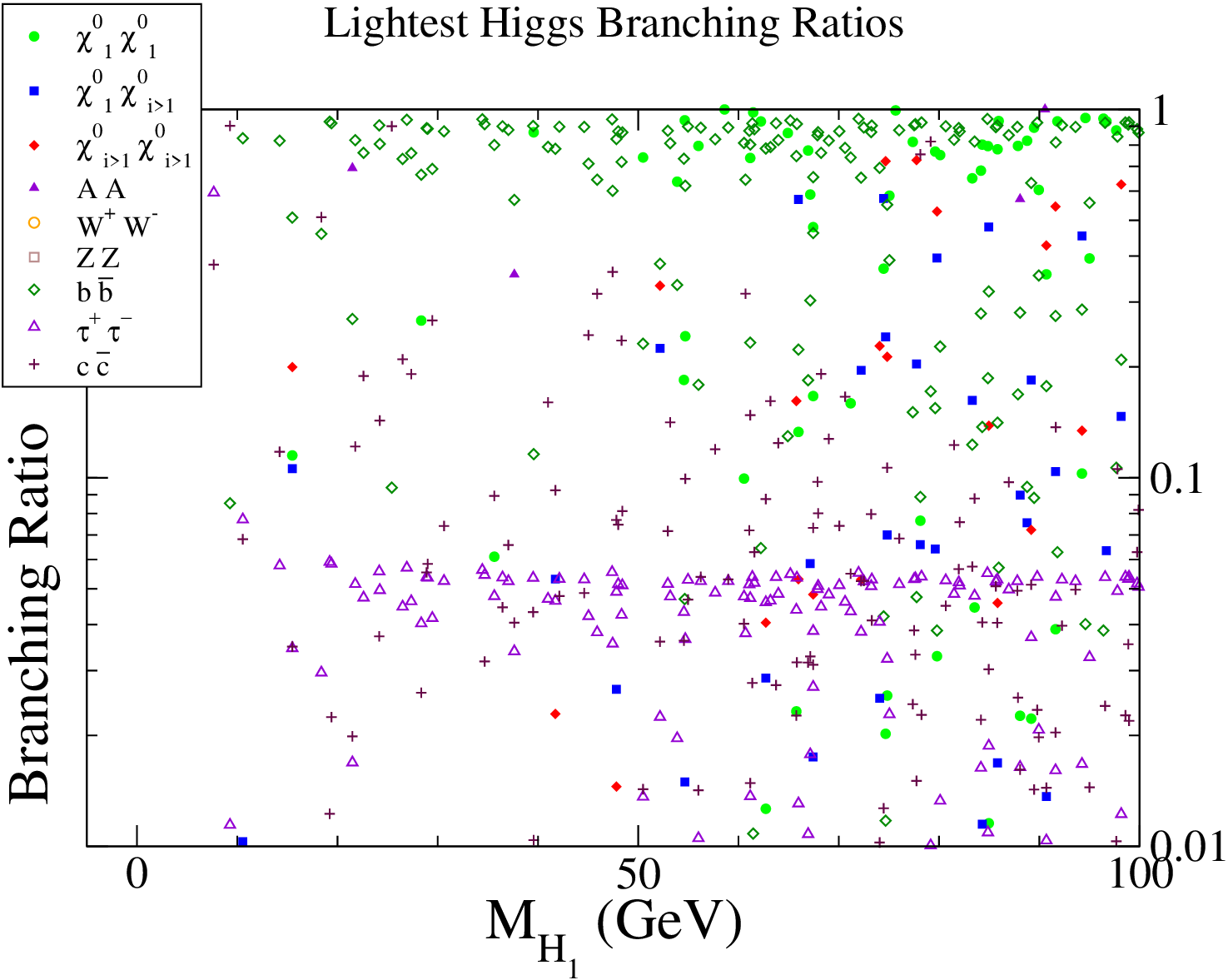}
    \caption {Branching ratios of the lightest CP-even Higgs in the low
    mass region $M_{H_1} < 100$ GeV}
    \label{lightestbrs}
\end{figure}

The main decay modes and corresponding branching fractions 
for the lightest CP-even Higgs $H_1$ are presented in Fig.~\ref{lightestbrs}. 
For lightest Higgs masses below approximately 100 GeV, 
the LEP2 constraint is very tight, and the lightest Higgs must be mostly singlet.  
Thus, the
decay modes to $A_1 A_1$ and $\chi^0_1 \chi^0_1$ are dominant when they
are kinematically allowed,
due to the presence of the extra $U(1)^\prime$ gauge coupling and trilinear 
superpotential terms proportional to $h$ and $\lambda$.  
When those modes are not kinematically accessible, the decays are very
similar to the MSSM modulo an eigenvector factor that is essentially how
much of $H_2$ and $H_1$ are in the lightest state.  Therefore $b
\bar{b}$, $c \bar{c}$ and $\tau^+ \tau^-$ decays dominate, with $c
\bar{c}$ and $\tau^+ \tau^-$ approximately an order of magnitude smaller
than $b \bar{b}$, due to the difference in their Yukawa couplings.  
Examples of this kind can be seen in Appendices~(\ref{h1smdominant},~\ref{maxamix}).
Since $\tan \beta \approx 1$, the $c \bar{c}$ mode can be competitive with
both $\tau^+ \tau^-$ and $b \bar{b}$ since their masses are similar.  
In the MSSM the $c
\bar{c}$ mode is suppressed because $\tan \beta$ is expected to be
larger.

\begin{figure}[tb]
    \includegraphics[scale=0.65]{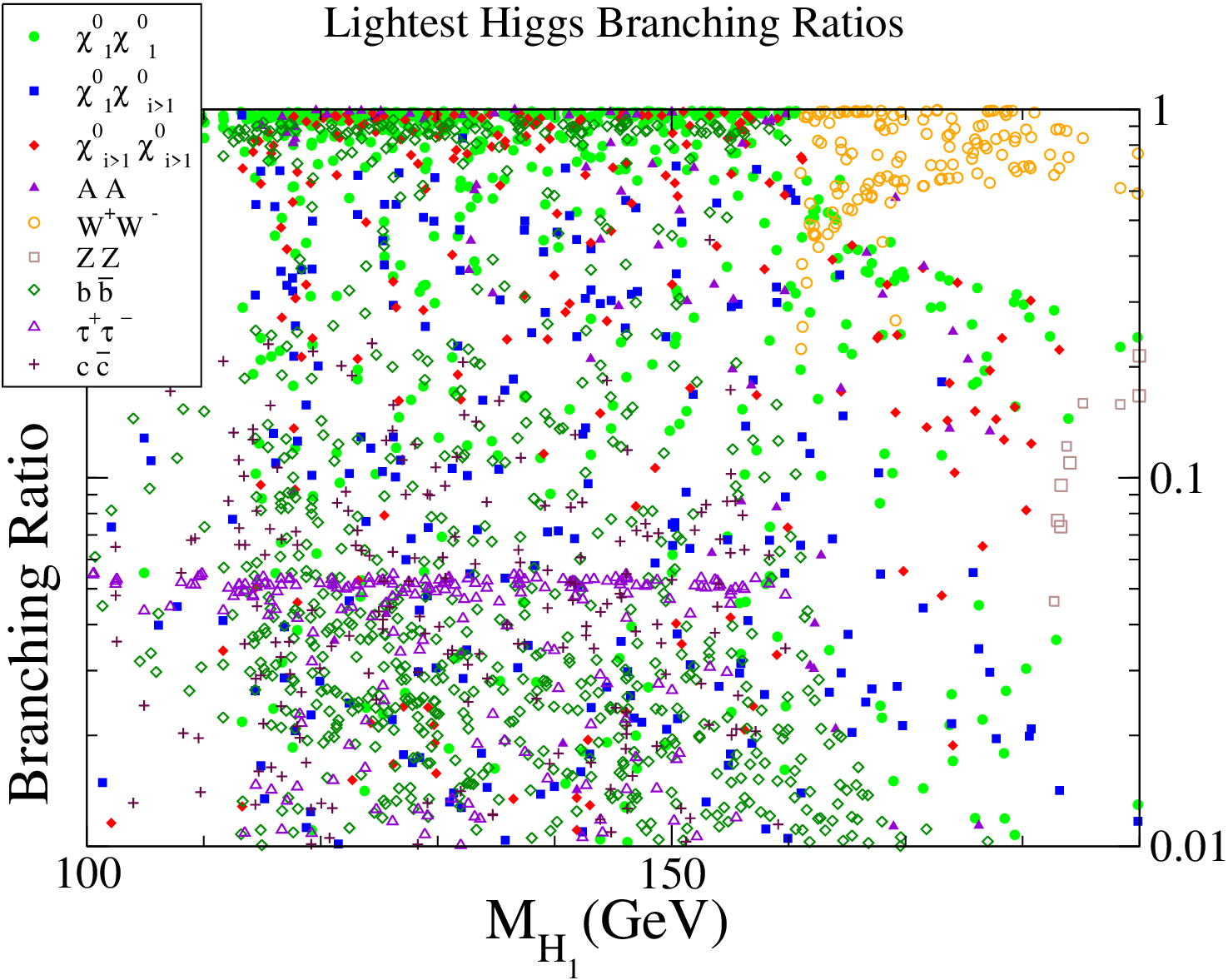}
    \caption {Branching ratios of the lightest CP-even Higgs in the high
    mass region $M_{H_1} > 100$ GeV}
    \label{lightestbrs_high}
\end{figure}

When the lightest Higgs is heavier than the LEP2 bound, it does not
need to be mostly singlet, and there can be a continuum of branching
ratios to $A_1 A_1$, $\chi_1^0 \chi_1^0$ or SM particles, depending
on how much singlet is in the lightest state.  This is indeed seen in 
 Fig.~\ref{lightestbrs_high} for a heavier $H_1$ where the modes 
 $H_1\to W^+ W^-,\ ZZ$ become substantial. 

A striking feature of this graph is that the usual ``discovery'' modes for
$M_{H_1} < 140$, $H_1 \rightarrow b \bar{b},\  \tau^+ \tau^-$ 
are often strongly suppressed by decays to $A_1$ and $\chi^0_1$. 
Only $H_1 \rightarrow W^+ W^-,\  Z Z$ decays are able to
compete with the new $A_1$ and $\chi_1^0$ decays, which are all of gauge
strength.  A striking example of this is Appendix~(\ref{a1gaugino}) 
and (\ref{h1a1a1dominant}).  One can see that the
traditional shape of the $W^+ W^-$ and $ZZ$ threshold is obscured by the
presence of $\chi_1^0$ and $A$ decays, depending on what is kinematically
accessible.  For a $H_1$ heavy enough for these decay modes to be open,
however, the coupling $h$ is typically greater than 0.8, large enough that it will become
non-perturbative before the Planck scale unless new thresholds enter at a
lower scale to modify its running.  Such examples can be seen in
Appendices (\ref{h1lightest},~\ref{h1a1a1dominant},~\ref{a1invisible}).

The $A_1$ or $H_1$ can be lighter than the $\chi^0_1$. However, we assume
R-parity is conserved. Therefore, decays of $\chi^0_1$ to $A_1$ or $H_1$ are
not allowed and the lightest neutralino is  assumed to
be the (stable) LSP.  We do not analyze the sfermion sector,
which can produce a sfermion LSP in some regions of parameter space, but
these scenarios are phenomenologically disfavored.  We therefore assume
$H$ and $A$ decays to $\chi^0_1$ are invisible at a collider.  We
separate the heavier neutralinos $\chi^0_{i>1}$ which may decay 
visibly~\cite{neutralinodecays}.

\subsection{CP-Odd}

\begin{figure}[tb]
    \includegraphics[scale=0.65]{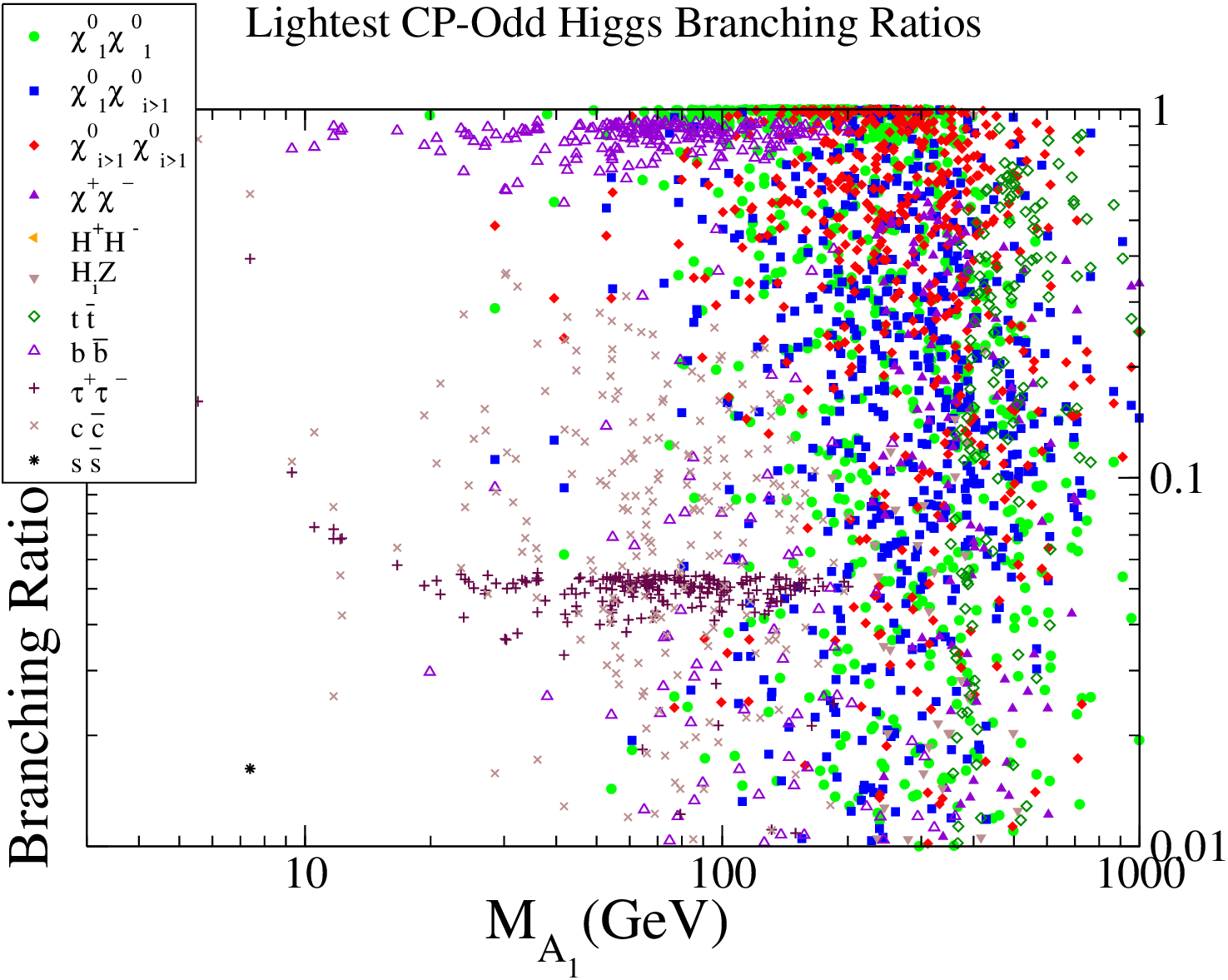}
    \caption {Branching ratios for the lightest CP-odd Higgs}
    \label{BRma}
\end{figure}

The decays of the CP-odd Higgs bosons are presented in Fig.~\ref{BRma}. 
The light $A_1$ will decay dominantly to neutralinos when it is
kinematically possible.  When it is not, it decays dominantly into the
nearest mass SM fermion, which is usually $b$ unless the $A_1$ is
lighter than the $b \bar{b}$ pair mass.  Charm and tau decays can also
be significant, depending on the value of $\tan \beta$.  The $c \bar c$
decays are about 3 times more likely than the $\tau^+ \tau^-$ due to the
color factor.  However, for larger $\tan \beta$ the $\tau^+ \tau^-$
dominates.

For heavy $A_1 \gsim 200$ GeV, decays to neutralinos and charginos
universally dominate due to their gauge strength, suppressing the $b
\bar b$ mode below 10\%.

The lightest $A$ can decay only into light SM fermions, the
photon, and neutralinos.  Hadronic
bottom and charm decays are difficult to separate from background, and
$\tau$'s are obscured by missing energy and hadronic background.

\subsection{The Higgs signatures at a linear collider}
\label{pheno}

\begin{figure}[tb]
    \includegraphics[scale=0.33]{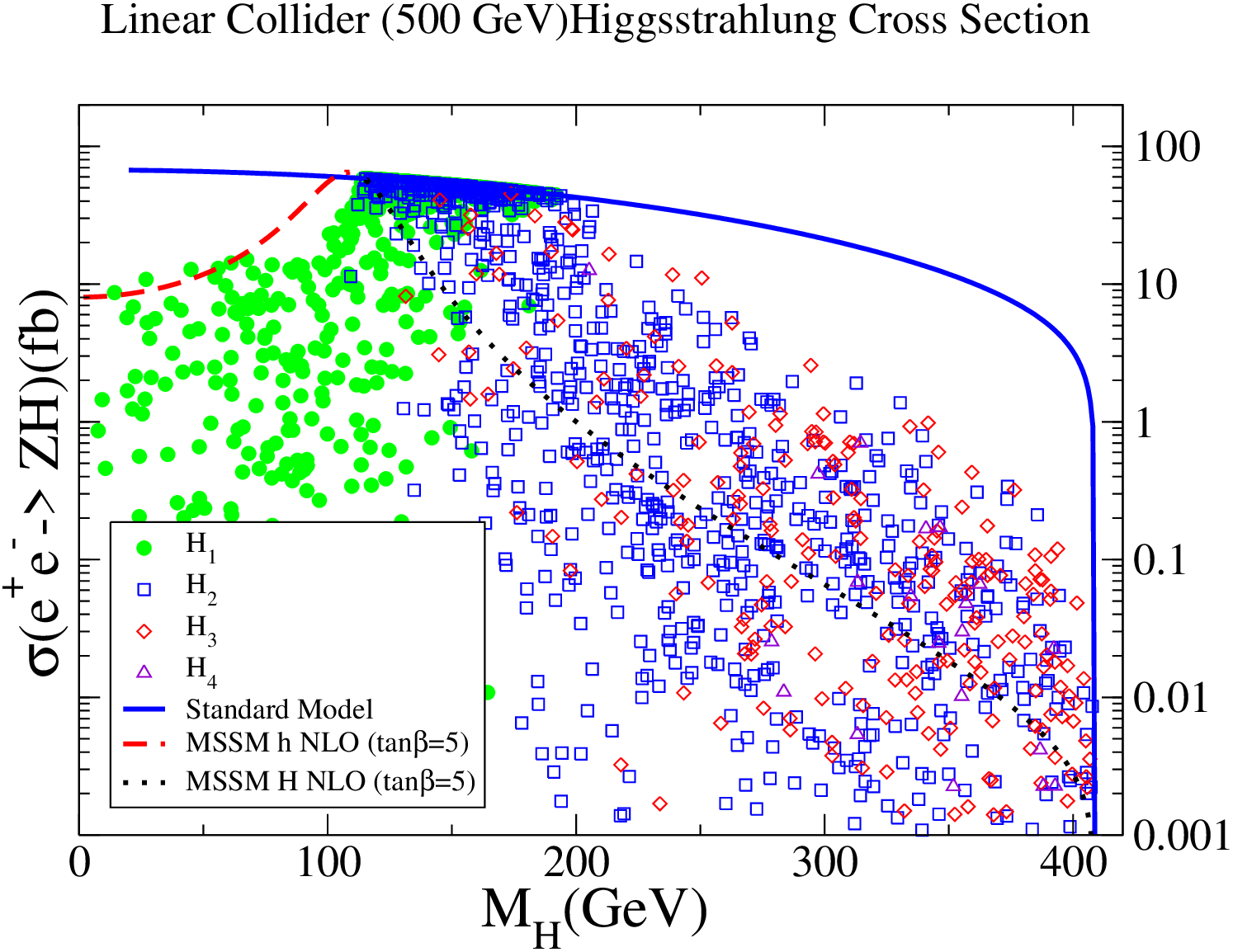}
    \includegraphics[scale=0.33]{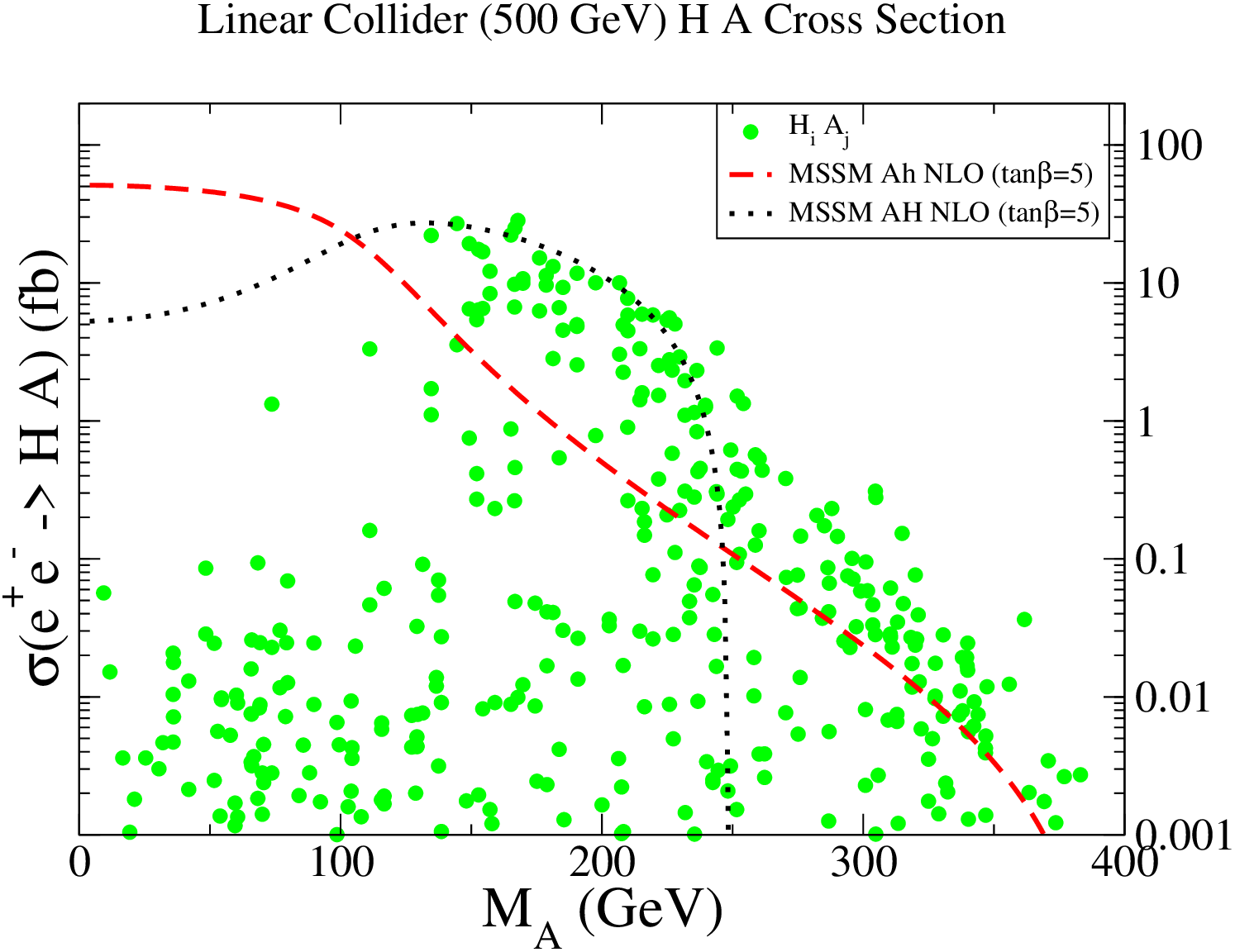}
    \caption {Cross section at a 500 $\gev$ linear
    collider (a) for $ZH_i$ production, and (b) for $A_i H_j$ production 
versus the corresponding Higgs boson mass.  In (a), the solid curve 
is the SM production, and the dashed and
dot-dashed for MSSM $h$ and $H$ production with $\tan\beta=5$.}
    \label{Higgsstrahlung}
\end{figure}

\begin{figure}[tb]
    \includegraphics[scale=0.33]{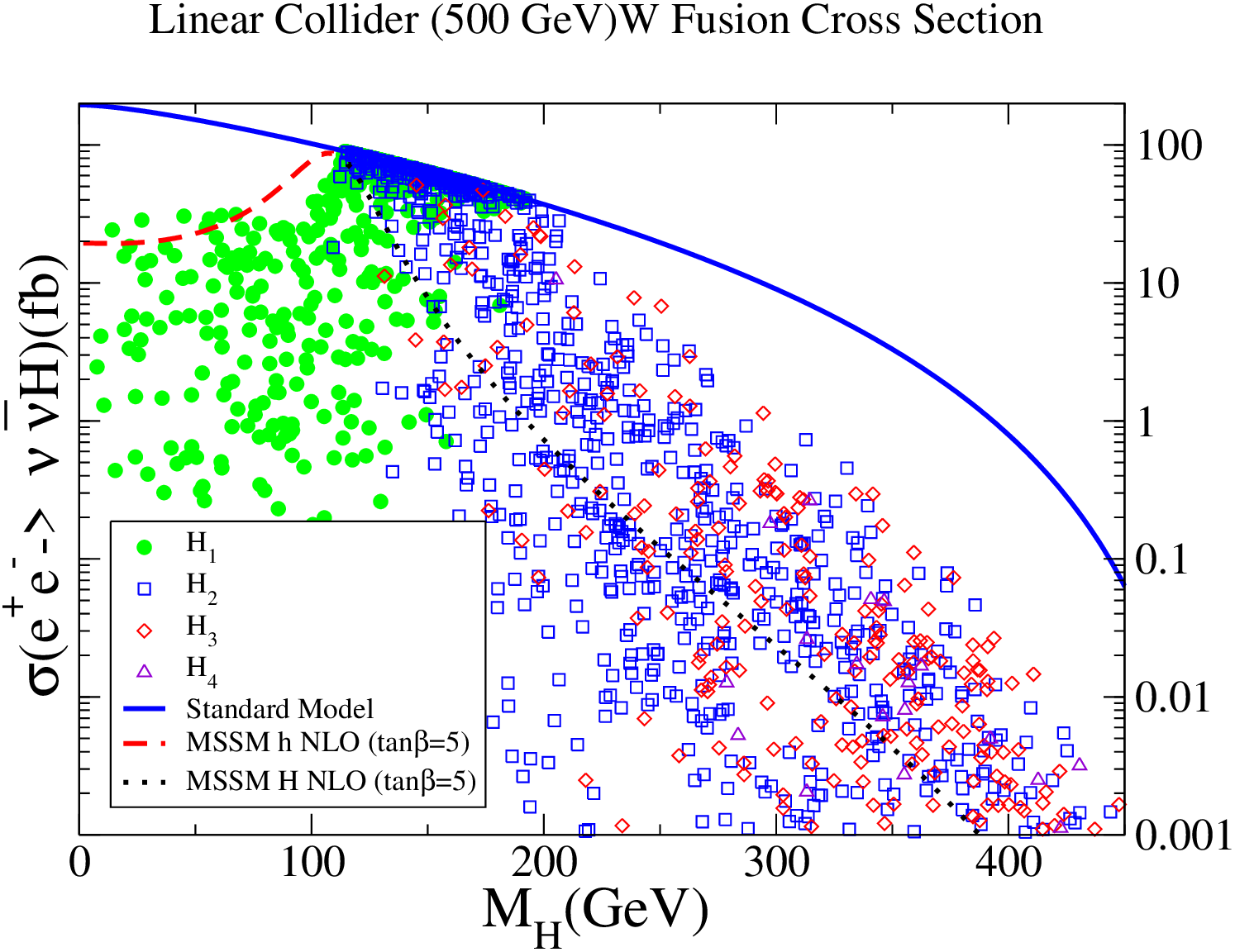}
    \includegraphics[scale=0.33]{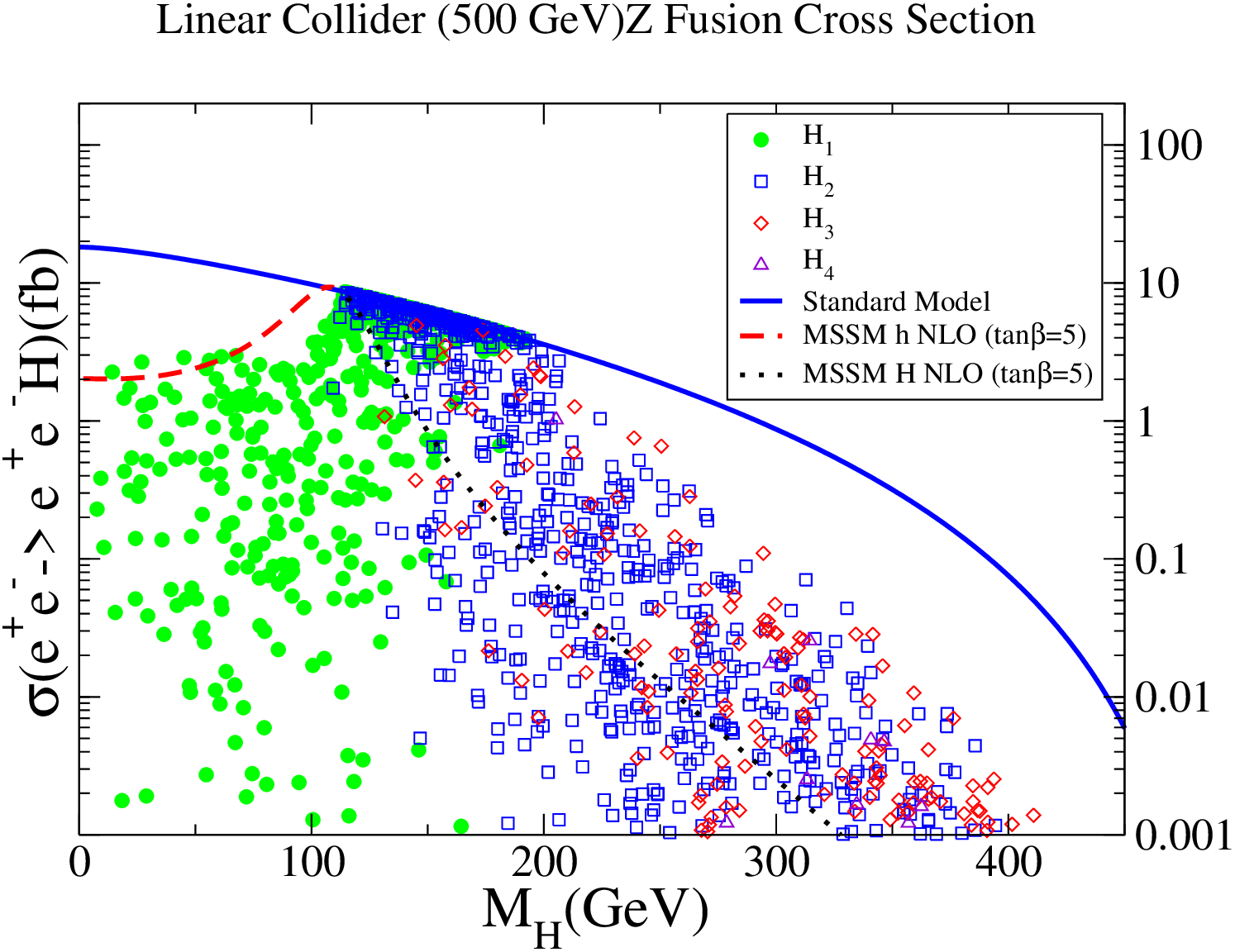}
    \caption {Cross section at a 500 $\gev$ linear
    collider for $H_i$ production in (a) $WW$ fusion, and (b) $ZZ$ fusion
versus the corresponding Higgs boson mass.  The solid curves
are the SM production, and the dashed and
dot-dashed for MSSM $h$ and $H$ production with $\tan\beta=5$.
In both graphs, interference from the $Z\rightarrow \nu \overline{\nu}$ and
$Z\rightarrow e^+e^-$ decays are neglected.}
    \label{WBF}
\end{figure}

The production via radiation of a Higgs from a virtual $Z$ boson is the
dominant mechanism for CP-even Higgs production at a linear collider.
We show this cross section in Fig. \ref{Higgsstrahlung}, where each point
is a viable model solution satisfying all the constraints.
The curves present the SM and MSSM cross sections for comparison.  
Model points with $M_H < 114.1$ are only those with suppressed
coupling to the $Z$, and those with large MSSM fraction are
removed by the LEP2 bounds discussed in Section \ref{hzha}.
As can be seen from Eq.~(\ref{hzxsect}) the ratio between the Standard
Model cross section and that for any model point simply reflects the
amount of mixing into the SM-like or MSSM-like Higgs for a given Higgs
state.  Since this ratio comes entirely from the $HZZ $ vertex,
Fig.~\ref{Higgsstrahlung}(a) and Fig.~\ref{WBF} are just 
Fig.~\ref{ZZHcoupling} times the SM curve, as
given in Eq.~(\ref{hzxsect}).

The production cross sections for the heavier Higgs particles are very small.
One can see the coupling to $ZZH$ in Fig.~\ref{ZZHcoupling}.  
For heavy states (that correspond to  the $H$ in MSSM),
$\cos(\alpha-\beta) \rightarrow 0$ as the $H$ gets heavier.
In this decoupling limit of the MSSM the heavy $H$ has
no coupling to the $Z$.

In supersymmetric models if both the $A$ and $H$ are light enough they
can be produced by the process $e^+ e^- \rightarrow H A$ at a lepton
collider.  We present this cross section in Fig.~\ref{Higgsstrahlung}(b).  In this
model the $HA$ cross sections do not provide additional constraint, 
and few model points are removed, unlike
the $HZ$ cross sections.  The $H A$ cross section is normally much
smaller than the $H Z$ cross section unless the center of mass energy is above and 
close to $M_H + M_A$.  As can be seen in Fig.~\ref{Higgsstrahlung}(b), 
the cross section is largest in this channel when $M_A \simeq M_H \simeq
\sqrt{s}/2$, and they have large $\xi_{\rm MSSM}$.
This is confirmed by seeing the MSSM curves as shown in 
 Fig.~\ref{Higgsstrahlung}(b).  

At 500 GeV the weak boson fusion production modes $e^+ e^- \rightarrow
\nu \bar{\nu}H,\ e^+e^- H$ 
as shown in Fig.~\ref{WBF} are comparable in size to the
Higgsstrahlung mode.  At higher energies, the weak boson fusion becomes
larger than Higgsstrahlung and is the most important production mode.
These curves are similar to Fig.~\ref{Higgsstrahlung}(a), reflecting
that all of these single Higgs production modes are simply a mixing
factor times the Standard Model curve.  It is particularly interesting
to note that the $ZZ$ fusion channel $e^+e^- \to e^+e^- H$ can serve as a
model-independent process to measure the $ZZH$ coupling regardless the
decay of $H$, even if $H$ is invisible \cite{Gunion:1998jc}.

As anticipated for the next generation linear collider with $\sqrt s=500$ GeV
and an integrated luminosity of the order of $500-1000$ fb$^{-1}$, one should
be able to cover a substantial region of the parameter space. For instance, with
a cross section of the order of 0.1 fb, this may lead to about 50$-$100 events.
As for further exploration of signal searches, it depends on specific model
parameters. While we have provided a comprehensive list of representative
models in the Appendices, we discuss a few of them for the purpose of 
illustration.

\begin{itemize}
    \item{\bf MSSM-like:} 
Examples of this type are presented in Appendices~(\ref{h1wwdominant},~
\ref{h1smdominant},~\ref{lightcharged},~\ref{maxamix},~\ref{h1heaviest}).  When
the MSSM fractions are close to one, the model is MSSM-like and their
mass relations approximately hold.  The standard MSSM or SM discovery
modes are present for the lightest CP-even state, even if with reduced
rates. As long as the Higgs boson mass is not nearly degenerate with
$M_Z$, the signal observation should be quite feasible at the LC, as
well as at the LHC.
    \item{\bf $H \rightarrow$ multi-jets:} 
Examples of this type are presented in Appendices~(\ref{a1lightest},~
\ref{h1a1a1dominant}). Certain parameters may lead to the dominant
decay modes $H_i \rightarrow H_j H_j$ or $H_i \rightarrow A_j A_j$ with
the $H_j$ or $A_j$ decaying hadronically, or to $\tau$'s.  This scenario
would make the signal search nearly impossible at hadron colliders due
to the overwhelming QCD backgrounds.  This is the typical difficult
scenario studied for the NMSSM~\cite{nmssmpheno}.  At a LC, however, the
reconstruction of the Higgs mass peak from jets is still possible. In
particular, if the Higgsstrahlung process yields a sizable cross
section, the signal could be picked up from the recoiled mass against
the distinctive signature of $Z\to e^+e^-, \mu^+\mu^-$.
    \item{\bf Invisible:} 
Examples of this type are presented in Appendices (\ref{a1gaugino},
\ref{a1invisible}, \ref{a1heaviest}).  MSSM and SM detection modes
are heavily suppressed by $H_i\rightarrow \chi^0 \chi^0, A_j A_j, H_j
H_j$ and dominant decays eventually produce neutralinos.  It is possible
to discover a Higgs in this mode at the LHC if its cross section is
large enough~\cite{Cavalli:2002vs}. 
At an  $e^+e^-$ linear collider on the other hand,
the Higgsstrahlung process with $Z\to \ell^+\ell^-$
and the $ZZ$ fusion process may yield
a sizable cross section and can make accurate measurements of this
invisible decay channel by detecting the recoiling $\ell^+\ell^-$ 
plus large missing energy.
    \item{\bf Neutralino:}
Examples of this type are presented in the Appendices (all appendices
have an example of this type).  One or more Higgs decays into heavy
neutralinos, which can then decay via cascade to the LSP, producing visible
signals and large missing energy~\cite{neutralinodecays}.  
If the lightest neutralino is mostly singlino or $Z^\prime$-ino, the
usual MSSM limits on neutralino mass do not apply, and couplings between
it and Higgs bosons can be large.  This mode is
extremely important, and dominates the Higgs decays due to the presence
of Higgs-singlet interaction $h$, singlet-singlet interactions 
$\lambda$ and the $Z^\prime$ coupling.  This mode has received some
attention in the literature~\cite{singlinopheno} but clearly warrants
more due to its dominance in parameter space.
\end{itemize}

It is clear that the model studied in this paper presents very rich physics
in the Higgs sector. An $e^+e^-$ linear collider will be ideally
suited for the detailed exploration of the non-standard Higgs physics.
Analyses for the LHC should also be performed, particularly for the
non-MSSM modes~\cite{nmssmpheno}.

\section{Summary and Conclusions}
\label{concl}

We have considered the Higgs sector in an extension of the MSSM with extra
SM singlets. By exploiting an extra \upr \ gauge symmetry, the domain-wall
problem is avoided. 
An effective $\mu$ parameter can be generated by a singlet VEV, which can
be decoupled from the new gauge boson \zpr \ mass.

The model involves a rich Higgs structure very different from that of the MSSM.
In particular, there are large mixings between Higgs doublets and SM singlets.
The lightest CP even Higgs boson can have a mass up to about 170 GeV.
Higgs bosons  considerably lighter than the LEP2 bound  are allowed.
The parameter $\tan \beta \sim 1$ is both allowed and theoretically favored. 

We parameterize the Higgs coupling strengths relative to the MSSM, called
the MSSM fraction  $\xi_{\rm MSSM}$.
We find that besides the typical SM-like and MSSM-like Higgs bosons,
there are model points leading to very different signatures from those. 
One of the features for the Higgs
boson decay is to have possibly a large invisible decay mode to LSP.
We present a comprehensive list of model scenarios in the Appendices.

Concentrating on a future $e^+e^-$ linear collider with $\sqrt s=500$ GeV,
we found that in a large parameter region the Higgs bosons are
accessible through the production channels $e^+ e^-\to ZH_i,\ H_i A_j$
as well as $WW$ and $ZZ$ fusion.
We outlined the searching strategy for some representative scenarios
at a future linear collider.

We find that this model has a large parameter space where the Higgs bosons
decay hadronically or invisibly.  As these modes are very difficult at
the LHC, effort should be invested in ways to discover or exclude such
modes at the LHC.  If discovery is not possible, a linear collider will
absolutely be required.

We emphasize the importance of neutralino decays, which dominate the
parameter space due to Higgs self-interactions and the $U(1)^\prime$
gauge coupling.  These decays are generically present and dominant in
extended models~\cite{singlinopheno}, and thus should be paid more
attention by phenomenologists and experimentalists.

\vskip 0.5in
\vbox{
\noindent{ {\bf Acknowledgments} } \\
\noindent
We thank V. Barger, J. Erler, J. Gunion, T. Li, and J. Wells, for useful
discussions.
This work was supported in part by DOE grants
DE--FG02--95ER--40896, DE--FG03--91ER--40674 and 
DOE--EY--76--02--3071; 
also in part by the Wisconsin Alumni Research Foundation, 
the National Natural Science Foundation of China, the 
Davis Institute for High Energy Physics, and the U.C. Davis Dean's office.}

\pagebreak
\renewcommand{\theequation}{A.\arabic{equation}}
\renewcommand{\thesection}{A-\arabic{section}}
% redefine the command that creates the equation no.
\setcounter{equation}{0}  % reset counter 
\setcounter{section}{0}
    % This is DATA[726]
\section{Lightest $A_1$}
\label{a1lightest}
\begin{flushleft}
\begin{tabular}[b]{l|c|c|c|c|c|c|c|c|c}
  \multicolumn{2}{c|}{$\tan{\beta} = 0.522$} &   \multicolumn{2}{c|}{$M_{Z^\prime} = 2415$ GeV} &   \multicolumn{2}{c|}{$M_{H^+} = 826$ GeV} &   \multicolumn{3}{c}{$\alpha_{ZZ^\prime} = $-$2.7\!\cdot\!10^{-4}$} \\
\hline 
  $M_{H}$ & 118 & 654 & 843 & 1731 & 2330 & 7839 &&& \\
  $\xi_{\rm MSSM}$ & 1 & $1.2\!\cdot\!10^{-3}$ & 1 & $3.3\!\cdot\!10^{-3}$ & $2.4\!\cdot\!10^{-4}$ & 0 &&& \\
  $\sigma(H_i Z)$ &  58 & & & & & & & & \\
  $\sigma(H_i \nu\overline{\nu})$ &  87& & & & & & & & \\
  $\sigma(H_ie^+e^-)$ & 8.3& & & & & & & & \\
\hline 
  $M_{A}$ & 6 & 821 & 1741 & 7839 & 0 & 0 &&& \\
  $\xi_{\rm MSSM}$ & $3.2\!\cdot\!10^{-4}$ & 0.99 & $5.9\!\cdot\!10^{-3}$ & 0 & 1 & $5.4\!\cdot\!10^{-4}$ &&& \\
$\sigma(H_1 A)$  & $4.8\!\cdot\!10^{-6}$ & & & & & & & & \\
\hline 
  $M_{\chi^0}$               & 42 & 165 & 213 & 284 & 470 & 774 & 778 & 1834 & 3222 \\
  $\xi_{\rm MSSM}$          & 0.24 & 0.95 & 0.81 & 1 & 1 & 0 & $2.1\!\cdot\!10^{-5}$ & $9.7\!\cdot\!10^{-5}$ & $4.5\!\cdot\!10^{-5}$ \\
  $\xi_{\tilde{s}}$         & 0.76 & 0.047 & 0.19 & $1.1\!\cdot\!10^{-3}$ & $9.2\!\cdot\!10^{-5}$ & 0.99 & 0.99 & 0.65 & 0.37 \\
  $\xi_{\tilde{Z^\prime}}$ & 0 & 0 & 0 & 0 & 0 & $5.3\!\cdot\!10^{-3}$ & 0.012 & 0.35 & 0.63 \\
\hline 
  $M_{\chi^+}$ & 173 & 470 &&&&&&& \\
\end{tabular}
\end{flushleft}
\vspace{-10pt}
Cross sections quoted are in fb for a linear $e^+ e^-$ collider at center-of-mass energy 500 GeV.\vspace{-10pt}
\begin{eqnarray*}
  v_2       = 80 \gev   & v_1       = 154 \gev   & v_s       = 272 \gev   \\
               v_{s1}    = 117 \gev   & v_{s2}    = 3504 \gev   & v_{s3}    = 3256 \gev   \\
               m_{H_u}^2 = (849 \gev)^2 & m_{H_d}^2 = (595 \gev)^2 & m_{S}^2   = (1585 \gev)^2 \\
               m_{S_1}^2 = (7817 \gev)^2 & m_{S_2}^2 = (496 \gev)^2 & m_{S_3}^2 = -(1115 \gev)^2 \\
               h = 0.608                 & A_h       = 1704 \gev   & \mu = h v_s = 166 \gev \\
               \lambda = 0.173          & A_\lambda = 3622 \gev  & \\
               M_1      = -267  \gev   & M_1^\prime = 1385 \gev & M_2 = 459 \gev \\
               m_{S S_1}^2 = -(48 \gev)^2 & m_{S S_2}^2 = -(479 \gev)^2 &
\end{eqnarray*}
Branching Ratios for dominant decay modes (greater than 1\% excluding model-dependent squark, slepton, $Z^\prime$ and exotic decays; $\chi^0_{i>1}$ are summed): 
\begin{flushleft}\begin{tabular}{c|lr|lr|lr|lr|lr|lr}
$H_1$ & $A_1 A_1$ & 81\% & $\chi^0_1 \chi^0_1$ & 18\% & $b\bar{b}$ & 1\% &&&&&&\\
$H_2$ & $\chi^+_1 \chi^-_1$ & 25\% & $\chi^0_1 \chi^0_{i>1}$ & 17\% & $W^+W^-$ & 14\% & $\chi^0_{i>1} \chi^0_{i>1}$ & 14\% & $t\bar{t}$ & 13\% & $H_1 H_1$ & 11\% \\
$H_3$ & $t\bar{t}$ & 77\% & $\chi^0_{i>1} \chi^0_{i>1}$ & 8\% & $\chi^0_1 \chi^0_{i>1}$ & 8\% & $\chi^0_1 \chi^0_1$ & 3\% & $\chi^+_1 \chi^-_2$ & 2\% & $H_1 H_1$ & 1\% \\
$H_4$ & $A_1 A_2$ & 27\% & $\chi^+_1 \chi^-_1$ & 23\% & $\chi^0_{i>1} \chi^0_{i>1}$ & 18\% & $H_1 H_3$ & 7\% & $\chi^0_1 \chi^0_{i>1}$ & 6\% & $W^+W^-$ & 4\% \\
$H_5$ & $W^+W^-$ & 24\% & $H_1 H_1$ & 24\% & $Z Z$ & 12\% & $H_2 H_2$ & 9\% & $\chi^0_{i>1} \chi^0_{i>1}$ & 7\% & $\chi^+_1 \chi^-_1$ & 6\% \\
$H_6$ & $\chi^0_{i>1} \chi^0_{i>1}$ & 95\% & $H_2 H_5$ & 3\% & $H_2 H_2$ & 1\% &&&&&&\\
\hline
$A_1$ & $c\bar{c}$ & 83\% & $\tau^+ \tau^-$ & 16\% & $s\bar{s}$ & 1\% &&&&&&\\
$A_2$ & $t\bar{t}$ & 79\% & $\chi^0_{i>1} \chi^0_{i>1}$ & 9\% & $\chi^0_1 \chi^0_1$ & 6\% & $\chi^0_1, \chi^0_{i>1}$ & 5\% & $\chi^+_1 \chi^-_1$ & 1\% &&\\
$A_3$ & $H_2 Z$ & 97\% & $\chi^+_1 \chi^-_1$ & 1\% & $\chi^0_{i>1} \chi^0_{i>1}$ & 1\% &&&&&&\\
$A_4$ & $H_2 Z$ & 64\% & $H_5 Z$ & 31\% & $\chi^0_{i>1} \chi^0_{i>1}$ & 3\% & $A_1 H_2$ & 1\% &&&&\\
\end{tabular}
\end{flushleft}
\newpage

Eigenvectors/rotation matrices 
 \[R_H = 
 \left[ \begin {array}{cccccc}\scriptstyle  0.45&\scriptstyle 0.89&\scriptstyle 0.038&\scriptstyle 0.00063&\scriptstyle 0.0045&\scriptstyle
 0.014\\\noalign{\medskip}\scriptstyle 0.022&\scriptstyle\scriptstyle- 0.026&\scriptstyle\scriptstyle 0.086&\scriptstyle\scriptstyle 0.045&\scriptstyle\scriptstyle 0.90&\scriptstyle\scriptstyle 0.42
\\\noalign{\medskip}\scriptstyle 0.89&\scriptstyle- 0.45&\scriptstyle 0.043&\scriptstyle- 0.0016&\scriptstyle- 0.031&\scriptstyle- 0.017
\\\noalign{\medskip}\scriptstyle 0.056&\scriptstyle 0.011&\scriptstyle- 0.99&\scriptstyle 0.0066&\scriptstyle 0.023&\scriptstyle 0.15
\\\noalign{\medskip}\scriptstyle 0.0096&\scriptstyle 0.012&\scriptstyle- 0.13&\scriptstyle- 0.021&\scriptstyle 0.42&\scriptstyle- 0.90
\\\noalign{\medskip}\scriptstyle- 0.000018&\scriptstyle- 0.000034&\scriptstyle 0.000043&\scriptstyle- 1.0&\scriptstyle 0.032&\scriptstyle
 0.039\end {array} \right] 
\]
\[ R_A = 
 \left[ \begin {array}{cccccc}\scriptstyle - 0.016&\scriptstyle- 0.0083&\scriptstyle 0.068&\scriptstyle 0.046&\scriptstyle- 0.88&\scriptstyle-
 0.47\\\noalign{\medskip}\scriptstyle 0.88&\scriptstyle\scriptstyle 0.46&\scriptstyle\scriptstyle- 0.075&\scriptstyle\scriptstyle 0.00088&\scriptstyle\scriptstyle- 0.026&\scriptstyle\scriptstyle
 0.000040\\\noalign{\medskip}\scriptstyle- 0.068&\scriptstyle\scriptstyle- 0.035&\scriptstyle\scriptstyle- 0.99&\scriptstyle\scriptstyle 0.0027&\scriptstyle\scriptstyle- 0.076&\scriptstyle\scriptstyle
 0.00012\\\noalign{\medskip}\scriptstyle 0.00000045&\scriptstyle\scriptstyle 0.00000023&\scriptstyle\scriptstyle 0.00017&\scriptstyle\scriptstyle 1.0&\scriptstyle\scriptstyle
 0.033&\scriptstyle\scriptstyle 0.036\\\noalign{\medskip}\scriptstyle 0.46&\scriptstyle\scriptstyle- 0.89&\scriptstyle\scriptstyle 0.00078&\scriptstyle\scriptstyle- 0.00033&\scriptstyle\scriptstyle-
 0.010&\scriptstyle\scriptstyle 0.019\\\noalign{\medskip}\scriptstyle 0.018&\scriptstyle\scriptstyle- 0.014&\scriptstyle\scriptstyle- 0.037&\scriptstyle\scriptstyle 0.016&\scriptstyle\scriptstyle 0.47&\scriptstyle\scriptstyle-
 0.88\end {array} \right] 
\]
\[ R_{\chi^0} = 
 \left[ \begin {array}{ccccccccc}\scriptstyle - 0.0011&\scriptstyle 0.064&\scriptstyle 0.053&\scriptstyle- 0.45&\scriptstyle- 0.15
&\scriptstyle\scriptstyle 0.87&\scriptstyle\scriptstyle 0.00096&\scriptstyle\scriptstyle 0.030&\scriptstyle\scriptstyle- 0.028\\\noalign{\medskip}\scriptstyle- 0.00051&\scriptstyle\scriptstyle- 0.36&\scriptstyle\scriptstyle-
 0.11&\scriptstyle\scriptstyle 0.58&\scriptstyle\scriptstyle- 0.69&\scriptstyle\scriptstyle 0.22&\scriptstyle\scriptstyle 0.00048&\scriptstyle\scriptstyle 0.0046&\scriptstyle\scriptstyle- 0.0042\\\noalign{\medskip}\scriptstyle
 0.00052&\scriptstyle- 0.020&\scriptstyle- 0.072&\scriptstyle 0.60&\scriptstyle 0.67&\scriptstyle 0.43&\scriptstyle- 0.00055&\scriptstyle 0.0046&\scriptstyle- 0.0041
\\\noalign{\medskip}\scriptstyle 0.000051&\scriptstyle 0.93&\scriptstyle- 0.038&\scriptstyle 0.27&\scriptstyle- 0.24&\scriptstyle 0.033&\scriptstyle-
 0.000052&\scriptstyle\scriptstyle- 0.00029&\scriptstyle\scriptstyle 0.00025\\\noalign{\medskip}\scriptstyle 0.00020&\scriptstyle\scriptstyle 0.0082&\scriptstyle\scriptstyle-
 0.99&\scriptstyle- 0.14&\scriptstyle 0.027&\scriptstyle- 0.0095&\scriptstyle- 0.00030&\scriptstyle 0.00087&\scriptstyle- 0.00057
\\\noalign{\medskip}\scriptstyle- 0.073&\scriptstyle 0.000063&\scriptstyle- 0.00038&\scriptstyle- 0.0020&\scriptstyle- 0.00066&\scriptstyle
 0.0081&\scriptstyle\scriptstyle 0.71&\scriptstyle\scriptstyle- 0.61&\scriptstyle\scriptstyle- 0.34\\\noalign{\medskip}\scriptstyle- 0.11&\scriptstyle\scriptstyle 0.00019&\scriptstyle\scriptstyle-
 0.00014&\scriptstyle\scriptstyle 0.0038&\scriptstyle\scriptstyle 0.0026&\scriptstyle\scriptstyle- 0.013&\scriptstyle\scriptstyle 0.69&\scriptstyle\scriptstyle 0.67&\scriptstyle\scriptstyle 0.26\\\noalign{\medskip}\scriptstyle
 0.59&\scriptstyle- 0.00015&\scriptstyle 0.00018&\scriptstyle- 0.0085&\scriptstyle- 0.0050&\scriptstyle 0.029&\scriptstyle 0.13&\scriptstyle- 0.32&\scriptstyle 0.73
\\\noalign{\medskip}\scriptstyle- 0.79&\scriptstyle 0.000050&\scriptstyle- 0.00012&\scriptstyle- 0.0060&\scriptstyle- 0.0029&\scriptstyle
 0.022&\scriptstyle- 0.063&\scriptstyle- 0.27&\scriptstyle 0.54\end {array} \right] 
\]
\[ R_{\chi^+} = 
 \left[ \begin {array}{cccc}\scriptstyle  0.027&\scriptstyle- 0.71&\scriptstyle- 0.14&\scriptstyle 0.69
\\\noalign{\medskip}\scriptstyle 0.027&\scriptstyle- 0.71&\scriptstyle 0.14&\scriptstyle- 0.69\\\noalign{\medskip}\scriptstyle
 0.71&\scriptstyle\scriptstyle 0.027&\scriptstyle\scriptstyle- 0.69&\scriptstyle\scriptstyle- 0.14\\\noalign{\medskip}\scriptstyle 0.71&\scriptstyle\scriptstyle 0.027&\scriptstyle\scriptstyle 0.69&\scriptstyle\scriptstyle 0.14
\end {array} \right] 
\]
where $H_i^{\rm mass} = R_H \Re H^{\rm gauge}, A_i^{\rm mass} = R_A \Im H^{\rm gauge}$ with $H^{\rm gauge} = [H_1^0, H_2^0, S, S_1, S_2, S_3]^T$; $A_5$ and $A_6$ are goldstone bosons corresponding to the $Z$ and $Z^\prime$.  $\chi^0_{\rm mass} = R_{\chi^0} [\tilde{Z}^\prime, \tilde{B}, \tilde{W}^0, \tilde{H}_2^0,  \tilde{H}_1^0, \tilde{S}, \tilde{S}_1, \tilde{S}_2, \tilde{S}_3]^T$, and $\chi^+_{\rm mass} =  R_{\chi^+} [\tilde{W}^+, \tilde{H}_1^+, \tilde{W}^-, \tilde{H}_2^-]^T$.

    % This is DATA[580]
\section{Lightest $H_1$}
\label{h1lightest}
\begin{flushleft}
\begin{tabular}[b]{l|c|c|c|c|c|c|c|c|c}
  \multicolumn{2}{c|}{$\tan{\beta} = 0.614$} &   \multicolumn{2}{c|}{$M_{Z^\prime} = 626$ GeV} &   \multicolumn{2}{c|}{$M_{H^+} = 444$ GeV} &   \multicolumn{3}{c}{$\alpha_{ZZ^\prime} = $-$3.3\!\cdot\!10^{-3}$} \\
\hline 
  $M_{H}$ & 8 & 127 & 471 & 627 & 1071 & 1762 &&& \\
  $\xi_{\rm MSSM}$ & 0.013 & 0.99 & 1 & $1.4\!\cdot\!10^{-3}$ & $4.5\!\cdot\!10^{-5}$ & $3.6\!\cdot\!10^{-4}$ &&& \\
  $\sigma(H_i Z)$ & 0.86 &  55 & & & & & & & \\
  $\sigma(H_i \nu\overline{\nu})$ & 2.5&  78& $3.5\!\cdot\!10^{-5}$& & & & & & \\
  $\sigma(H_ie^+e^-)$ & 0.23& 7.5& $3.2\!\cdot\!10^{-6}$& & & & & & \\
\hline 
  $M_{A}$ & 172 & 446 & 1067 & 1757 & 0 & 0 &&& \\
  $\xi_{\rm MSSM}$ & $2.0\!\cdot\!10^{-4}$ & 1 & $1.4\!\cdot\!10^{-4}$ & $6.2\!\cdot\!10^{-4}$ & 1 & $1.7\!\cdot\!10^{-3}$ &&& \\
$\sigma(H_1 A)$  & $1.1\!\cdot\!10^{-7}$ & $6.8\!\cdot\!10^{-6}$ & & & & & & & \\
$\sigma(H_2 A)$  & $2.3\!\cdot\!10^{-5}$ & & & & & & & & \\
\hline 
  $M_{\chi^0}$               & 44 & 47 & 58 & 62 & 120 & 170 & 286 & 556 & 1357 \\
  $\xi_{\rm MSSM}$          & 0.98 & 0.3 & $2.3\!\cdot\!10^{-4}$ & $4.4\!\cdot\!10^{-4}$ & 0.94 & 0.77 & $3.2\!\cdot\!10^{-3}$ & 1 & $3.1\!\cdot\!10^{-4}$ \\
  $\xi_{\tilde{s}}$         & 0.015 & 0.7 & 1 & 1 & 0.059 & 0.23 & 0.82 & 0 & 0.18 \\
  $\xi_{\tilde{Z^\prime}}$ & 0 & $1.2\!\cdot\!10^{-4}$ & $3.4\!\cdot\!10^{-4}$ & $5.3\!\cdot\!10^{-5}$ & $4.7\!\cdot\!10^{-5}$ & 0 & 0.18 & 0 & 0.82 \\
\hline 
  $M_{\chi^+}$ & 103 & 556 &&&&&&& \\
\end{tabular}
\end{flushleft}
\vspace{-10pt}
Cross sections quoted are in fb for a linear $e^+ e^-$ collider at center-of-mass energy 500 GeV.\vspace{-10pt}
\begin{eqnarray*}
  v_2       = 91 \gev   & v_1       = 148 \gev   & v_s       = 212 \gev   \\
               v_{s1}    = 1726 \gev   & v_{s2}    = 120 \gev   & v_{s3}    = 397 \gev   \\
               m_{H_u}^2 = (194 \gev)^2 & m_{H_d}^2 = -(184 \gev)^2 & m_{S}^2   = (1770 \gev)^2 \\
               m_{S_1}^2 = -(310 \gev)^2 & m_{S_2}^2 = (1001 \gev)^2 & m_{S_3}^2 = (580 \gev)^2 \\
               h = -0.552                 & A_h       = 764 \gev   & \mu = h v_s = -117 \gev \\
               \lambda = 0.035          & A_\lambda = 3234 \gev  & \\
               M_1      = -31  \gev   & M_1^\prime = -1067 \gev & M_2 = 542 \gev \\
               m_{S S_1}^2 = -(587 \gev)^2 & m_{S S_2}^2 = -(536 \gev)^2 &
\end{eqnarray*}
Branching Ratios for dominant decay modes (greater than 1\% excluding model-dependent squark, slepton, $Z^\prime$ and exotic decays; $\chi^0_{i>1}$ are summed): 
\begin{flushleft}\begin{tabular}{c|lr|lr|lr|lr|lr|lr}
$H_1$ & $\tau^+\tau^-$ & 59\% & $c\bar{c}$ & 38\% & $s\bar{s}$ & 2\% &&&&&&\\
$H_2$ & $\chi^0_{i>1} \chi^0_{i>1}$ & 60\% & $\chi^0_1 \chi^0_{i>1}$ & 17\% & $\chi^0_1 \chi^0_1$ & 13\% & $H_1 H_1$ & 6\% & $b\bar{b}$ & 4\% &&\\
$H_3$ & $t\bar{t}$ & 66\% & $\chi^0_{i>1} \chi^0_{i>1}$ & 20\% & $\chi^0_1 \chi^0_{i>1}$ & 8\% & $H_2 H_2$ & 4\% & $W^+W^-$ & 1\% & $Z Z$ & 1\% \\
$H_4$ & $A_1 A_1$ & 41\% & $H_1 H_1$ & 35\% & $\chi^0_{i>1} \chi^0_{i>1}$ & 12\% & $H_2 H_2$ & 3\% & $W^+W^-$ & 3\% & $\chi^+_1 \chi^-_1$ & 2\% \\
$H_5$ & $\chi^0_{i>1} \chi^0_{i>1}$ & 59\% & $\chi^+_1 \chi^-_1$ & 13\% & $H_1 H_1$ & 8\% & $A_1 A_1$ & 4\% & $A_2 A_2$ & 4\% & $H_1 H_4$ & 3\% \\
$H_6$ & $\chi^0_{i>1} \chi^0_{i>1}$ & 41\% & $\chi^+_1 \chi^-_1$ & 38\% & $A_2 A_2$ & 5\% & $H_3 H_3$ & 4\% & $\chi^0_1 \chi^0_{i>1}$ & 3\% & $H_2 H_3$ & 2\% \\
\hline
$A_1$ & $\chi^0_{i>1} \chi^0_{i>1}$ & 98\% & $\chi^0_1 \chi^0_1$ & 2\% & $\chi^0_1, \chi^0_{i>1}$ & 1\% &&&&&&\\
$A_2$ & $t\bar{t}$ & 62\% & $\chi^0_{i>1} \chi^0_{i>1}$ & 26\% & $\chi^0_1, \chi^0_{i>1}$ & 7\% & $\chi^+_1 \chi^-_1$ & 2\% & $H_2 Z$ & 1\% & $\chi^0_1 \chi^0_1$ & 1\% \\
$A_3$ & $H_2 Z$ & 100\% &&&&&&&&&&\\
$A_4$ & $\chi^0_{i>1} \chi^0_{i>1}$ & 44\% & $\chi^+_1 \chi^-_1$ & 41\% & $H_2 Z$ & 4\% & $A_1 H_1$ & 3\% & $\chi^0_1, \chi^0_{i>1}$ & 3\% & $H_5 Z$ & 2\% \\
\end{tabular}
\end{flushleft}
\newpage

Eigenvectors/rotation matrices 
 \[R_H = 
 \left[ \begin {array}{cccccc}\scriptstyle  0.062&\scriptstyle 0.094&\scriptstyle 0.072&\scriptstyle 0.46&\scriptstyle 0.18&\scriptstyle 0.86
\\\noalign{\medskip}\scriptstyle 0.46&\scriptstyle 0.88&\scriptstyle 0.0025&\scriptstyle- 0.082&\scriptstyle- 0.017&\scriptstyle- 0.083
\\\noalign{\medskip}\scriptstyle- 0.88&\scriptstyle 0.47&\scriptstyle- 0.015&\scriptstyle- 0.010&\scriptstyle- 0.00038&\scriptstyle 0.020
\\\noalign{\medskip}\scriptstyle- 0.0028&\scriptstyle 0.037&\scriptstyle 0.13&\scriptstyle 0.87&\scriptstyle- 0.052&\scriptstyle- 0.47
\\\noalign{\medskip}\scriptstyle- 0.0067&\scriptstyle- 0.0011&\scriptstyle 0.14&\scriptstyle- 0.060&\scriptstyle 0.97&\scriptstyle- 0.18
\\\noalign{\medskip}\scriptstyle- 0.018&\scriptstyle- 0.0070&\scriptstyle 0.98&\scriptstyle- 0.14&\scriptstyle- 0.15&\scriptstyle 0.023
\end {array} \right] 
\]
\[ R_A = 
 \left[ \begin {array}{cccccc}\scriptstyle  0.012&\scriptstyle 0.0074&\scriptstyle- 0.033&\scriptstyle 0.42&\scriptstyle- 0.16&\scriptstyle
 0.89\\\noalign{\medskip}\scriptstyle 0.85&\scriptstyle\scriptstyle 0.52&\scriptstyle\scriptstyle- 0.021&\scriptstyle\scriptstyle- 0.045&\scriptstyle\scriptstyle 0.0074&\scriptstyle\scriptstyle 0.0058
\\\noalign{\medskip}\scriptstyle- 0.0099&\scriptstyle- 0.0061&\scriptstyle- 0.15&\scriptstyle- 0.0052&\scriptstyle 0.97&\scriptstyle 0.17
\\\noalign{\medskip}\scriptstyle 0.021&\scriptstyle 0.013&\scriptstyle 0.98&\scriptstyle 0.11&\scriptstyle 0.15&\scriptstyle 0.010
\\\noalign{\medskip}\scriptstyle 0.52&\scriptstyle- 0.85&\scriptstyle- 0.00057&\scriptstyle 0.0047&\scriptstyle 0.00032&\scriptstyle- 0.0021
\\\noalign{\medskip}\scriptstyle 0.032&\scriptstyle 0.026&\scriptstyle- 0.11&\scriptstyle 0.90&\scriptstyle 0.063&\scriptstyle- 0.41
\end {array} \right] 
\]
\[ R_{\chi^0} = 
 \left[ \begin {array}{ccccccccc}\scriptstyle  0.0012&\scriptstyle 0.95&\scriptstyle- 0.039&\scriptstyle 0.21&\scriptstyle- 0.20&\scriptstyle-
 0.12&\scriptstyle\scriptstyle- 0.019&\scriptstyle\scriptstyle 0.0091&\scriptstyle\scriptstyle- 0.0033\\\noalign{\medskip}\scriptstyle 0.011&\scriptstyle\scriptstyle 0.18&\scriptstyle\scriptstyle 0.051&\scriptstyle\scriptstyle
- 0.48&\scriptstyle\scriptstyle- 0.18&\scriptstyle\scriptstyle 0.82&\scriptstyle\scriptstyle 0.14&\scriptstyle\scriptstyle 0.021&\scriptstyle\scriptstyle- 0.022\\\noalign{\medskip}\scriptstyle- 0.019&\scriptstyle\scriptstyle
 0.012&\scriptstyle- 0.00096&\scriptstyle 0.0093&\scriptstyle 0.0016&\scriptstyle- 0.012&\scriptstyle 0.30&\scriptstyle- 0.71&\scriptstyle 0.63
\\\noalign{\medskip}\scriptstyle- 0.0073&\scriptstyle 0.0064&\scriptstyle 0.0023&\scriptstyle- 0.019&\scriptstyle- 0.0052&\scriptstyle 0.030&\scriptstyle-
 0.27&\scriptstyle\scriptstyle- 0.70&\scriptstyle\scriptstyle- 0.66\\\noalign{\medskip}\scriptstyle 0.0069&\scriptstyle\scriptstyle- 0.25&\scriptstyle\scriptstyle- 0.17&\scriptstyle\scriptstyle 0.58&\scriptstyle\scriptstyle-
 0.72&\scriptstyle\scriptstyle 0.24&\scriptstyle\scriptstyle 0.031&\scriptstyle\scriptstyle- 0.0013&\scriptstyle\scriptstyle- 0.014\\\noalign{\medskip}\scriptstyle 0.0029&\scriptstyle\scriptstyle- 0.057
&\scriptstyle 0.020&\scriptstyle- 0.60&\scriptstyle- 0.63&\scriptstyle- 0.48&\scriptstyle- 0.0095&\scriptstyle- 0.0017&\scriptstyle 0.0052
\\\noalign{\medskip}\scriptstyle- 0.42&\scriptstyle 0.0017&\scriptstyle 0.0039&\scriptstyle- 0.039&\scriptstyle- 0.040&\scriptstyle 0.12&\scriptstyle-
 0.82&\scriptstyle\scriptstyle- 0.021&\scriptstyle\scriptstyle 0.36\\\noalign{\medskip}\scriptstyle- 0.0019&\scriptstyle\scriptstyle 0.013&\scriptstyle\scriptstyle- 0.98&\scriptstyle\scriptstyle- 0.14&\scriptstyle\scriptstyle
 0.11&\scriptstyle\scriptstyle- 0.0023&\scriptstyle\scriptstyle- 0.0019&\scriptstyle\scriptstyle- 0.000088&\scriptstyle\scriptstyle 0.00085\\\noalign{\medskip}\scriptstyle- 0.91&\scriptstyle\scriptstyle
 0.00028&\scriptstyle- 0.00036&\scriptstyle 0.015&\scriptstyle 0.0085&\scriptstyle- 0.045&\scriptstyle 0.38&\scriptstyle 0.030&\scriptstyle- 0.18
\end {array} \right] 
\]
\[ R_{\chi^+} = 
 \left[ \begin {array}{cccc}\scriptstyle - 0.10&\scriptstyle 0.70&\scriptstyle- 0.14&\scriptstyle 0.69
\\\noalign{\medskip}\scriptstyle 0.10&\scriptstyle- 0.70&\scriptstyle- 0.14&\scriptstyle 0.69\\\noalign{\medskip}\scriptstyle 0.70
&\scriptstyle\scriptstyle 0.10&\scriptstyle\scriptstyle 0.69&\scriptstyle\scriptstyle 0.14\\\noalign{\medskip}\scriptstyle 0.70&\scriptstyle\scriptstyle 0.10&\scriptstyle\scriptstyle- 0.69&\scriptstyle\scriptstyle- 0.14
\end {array} \right] 
\]

    % This is DATA[2]
\section{Typical light $H_1 \rightarrow \rm{SM}$ dominant}
\label{h1smdominant}
\begin{flushleft}
\begin{tabular}[b]{l|c|c|c|c|c|c|c|c|c}
  \multicolumn{2}{c|}{$\tan{\beta} = 2.64$} &   \multicolumn{2}{c|}{$M_{Z^\prime} = 828$ GeV} &   \multicolumn{2}{c|}{$M_{H^+} = 792$ GeV} &   \multicolumn{3}{c}{$\alpha_{ZZ^\prime} = $$3.1\!\cdot\!10^{-3}$} \\
\hline 
  $M_{H}$ & 46 & 119 & 332 & 780 & 828 & 1558 &&& \\
  $\xi_{\rm MSSM}$ & $3.5\!\cdot\!10^{-3}$ & 0.99 & $1.9\!\cdot\!10^{-5}$ & 0.92 & 0.064 & 0.019 &&& \\
  $\sigma(H_i Z)$ & 0.23 &  57 & 0.00018 & & & & & & \\
  $\sigma(H_i \nu\overline{\nu})$ & 0.54&  85& $5.9\!\cdot\!10^{-5}$& & & & & & \\
  $\sigma(H_ie^+e^-)$ & 0.051& 8.2& $5.7\!\cdot\!10^{-6}$& & & & & & \\
\hline 
  $M_{A}$ & 59 & 337 & 774 & 1558 & 0 & 0 &&& \\
  $\xi_{\rm MSSM}$ & 0 & 0 & 0.97 & 0.026 & 0.99 & $5.7\!\cdot\!10^{-3}$ &&& \\
$\sigma(H_1 A)$  & $2.4\!\cdot\!10^{-9}$ & $1.6\!\cdot\!10^{-10}$ & & & & & & & \\
$\sigma(H_2 A)$  & $6.1\!\cdot\!10^{-8}$ & $1.7\!\cdot\!10^{-9}$ & & & & & & & \\
$\sigma(H_3 A)$  & $3.8\!\cdot\!10^{-10}$ & & & & & & & & \\
\hline 
  $M_{\chi^0}$               & 42 & 72 & 79 & 104 & 180 & 216 & 290 & 627 & 1102 \\
  $\xi_{\rm MSSM}$          & 0.8 & $4.2\!\cdot\!10^{-3}$ & $8.9\!\cdot\!10^{-5}$ & 0.72 & 0.8 & 0.68 & 0.99 & $9.6\!\cdot\!10^{-4}$ & $3.7\!\cdot\!10^{-4}$ \\
  $\xi_{\tilde{s}}$         & 0.2 & 0.99 & 1 & 0.28 & 0.2 & 0.32 & 0.01 & 0.64 & 0.36 \\
  $\xi_{\tilde{Z^\prime}}$ & $1.1\!\cdot\!10^{-5}$ & $1.8\!\cdot\!10^{-3}$ & $1.3\!\cdot\!10^{-3}$ & 0 & $1.5\!\cdot\!10^{-5}$ & 0 & 0 & 0.36 & 0.64 \\
\hline 
  $M_{\chi^+}$ & 124 & 289 &&&&&&& \\
\end{tabular}
\end{flushleft}
\vspace{-10pt}
Cross sections quoted are in fb for a linear $e^+ e^-$ collider at center-of-mass energy 500 GeV.\vspace{-10pt}
\begin{eqnarray*}
  v_2       = 163 \gev   & v_1       = 62 \gev   & v_s       = 128 \gev   \\
               v_{s1}    = 2434 \gev   & v_{s2}    = 679 \gev   & v_{s3}    = 96 \gev   \\
               m_{H_u}^2 = -(338 \gev)^2 & m_{H_d}^2 = (602 \gev)^2 & m_{S}^2   = (1640 \gev)^2 \\
               m_{S_1}^2 = -(577 \gev)^2 & m_{S_2}^2 = -(583 \gev)^2 & m_{S_3}^2 = (884 \gev)^2 \\
               h = -0.920                 & A_h       = 1818 \gev   & \mu = h v_s = -118 \gev \\
               \lambda = 0.035          & A_\lambda = 184 \gev  & \\
               M_1      = -88  \gev   & M_1^\prime = 482 \gev & M_2 = 268 \gev \\
               m_{S S_1}^2 = -(336 \gev)^2 & m_{S S_2}^2 = -(136 \gev)^2 &
\end{eqnarray*}
Branching Ratios for dominant decay modes (greater than 1\% excluding model-dependent squark, slepton, $Z^\prime$ and exotic decays; $\chi^0_{i>1}$ are summed): 
\begin{flushleft}\begin{tabular}{c|lr|lr|lr|lr|lr|lr}
$H_1$ & $b\bar{b}$ & 64\% & $c\bar{c}$ & 32\% & $\tau^+\tau^-$ & 4\% &&&&&&\\
$H_2$ & $b\bar{b}$ & 41\% & $H_1 H_1$ & 27\% & $c\bar{c}$ & 23\% & $\chi^0_1 \chi^0_1$ & 5\% & $\tau^+\tau^-$ & 2\% & $A_1 A_1$ & 1\% \\
$H_3$ & $\chi^0_{i>1} \chi^0_{i>1}$ & 72\% & $H_1 H_1$ & 21\% & $A_1 A_1$ & 4\% & $H_2 H_2$ & 1\% & $\chi^0_1 \chi^0_{i>1}$ & 1\% & $W^+W^-$ & 1\% \\
$H_4$ & $\chi^0_{i>1} \chi^0_{i>1}$ & 41\% & $t\bar{t}$ & 21\% & $\chi^0_1 \chi^0_{i>1}$ & 16\% & $\chi^0_1 \chi^0_1$ & 11\% & $\chi^+_1 \chi^-_1$ & 8\% & $H_2 H_2$ & 1\% \\
$H_5$ & $H_3 H_3$ & 33\% & $A_1 A_1$ & 16\% & $H_1 H_1$ & 14\% & $\chi^0_{i>1} \chi^0_{i>1}$ & 13\% & $t\bar{t}$ & 8\% & $\chi^0_1 \chi^0_{i>1}$ & 4\% \\
$H_6$ & $A_2 A_3$ & 25\% & $\chi^+_1 \chi^-_1$ & 18\% & $\chi^0_{i>1} \chi^0_{i>1}$ & 14\% & $H_2 H_4$ & 11\% & $\chi^0_1 \chi^0_{i>1}$ & 10\% & $A_3 Z$ & 7\% \\
\hline
$A_1$ & $b\bar{b}$ & 93\% & $\tau^+ \tau^-$ & 5\% & $c\bar{c}$ & 1\% &&&&&&\\
$A_2$ & $\chi^0_{i>1} \chi^0_{i>1}$ & 97\% & $A_1 H_1$ & 2\% & $\chi^0_1, \chi^0_{i>1}$ & 1\% &&&&&&\\
$A_3$ & $\chi^0_{i>1} \chi^0_{i>1}$ & 35\% & $\chi^0_1, \chi^0_{i>1}$ & 24\% & $t\bar{t}$ & 19\% & $\chi^+_1 \chi^-_1$ & 9\% & $\chi^+_1 \chi^-_2$ & 8\% & $\chi^0_1 \chi^0_1$ & 5\% \\
$A_4$ & $H_3 Z$ & 93\% & $\chi^+_1 \chi^-_1$ & 2\% & $\chi^0_{i>1} \chi^0_{i>1}$ & 2\% & $H_4 Z$ & 1\% & $A_2 H_5$ & 1\% & $\chi^0_1, \chi^0_{i>1}$ & 1\% \\
\end{tabular}
\end{flushleft}
\newpage

Eigenvectors/rotation matrices 
 \[R_H = 
 \left[ \begin {array}{cccccc}\scriptstyle  0.054&\scriptstyle 0.023&\scriptstyle 0.0090&\scriptstyle 0.26&\scriptstyle- 0.96&\scriptstyle-
 0.12\\\noalign{\medskip}\scriptstyle- 0.93&\scriptstyle\scriptstyle- 0.37&\scriptstyle\scriptstyle- 0.066&\scriptstyle\scriptstyle 0.034&\scriptstyle\scriptstyle- 0.051&\scriptstyle\scriptstyle- 0.012
\\\noalign{\medskip}\scriptstyle- 0.0042&\scriptstyle 0.0012&\scriptstyle 0.0063&\scriptstyle 0.13&\scriptstyle- 0.093&\scriptstyle 0.99
\\\noalign{\medskip}\scriptstyle 0.36&\scriptstyle- 0.89&\scriptstyle- 0.14&\scriptstyle- 0.23&\scriptstyle- 0.069&\scriptstyle 0.028
\\\noalign{\medskip}\scriptstyle 0.11&\scriptstyle- 0.23&\scriptstyle 0.053&\scriptstyle 0.92&\scriptstyle 0.26&\scriptstyle- 0.10
\\\noalign{\medskip}\scriptstyle- 0.018&\scriptstyle- 0.14&\scriptstyle 0.99&\scriptstyle- 0.083&\scriptstyle- 0.018&\scriptstyle 0.0033
\end {array} \right] 
\]
\[ R_A = 
 \left[ \begin {array}{cccccc}\scriptstyle - 0.00096&\scriptstyle- 0.0025&\scriptstyle 0.0060&\scriptstyle- 0.28&\scriptstyle 0.95
&\scriptstyle\scriptstyle- 0.13\\\noalign{\medskip}\scriptstyle- 0.00063&\scriptstyle\scriptstyle- 0.0017&\scriptstyle\scriptstyle 0.0033&\scriptstyle\scriptstyle- 0.038&\scriptstyle\scriptstyle- 0.14&\scriptstyle\scriptstyle-
 0.99\\\noalign{\medskip}\scriptstyle- 0.35&\scriptstyle\scriptstyle- 0.92&\scriptstyle\scriptstyle 0.16&\scriptstyle\scriptstyle 0.030&\scriptstyle\scriptstyle 0.0050&\scriptstyle\scriptstyle 0.00043
\\\noalign{\medskip}\scriptstyle 0.057&\scriptstyle 0.15&\scriptstyle 0.99&\scriptstyle 0.046&\scriptstyle 0.0075&\scriptstyle 0.00014
\\\noalign{\medskip}\scriptstyle 0.93&\scriptstyle- 0.35&\scriptstyle 0.0036&\scriptstyle- 0.069&\scriptstyle- 0.019&\scriptstyle 0.0054
\\\noalign{\medskip}\scriptstyle 0.075&\scriptstyle- 0.0042&\scriptstyle- 0.050&\scriptstyle 0.96&\scriptstyle 0.27&\scriptstyle- 0.075
\end {array} \right] 
\]
\[ R_{\chi^0} = 
 \left[ \begin {array}{ccccccccc}\scriptstyle  0.0033&\scriptstyle 0.60&\scriptstyle 0.16&\scriptstyle- 0.12&\scriptstyle 0.64&\scriptstyle-
 0.44&\scriptstyle\scriptstyle- 0.054&\scriptstyle\scriptstyle 0.027&\scriptstyle\scriptstyle- 0.020\\\noalign{\medskip}\scriptstyle- 0.042&\scriptstyle\scriptstyle 0.059&\scriptstyle\scriptstyle 0.0052
&\scriptstyle\scriptstyle- 0.020&\scriptstyle\scriptstyle 0.016&\scriptstyle\scriptstyle 0.0052&\scriptstyle\scriptstyle 0.27&\scriptstyle\scriptstyle- 0.68&\scriptstyle\scriptstyle 0.68\\\noalign{\medskip}\scriptstyle 0.036&\scriptstyle\scriptstyle
 0.0018&\scriptstyle 0.0029&\scriptstyle 0.0026&\scriptstyle 0.0084&\scriptstyle- 0.0080&\scriptstyle 0.11&\scriptstyle- 0.68&\scriptstyle- 0.73
\\\noalign{\medskip}\scriptstyle- 0.0030&\scriptstyle 0.73&\scriptstyle- 0.054&\scriptstyle- 0.23&\scriptstyle- 0.36&\scriptstyle 0.53&\scriptstyle 0.031&\scriptstyle
 0.038&\scriptstyle\scriptstyle- 0.040\\\noalign{\medskip}\scriptstyle 0.0039&\scriptstyle\scriptstyle 0.32&\scriptstyle\scriptstyle- 0.12&\scriptstyle\scriptstyle 0.70&\scriptstyle\scriptstyle- 0.44&\scriptstyle\scriptstyle-
 0.44&\scriptstyle\scriptstyle- 0.018&\scriptstyle\scriptstyle- 0.0079&\scriptstyle\scriptstyle 0.0074\\\noalign{\medskip}\scriptstyle 0.0017&\scriptstyle\scriptstyle 0.024&\scriptstyle\scriptstyle 0.25
&\scriptstyle\scriptstyle 0.66&\scriptstyle\scriptstyle 0.43&\scriptstyle\scriptstyle 0.57&\scriptstyle\scriptstyle 0.0062&\scriptstyle\scriptstyle 0.0022&\scriptstyle\scriptstyle 0.0011\\\noalign{\medskip}\scriptstyle 0.0018&\scriptstyle\scriptstyle
- 0.029&\scriptstyle 0.94&\scriptstyle- 0.080&\scriptstyle- 0.30&\scriptstyle- 0.10&\scriptstyle 0.0050&\scriptstyle 0.0016&\scriptstyle 0.00048
\\\noalign{\medskip}\scriptstyle- 0.60&\scriptstyle- 0.0016&\scriptstyle 0.0018&\scriptstyle 0.015&\scriptstyle 0.027&\scriptstyle- 0.046&\scriptstyle
 0.76&\scriptstyle\scriptstyle 0.22&\scriptstyle\scriptstyle- 0.12\\\noalign{\medskip}\scriptstyle- 0.80&\scriptstyle\scriptstyle- 0.00057&\scriptstyle\scriptstyle 0.0015&\scriptstyle\scriptstyle-
 0.0053&\scriptstyle- 0.018&\scriptstyle 0.029&\scriptstyle- 0.58&\scriptstyle- 0.16&\scriptstyle 0.021\end {array} \right] 
\]
\[ R_{\chi^+} = 
 \left[ \begin {array}{cccc}\scriptstyle  0.27&\scriptstyle- 0.65&\scriptstyle- 0.018&\scriptstyle- 0.71
\\\noalign{\medskip}\scriptstyle- 0.27&\scriptstyle 0.65&\scriptstyle- 0.018&\scriptstyle- 0.71\\\noalign{\medskip}\scriptstyle-
 0.65&\scriptstyle\scriptstyle- 0.27&\scriptstyle\scriptstyle 0.71&\scriptstyle\scriptstyle- 0.018\\\noalign{\medskip}\scriptstyle- 0.65&\scriptstyle\scriptstyle- 0.27&\scriptstyle\scriptstyle- 0.71&\scriptstyle\scriptstyle
 0.018\end {array} \right] 
\]

    % This is DATA[709]
\section{MSSM-like (singlets decoupled)}
\label{mssmlike}
\begin{flushleft}
\begin{tabular}[b]{l|c|c|c|c|c|c|c|c|c}
  \multicolumn{2}{c|}{$\tan{\beta} = 0.965$} &   \multicolumn{2}{c|}{$M_{Z^\prime} = 1558$ GeV} &   \multicolumn{2}{c|}{$M_{H^+} = 375$ GeV} &   \multicolumn{3}{c}{$\alpha_{ZZ^\prime} = $-$4.1\!\cdot\!10^{-5}$} \\
\hline 
  $M_{H}$ & 114 & 389 & 472 & 1498 & 2806 & 2887 &&& \\
  $\xi_{\rm MSSM}$ & 1 & 1 & $4.1\!\cdot\!10^{-4}$ & $3.8\!\cdot\!10^{-4}$ & 0 & 0 &&& \\
  $\sigma(H_i Z)$ &  58 & 0.019 & & & & & & & \\
  $\sigma(H_i \nu\overline{\nu})$ &  90& 0.0041& $2.9\!\cdot\!10^{-6}$& & & & & & \\
  $\sigma(H_ie^+e^-)$ & 8.6& 0.00039& $2.7\!\cdot\!10^{-7}$& & & & & & \\
\hline 
  $M_{A}$ & 374 & 486 & 2804 & 2855 & 0 & 0 &&& \\
  $\xi_{\rm MSSM}$ & 1 & 0 & 0 & 0 & 1 & $3.3\!\cdot\!10^{-4}$ &&& \\
$\sigma(H_1 A)$  & 0.0012 & & & & & & & & \\
\hline 
  $M_{\chi^0}$               & 6 & 22 & 226 & 250 & 743 & 746 & 826 & 970 & 2513 \\
  $\xi_{\rm MSSM}$          & 1 & 0.082 & 1 & 0.92 & 0 & $2.5\!\cdot\!10^{-5}$ & 1 & $3.3\!\cdot\!10^{-4}$ & $7.4\!\cdot\!10^{-5}$ \\
  $\xi_{\tilde{s}}$         & $8.6\!\cdot\!10^{-4}$ & 0.92 & $3.6\!\cdot\!10^{-5}$ & 0.081 & 1 & 0.99 & 0 & 0.73 & 0.28 \\
  $\xi_{\tilde{Z^\prime}}$ & 0 & 0 & 0 & 0 & $2.3\!\cdot\!10^{-3}$ & 0.011 & 0 & 0.27 & 0.72 \\
\hline 
  $M_{\chi^+}$ & 217 & 826 &&&&&&& \\
\end{tabular}
\end{flushleft}
\vspace{-10pt}
Cross sections quoted are in fb for a linear $e^+ e^-$ collider at center-of-mass energy 500 GeV.\vspace{-10pt}
\begin{eqnarray*}
  v_2       = 121 \gev   & v_1       = 125 \gev   & v_s       = 531 \gev   \\
               v_{s1}    = 318 \gev   & v_{s2}    = 4717 \gev   & v_{s3}    = 201 \gev   \\
               m_{H_u}^2 = -(761 \gev)^2 & m_{H_d}^2 = -(757 \gev)^2 & m_{S}^2   = (3017 \gev)^2 \\
               m_{S_1}^2 = (917 \gev)^2 & m_{S_2}^2 = -(1032 \gev)^2 & m_{S_3}^2 = (2720 \gev)^2 \\
               h = 0.429                 & A_h       = 306 \gev   & \mu = h v_s = 228 \gev \\
               \lambda = 0.160          & A_\lambda = 4721 \gev  & \\
               M_1      = -15  \gev   & M_1^\prime = 1540 \gev & M_2 = -816 \gev \\
               m_{S S_1}^2 = -(448 \gev)^2 & m_{S S_2}^2 = -(938 \gev)^2 &
\end{eqnarray*}
Branching Ratios for dominant decay modes (greater than 1\% excluding model-dependent squark, slepton, $Z^\prime$ and exotic decays; $\chi^0_{i>1}$ are summed): 
\begin{flushleft}\begin{tabular}{c|lr|lr|lr|lr|lr|lr}
$H_1$ & $\chi^0_{i>1} \chi^0_{i>1}$ & 77\% & $\chi^0_1 \chi^0_1$ & 19\% & $b\bar{b}$ & 4\% &&&&&&\\
$H_2$ & $t\bar{t}$ & 51\% & $\chi^0_{i>1} \chi^0_{i>1}$ & 21\% & $\chi^0_1 \chi^0_{i>1}$ & 14\% & $H_1 H_1$ & 10\% & $W^+W^-$ & 2\% & $Z Z$ & 1\% \\
$H_3$ & $H_1 H_1$ & 42\% & $W^+W^-$ & 35\% & $Z Z$ & 16\% & $t\bar{t}$ & 6\% &&&&\\
$H_4$ & $H_3 H_3$ & 25\% & $H_2 H_2$ & 16\% & $A_1 A_1$ & 16\% & $H_1 H_1$ & 13\% & $W^+W^-$ & 13\% & $Z Z$ & 6\% \\
$H_5$ & $\chi^0_{i>1} \chi^0_{i>1}$ & 53\% & $\chi^+_1 \chi^-_1$ & 22\% & $H_3 H_4$ & 22\% & $\chi^0_1 \chi^0_{i>1}$ & 1\% & $A_1 A_1$ & 1\% & $H_2 H_2$ & 1\% \\
$H_6$ & $\chi^0_{i>1} \chi^0_{i>1}$ & 52\% & $\chi^+_1 \chi^-_1$ & 38\% & $H_3 H_4$ & 8\% & $\chi^0_1 \chi^0_{i>1}$ & 2\% & $A_1 A_1$ & 1\% & $H_2 H_2$ & 1\% \\
\hline
$A_1$ & $t\bar{t}$ & 43\% & $\chi^0_{i>1} \chi^0_{i>1}$ & 35\% & $\chi^0_1, \chi^0_{i>1}$ & 16\% & $\chi^0_1 \chi^0_1$ & 3\% & $H_1 Z$ & 3\% & $b\bar{b}$ & 1\% \\
$A_2$ & $H_1 Z$ & 100\% &&&&&&&&&&\\
$A_3$ & $\chi^0_{i>1} \chi^0_{i>1}$ & 60\% & $\chi^+_1 \chi^-_1$ & 38\% & $\chi^0_1, \chi^0_{i>1}$ & 2\% &&&&&&\\
$A_4$ & $\chi^0_{i>1} \chi^0_{i>1}$ & 54\% & $\chi^+_1 \chi^-_1$ & 41\% & $A_1 H_1$ & 4\% & $\chi^0_1, \chi^0_{i>1}$ & 2\% &&&&\\
\end{tabular}
\end{flushleft}
\newpage

Eigenvectors/rotation matrices 
 \[R_H = 
 \left[ \begin {array}{cccccc}\scriptstyle  0.65&\scriptstyle 0.76&\scriptstyle- 0.0042&\scriptstyle- 0.015&\scriptstyle- 0.021&\scriptstyle-
 0.010\\\noalign{\medskip}\scriptstyle- 0.76&\scriptstyle\scriptstyle 0.65&\scriptstyle\scriptstyle 0.00022&\scriptstyle\scriptstyle 0.0025&\scriptstyle\scriptstyle 0.0014&\scriptstyle\scriptstyle
 0.0017\\\noalign{\medskip}\scriptstyle 0.015&\scriptstyle\scriptstyle 0.013&\scriptstyle\scriptstyle 0.034&\scriptstyle\scriptstyle 0.83&\scriptstyle\scriptstyle 0.082&\scriptstyle\scriptstyle 0.55
\\\noalign{\medskip}\scriptstyle- 0.014&\scriptstyle- 0.014&\scriptstyle- 0.19&\scriptstyle 0.11&\scriptstyle- 0.98&\scriptstyle- 0.0013
\\\noalign{\medskip}\scriptstyle 0.00039&\scriptstyle 0.00049&\scriptstyle 0.54&\scriptstyle 0.45&\scriptstyle- 0.056&\scriptstyle- 0.71
\\\noalign{\medskip}\scriptstyle 0.00055&\scriptstyle 0.00046&\scriptstyle- 0.82&\scriptstyle 0.31&\scriptstyle 0.19&\scriptstyle- 0.44
\end {array} \right] 
\]
\[ R_A = 
 \left[ \begin {array}{cccccc}\scriptstyle  0.72&\scriptstyle 0.69&\scriptstyle- 0.00086&\scriptstyle 0.00073&\scriptstyle- 0.019&\scriptstyle
 0.00033\\\noalign{\medskip}\scriptstyle- 0.0017&\scriptstyle\scriptstyle- 0.0016&\scriptstyle\scriptstyle- 0.0099&\scriptstyle\scriptstyle 0.83&\scriptstyle\scriptstyle- 0.10&\scriptstyle\scriptstyle-
 0.54\\\noalign{\medskip}\scriptstyle 0.0013&\scriptstyle\scriptstyle 0.0013&\scriptstyle\scriptstyle 0.62&\scriptstyle\scriptstyle- 0.42&\scriptstyle\scriptstyle 0.042&\scriptstyle\scriptstyle- 0.66
\\\noalign{\medskip}\scriptstyle 0.0016&\scriptstyle 0.0015&\scriptstyle 0.78&\scriptstyle 0.36&\scriptstyle 0.11&\scriptstyle 0.51
\\\noalign{\medskip}\scriptstyle 0.69&\scriptstyle- 0.72&\scriptstyle 0.0000070&\scriptstyle- 0.0000042&\scriptstyle- 0.000062&\scriptstyle
 0.0000053\\\noalign{\medskip}\scriptstyle- 0.013&\scriptstyle\scriptstyle- 0.013&\scriptstyle\scriptstyle 0.11&\scriptstyle\scriptstyle- 0.067&\scriptstyle\scriptstyle- 0.99&\scriptstyle\scriptstyle
 0.084\end {array} \right] 
\]
\[ R_{\chi^0} = 
 \left[ \begin {array}{ccccccccc}\scriptstyle  0.000023&\scriptstyle- 0.98&\scriptstyle 0.020&\scriptstyle 0.15&\scriptstyle- 0.14
&\scriptstyle\scriptstyle- 0.029&\scriptstyle\scriptstyle 0.00023&\scriptstyle\scriptstyle- 0.0035&\scriptstyle\scriptstyle 0.00015\\\noalign{\medskip}\scriptstyle- 0.0020&\scriptstyle\scriptstyle-
 0.030&\scriptstyle- 0.00049&\scriptstyle- 0.20&\scriptstyle- 0.20&\scriptstyle 0.95&\scriptstyle- 0.0075&\scriptstyle 0.11&\scriptstyle- 0.0054
\\\noalign{\medskip}\scriptstyle 0.000068&\scriptstyle- 0.20&\scriptstyle- 0.13&\scriptstyle- 0.69&\scriptstyle 0.69&\scriptstyle- 0.0060&\scriptstyle
 0.000030&\scriptstyle\scriptstyle- 0.00045&\scriptstyle\scriptstyle 0.000038\\\noalign{\medskip}\scriptstyle- 0.0021&\scriptstyle\scriptstyle 0.0028&\scriptstyle\scriptstyle-
 0.0013&\scriptstyle- 0.68&\scriptstyle- 0.68&\scriptstyle- 0.28&\scriptstyle 0.0013&\scriptstyle- 0.013&\scriptstyle- 0.00016
\\\noalign{\medskip}\scriptstyle- 0.048&\scriptstyle 0.0000034&\scriptstyle- 0.0000030&\scriptstyle- 0.00076&\scriptstyle- 0.00070
&\scriptstyle\scriptstyle 0.011&\scriptstyle\scriptstyle 0.71&\scriptstyle\scriptstyle- 0.081&\scriptstyle\scriptstyle- 0.70\\\noalign{\medskip}\scriptstyle 0.11&\scriptstyle\scriptstyle- 0.0000065&\scriptstyle\scriptstyle-
 0.00012&\scriptstyle- 0.0035&\scriptstyle- 0.0035&\scriptstyle 0.025&\scriptstyle 0.69&\scriptstyle- 0.14&\scriptstyle 0.70
\\\noalign{\medskip}\scriptstyle 0.00013&\scriptstyle 0.0072&\scriptstyle- 0.99&\scriptstyle 0.095&\scriptstyle- 0.093&\scriptstyle 0.00012&\scriptstyle-
 0.00013&\scriptstyle\scriptstyle- 0.00026&\scriptstyle\scriptstyle- 0.00011\\\noalign{\medskip}\scriptstyle- 0.52&\scriptstyle\scriptstyle 0.000025&\scriptstyle\scriptstyle-
 0.00028&\scriptstyle\scriptstyle 0.013&\scriptstyle\scriptstyle 0.013&\scriptstyle\scriptstyle- 0.094&\scriptstyle\scriptstyle 0.16&\scriptstyle\scriptstyle 0.83&\scriptstyle\scriptstyle 0.094\\\noalign{\medskip}\scriptstyle-
 0.85&\scriptstyle 0.0000059&\scriptstyle- 0.0000081&\scriptstyle- 0.0062&\scriptstyle- 0.0060&\scriptstyle 0.058&\scriptstyle- 0.049&\scriptstyle- 0.52&\scriptstyle
 0.069\end {array} \right] 
\]
\[ R_{\chi^+} = 
 \left[ \begin {array}{cccc}\scriptstyle - 0.093&\scriptstyle- 0.70&\scriptstyle 0.095&\scriptstyle 0.70
\\\noalign{\medskip}\scriptstyle- 0.093&\scriptstyle- 0.70&\scriptstyle- 0.095&\scriptstyle- 0.70\\\noalign{\medskip}\scriptstyle
 0.70&\scriptstyle\scriptstyle- 0.093&\scriptstyle\scriptstyle- 0.70&\scriptstyle\scriptstyle 0.095\\\noalign{\medskip}\scriptstyle- 0.70&\scriptstyle\scriptstyle 0.093&\scriptstyle\scriptstyle- 0.70&\scriptstyle\scriptstyle
 0.095\end {array} \right] 
\]

    % This is DATA[389]
\section{Large mixing among CP-odd higgses}
\label{maxamix}
\begin{flushleft}
\begin{tabular}[b]{l|c|c|c|c|c|c|c|c|c}
  \multicolumn{2}{c|}{$\tan{\beta} = 1.05$} &   \multicolumn{2}{c|}{$M_{Z^\prime} = 996$ GeV} &   \multicolumn{2}{c|}{$M_{H^+} = 368$ GeV} &   \multicolumn{3}{c}{$\alpha_{ZZ^\prime} = $$1.3\!\cdot\!10^{-4}$} \\
\hline 
  $M_{H}$ & 62 & 161 & 381 & 472 & 998 & 1510 &&& \\
  $\xi_{\rm MSSM}$ & 0.12 & 0.87 & 1 & $1.8\!\cdot\!10^{-3}$ & $8.9\!\cdot\!10^{-4}$ & 0 &&& \\
  $\sigma(H_i Z)$ &   8 &  44 & 0.025 & & & & & & \\
  $\sigma(H_i \nu\overline{\nu})$ &  17&  49& 0.0056& $8.8\!\cdot\!10^{-6}$& & & & & \\
  $\sigma(H_ie^+e^-)$ & 1.6& 4.8& 0.00054& $8.2\!\cdot\!10^{-7}$& & & & & \\
\hline 
  $M_{A}$ & 262 & 332 & 445 & 1510 & 0 & 0 &&& \\
  $\xi_{\rm MSSM}$ & 0.18 & 0.34 & 0.48 & $2.3\!\cdot\!10^{-5}$ & $2.8\!\cdot\!10^{-3}$ & 1 &&& \\
$\sigma(H_1 A)$  & 0.0026 & 0.002 & & & & & & & \\
$\sigma(H_2 A)$  & 0.0039 & 0.00017 & & & & & & & \\
\hline 
  $M_{\chi^0}$               & 67 & 162 & 164 & 187 & 237 & 250 & 309 & 568 & 1746 \\
  $\xi_{\rm MSSM}$          & 0.22 & 0 & $3.8\!\cdot\!10^{-5}$ & 1 & 0.78 & 1 & 1 & $1.0\!\cdot\!10^{-3}$ & $1.6\!\cdot\!10^{-4}$ \\
  $\xi_{\tilde{s}}$         & 0.78 & 1 & 1 & $3.6\!\cdot\!10^{-4}$ & 0.22 & $3.6\!\cdot\!10^{-4}$ & $4.2\!\cdot\!10^{-5}$ & 0.75 & 0.25 \\
  $\xi_{\tilde{Z^\prime}}$ & $4.0\!\cdot\!10^{-5}$ & $3.2\!\cdot\!10^{-5}$ & $2.7\!\cdot\!10^{-4}$ & 0 & 0 & 0 & 0 & 0.25 & 0.75 \\
\hline 
  $M_{\chi^+}$ & 183 & 308 &&&&&&& \\
\end{tabular}
\end{flushleft}
\vspace{-10pt}
Cross sections quoted are in fb for a linear $e^+ e^-$ collider at center-of-mass energy 500 GeV.\vspace{-10pt}
\begin{eqnarray*}
  v_2       = 126 \gev   & v_1       = 120 \gev   & v_s       = 233 \gev   \\
               v_{s1}    = 260 \gev   & v_{s2}    = 87 \gev   & v_{s3}    = 1516 \gev   \\
               m_{H_u}^2 = (379 \gev)^2 & m_{H_d}^2 = (400 \gev)^2 & m_{S}^2   = -(207 \gev)^2 \\
               m_{S_1}^2 = (610 \gev)^2 & m_{S_2}^2 = (1535 \gev)^2 & m_{S_3}^2 = -(695 \gev)^2 \\
               h = -0.727                 & A_h       = 426 \gev   & \mu = h v_s = -170 \gev \\
               \lambda = 0.106          & A_\lambda = 2202 \gev  & \\
               M_1      = 246  \gev   & M_1^\prime = 1176 \gev & M_2 = 294 \gev \\
               m_{S S_1}^2 = -(202 \gev)^2 & m_{S S_2}^2 = -(635 \gev)^2 &
\end{eqnarray*}
Branching Ratios for dominant decay modes (greater than 1\% excluding model-dependent squark, slepton, $Z^\prime$ and exotic decays; $\chi^0_{i>1}$ are summed): 
\begin{flushleft}\begin{tabular}{c|lr|lr|lr|lr|lr|lr}
$H_1$ & $b\bar{b}$ & 88\% & $c\bar{c}$ & 6\% & $\tau^+\tau^-$ & 5\% &&&&&&\\
$H_2$ & $H_1 H_1$ & 49\% & $\chi^0_1 \chi^0_1$ & 29\% & $W^+W^-$ & 21\% & $b\bar{b}$ & 1\% &&&&\\
$H_3$ & $\chi^0_1 \chi^0_{i>1}$ & 49\% & $t\bar{t}$ & 34\% & $H_2 H_2$ & 7\% & $A_1 Z$ & 5\% & $W^+W^-$ & 2\% & $H_1 H_2$ & 1\% \\
$H_4$ & $\chi^+_1 \chi^-_1$ & 38\% & $\chi^0_1 \chi^0_{i>1}$ & 20\% & $\chi^0_{i>1} \chi^0_{i>1}$ & 17\% & $H_2 H_2$ & 7\% & $H_1 H_1$ & 6\% & $\chi^0_1 \chi^0_1$ & 5\% \\
$H_5$ & $H_1 H_4$ & 33\% & $A_1 A_2$ & 17\% & $H_2 H_2$ & 9\% & $W^+W^-$ & 7\% & $\chi^0_{i>1} \chi^0_{i>1}$ & 6\% & $A_2 A_3$ & 6\% \\
$H_6$ & $\chi^0_{i>1} \chi^0_{i>1}$ & 53\% & $\chi^+_1 \chi^-_1$ & 19\% & $H_1 H_5$ & 14\% & $\chi^0_1 \chi^0_{i>1}$ & 4\% & $H_2 H_5$ & 2\% & $H_3 H_3$ & 2\% \\
\hline
$A_1$ & $\chi^0_1 \chi^0_1$ & 99\% &&&&&&&&&&\\
$A_2$ & $\chi^0_1 \chi^0_1$ & 86\% & $\chi^0_1, \chi^0_{i>1}$ & 8\% & $A_1 H_1$ & 5\% & $H_1 Z$ & 1\% & $H_2 Z$ & 1\% &&\\
$A_3$ & $H_2 Z$ & 41\% & $H_1 Z$ & 32\% & $t\bar{t}$ & 11\% & $\chi^0_{i>1} \chi^0_{i>1}$ & 7\% & $\chi^+_1 \chi^-_1$ & 6\% & $\chi^0_1 \chi^0_1$ & 3\% \\
$A_4$ & $\chi^0_{i>1} \chi^0_{i>1}$ & 22\% & $A_2 H_4$ & 20\% & $A_2 H_3$ & 17\% & $\chi^+_1 \chi^-_1$ & 12\% & $A_2 H_2$ & 9\% & $H_3 Z$ & 5\% \\
\end{tabular}
\end{flushleft}
\newpage

Eigenvectors/rotation matrices 
 \[R_H = 
 \left[ \begin {array}{cccccc}\scriptstyle  0.23&\scriptstyle 0.26&\scriptstyle 0.58&\scriptstyle 0.70&\scriptstyle 0.23&\scriptstyle 0.054
\\\noalign{\medskip}\scriptstyle- 0.64&\scriptstyle- 0.68&\scriptstyle 0.18&\scriptstyle 0.29&\scriptstyle 0.084&\scriptstyle- 0.0052
\\\noalign{\medskip}\scriptstyle- 0.73&\scriptstyle 0.68&\scriptstyle 0.0098&\scriptstyle- 0.017&\scriptstyle- 0.0012&\scriptstyle- 0.0050
\\\noalign{\medskip}\scriptstyle- 0.012&\scriptstyle- 0.040&\scriptstyle 0.77&\scriptstyle- 0.62&\scriptstyle 0.043&\scriptstyle- 0.15
\\\noalign{\medskip}\scriptstyle 0.022&\scriptstyle 0.021&\scriptstyle- 0.075&\scriptstyle 0.14&\scriptstyle- 0.057&\scriptstyle- 0.99
\\\noalign{\medskip}\scriptstyle- 0.0014&\scriptstyle- 0.0016&\scriptstyle 0.19&\scriptstyle 0.15&\scriptstyle- 0.97&\scriptstyle 0.064
\end {array} \right] 
\]
\[ R_A = 
 \left[ \begin {array}{cccccc}\scriptstyle  0.29&\scriptstyle 0.31&\scriptstyle- 0.60&\scriptstyle- 0.65&\scriptstyle 0.23&\scriptstyle 0.0092
\\\noalign{\medskip}\scriptstyle 0.40&\scriptstyle 0.42&\scriptstyle- 0.37&\scriptstyle 0.71&\scriptstyle- 0.053&\scriptstyle 0.10
\\\noalign{\medskip}\scriptstyle- 0.48&\scriptstyle- 0.50&\scriptstyle- 0.68&\scriptstyle 0.21&\scriptstyle 0.10&\scriptstyle 0.054
\\\noalign{\medskip}\scriptstyle 0.0033&\scriptstyle 0.0035&\scriptstyle 0.19&\scriptstyle 0.17&\scriptstyle 0.97&\scriptstyle 0.027
\\\noalign{\medskip}\scriptstyle- 0.052&\scriptstyle 0.0099&\scriptstyle 0.076&\scriptstyle- 0.085&\scriptstyle- 0.028&\scriptstyle 0.99
\\\noalign{\medskip}\scriptstyle 0.72&\scriptstyle- 0.69&\scriptstyle 0.0034&\scriptstyle- 0.0038&\scriptstyle- 0.0013&\scriptstyle 0.044
\end {array} \right] 
\]
\[ R_{\chi^0} = 
 \left[ \begin {array}{ccccccccc}\scriptstyle - 0.0063&\scriptstyle 0.0041&\scriptstyle- 0.0065&\scriptstyle- 0.32&\scriptstyle-
 0.34&\scriptstyle\scriptstyle 0.88&\scriptstyle\scriptstyle 0.018&\scriptstyle\scriptstyle 0.0090&\scriptstyle\scriptstyle- 0.087\\\noalign{\medskip}\scriptstyle- 0.0056&\scriptstyle\scriptstyle
 0.000026&\scriptstyle- 0.000031&\scriptstyle- 0.0012&\scriptstyle- 0.0012&\scriptstyle 0.0014&\scriptstyle- 0.70&\scriptstyle 0.71&\scriptstyle- 0.046
\\\noalign{\medskip}\scriptstyle- 0.016&\scriptstyle- 0.00011&\scriptstyle 0.00019&\scriptstyle 0.0036&\scriptstyle 0.0049&\scriptstyle- 0.012
&\scriptstyle\scriptstyle 0.71&\scriptstyle\scriptstyle 0.71&\scriptstyle\scriptstyle 0.060\\\noalign{\medskip}\scriptstyle 0.00024&\scriptstyle\scriptstyle 0.099&\scriptstyle\scriptstyle- 0.16&\scriptstyle\scriptstyle 0.70&\scriptstyle\scriptstyle-
 0.69&\scriptstyle\scriptstyle- 0.019&\scriptstyle\scriptstyle 0.00037&\scriptstyle\scriptstyle 0.00048&\scriptstyle\scriptstyle 0.0013\\\noalign{\medskip}\scriptstyle- 0.0021&\scriptstyle\scriptstyle
 0.045&\scriptstyle- 0.013&\scriptstyle- 0.63&\scriptstyle- 0.62&\scriptstyle- 0.47&\scriptstyle- 0.00041&\scriptstyle- 0.0010&\scriptstyle 0.0088
\\\noalign{\medskip}\scriptstyle- 0.000057&\scriptstyle- 0.99&\scriptstyle- 0.14&\scriptstyle 0.026&\scriptstyle- 0.082&\scriptstyle- 0.019&\scriptstyle-
 0.000013&\scriptstyle\scriptstyle- 0.000024&\scriptstyle\scriptstyle 0.00023\\\noalign{\medskip}\scriptstyle- 0.000045&\scriptstyle\scriptstyle 0.12&\scriptstyle\scriptstyle-
 0.98&\scriptstyle- 0.11&\scriptstyle 0.14&\scriptstyle 0.0065&\scriptstyle- 0.000011&\scriptstyle- 0.000012&\scriptstyle 0.00015
\\\noalign{\medskip}\scriptstyle- 0.50&\scriptstyle- 0.000056&\scriptstyle 0.000097&\scriptstyle 0.022&\scriptstyle 0.023&\scriptstyle- 0.074&\scriptstyle
 0.060&\scriptstyle\scriptstyle- 0.00011&\scriptstyle\scriptstyle- 0.86\\\noalign{\medskip}\scriptstyle- 0.87&\scriptstyle\scriptstyle- 0.000019&\scriptstyle\scriptstyle 0.000036
&\scriptstyle- 0.0087&\scriptstyle- 0.0092&\scriptstyle 0.037&\scriptstyle- 0.043&\scriptstyle- 0.018&\scriptstyle 0.49\end {array} \right] 
\]
\[ R_{\chi^+} = 
 \left[ \begin {array}{cccc}\scriptstyle  0.13&\scriptstyle- 0.70&\scriptstyle 0.11&\scriptstyle- 0.70
\\\noalign{\medskip}\scriptstyle- 0.13&\scriptstyle 0.70&\scriptstyle 0.11&\scriptstyle- 0.70\\\noalign{\medskip}\scriptstyle 0.70
&\scriptstyle\scriptstyle 0.13&\scriptstyle\scriptstyle 0.70&\scriptstyle\scriptstyle 0.11\\\noalign{\medskip}\scriptstyle 0.70&\scriptstyle\scriptstyle 0.13&\scriptstyle\scriptstyle- 0.70&\scriptstyle\scriptstyle- 0.11
\end {array} \right] 
\]

    % This is DATA[69]
\section{Typical heavy $A_1 \rightarrow $ gaugino dominant}
\label{a1gaugino}
\begin{flushleft}
\begin{tabular}[b]{l|c|c|c|c|c|c|c|c|c}
  \multicolumn{2}{c|}{$\tan{\beta} = 1$} &   \multicolumn{2}{c|}{$M_{Z^\prime} = 564$ GeV} &   \multicolumn{2}{c|}{$M_{H^+} = 450$ GeV} &   \multicolumn{3}{c}{$\alpha_{ZZ^\prime} = $$4.3\!\cdot\!10^{-5}$} \\
\hline 
  $M_{H}$ & 139 & 369 & 461 & 466 & 1576 & 2234 &&& \\
  $\xi_{\rm MSSM}$ & 1 & $1.6\!\cdot\!10^{-4}$ & 0.99 & $7.8\!\cdot\!10^{-3}$ & 0 & $5.6\!\cdot\!10^{-5}$ &&& \\
  $\sigma(H_i Z)$ &  54 & 0.0013 & & & & & & & \\
  $\sigma(H_i \nu\overline{\nu})$ &  71& 0.00032& $3.6\!\cdot\!10^{-5}$& $2.6\!\cdot\!10^{-5}$& & & & & \\
  $\sigma(H_ie^+e^-)$ & 6.8& $3.0\!\cdot\!10^{-5}$& $3.4\!\cdot\!10^{-6}$& $2.4\!\cdot\!10^{-6}$& & & & & \\
\hline 
  $M_{A}$ & 333 & 455 & 1562 & 2226 & 0 & 0 &&& \\
  $\xi_{\rm MSSM}$ & $4.6\!\cdot\!10^{-4}$ & 1 & 0 & $2.0\!\cdot\!10^{-4}$ & 0.021 & 0.98 &&& \\
$\sigma(H_1 A)$  & $1.1\!\cdot\!10^{-6}$ & & & & & & & & \\
\hline 
  $M_{\chi^0}$               & 16 & 62 & 64 & 89 & 152 & 217 & 343 & 489 & 828 \\
  $\xi_{\rm MSSM}$          & 0.087 & 1 & 0.01 & 0.098 & 1 & 0.8 & 1 & $1.3\!\cdot\!10^{-3}$ & $3.9\!\cdot\!10^{-4}$ \\
  $\xi_{\tilde{s}}$         & 0.85 & $1.9\!\cdot\!10^{-5}$ & 0.93 & 0.81 & 0 & 0.2 & 0 & 0.77 & 0.44 \\
  $\xi_{\tilde{Z^\prime}}$ & 0.065 & 0 & 0.059 & 0.089 & 0 & $2.1\!\cdot\!10^{-5}$ & 0 & 0.23 & 0.56 \\
\hline 
  $M_{\chi^+}$ & 128 & 342 &&&&&&& \\
\end{tabular}
\end{flushleft}
\vspace{-10pt}
Cross sections quoted are in fb for a linear $e^+ e^-$ collider at center-of-mass energy 500 GeV.\vspace{-10pt}
\begin{eqnarray*}
  v_2       = 123 \gev   & v_1       = 123 \gev   & v_s       = 268 \gev   \\
               v_{s1}    = 1064 \gev   & v_{s2}    = 1283 \gev   & v_{s3}    = 181 \gev   \\
               m_{H_u}^2 = -(103 \gev)^2 & m_{H_d}^2 = (27 \gev)^2 & m_{S}^2   = (2226 \gev)^2 \\
               m_{S_1}^2 = -(173 \gev)^2 & m_{S_2}^2 = (106 \gev)^2 & m_{S_3}^2 = (1590 \gev)^2 \\
               h = -0.613                 & A_h       = 631 \gev   & \mu = h v_s = -164 \gev \\
               \lambda = 0.177          & A_\lambda = 1745 \gev  & \\
               M_1      = -87  \gev   & M_1^\prime = 350 \gev & M_2 = -306 \gev \\
               m_{S S_1}^2 = -(620 \gev)^2 & m_{S S_2}^2 = -(827 \gev)^2 &
\end{eqnarray*}
Branching Ratios for dominant decay modes (greater than 1\% excluding model-dependent squark, slepton, $Z^\prime$ and exotic decays; $\chi^0_{i>1}$ are summed): 
\begin{flushleft}\begin{tabular}{c|lr|lr|lr|lr|lr|lr}
$H_1$ & $\chi^0_1 \chi^0_1$ & 44\% & $\chi^0_1 \chi^0_{i>1}$ & 41\% & $\chi^0_{i>1} \chi^0_{i>1}$ & 12\% & $b\bar{b}$ & 3\% &&&&\\
$H_2$ & $\chi^0_{i>1} \chi^0_{i>1}$ & 57\% & $\chi^0_1 \chi^0_{i>1}$ & 25\% & $\chi^0_1 \chi^0_1$ & 17\% &&&&&&\\
$H_3$ & $t\bar{t}$ & 57\% & $\chi^0_{i>1} \chi^0_{i>1}$ & 20\% & $\chi^0_1 \chi^0_{i>1}$ & 19\% & $H_1 H_1$ & 3\% & $W^+W^-$ & 1\% &&\\
$H_4$ & $\chi^0_{i>1} \chi^0_{i>1}$ & 20\% & $H_1 H_1$ & 20\% & $t\bar{t}$ & 16\% & $\chi^0_1 \chi^0_{i>1}$ & 14\% & $W^+W^-$ & 14\% & $\chi^+_1 \chi^-_1$ & 7\% \\
$H_5$ & $\chi^0_{i>1} \chi^0_{i>1}$ & 54\% & $\chi^0_1 \chi^0_{i>1}$ & 28\% & $H_4 H_4$ & 10\% & $H_2 H_2$ & 3\% & $\chi^0_1 \chi^0_1$ & 2\% & $H_2 H_4$ & 1\% \\
$H_6$ & $\chi^0_{i>1} \chi^0_{i>1}$ & 45\% & $\chi^+_1 \chi^-_1$ & 29\% & $A_1 A_2$ & 7\% & $\chi^+_1 \chi^-_2$ & 6\% & $\chi^0_1 \chi^0_{i>1}$ & 5\% & $H_3 H_3$ & 3\% \\
\hline
$A_1$ & $\chi^0_{i>1} \chi^0_{i>1}$ & 55\% & $\chi^0_1, \chi^0_{i>1}$ & 24\% & $\chi^0_1 \chi^0_1$ & 22\% &&&&&&\\
$A_2$ & $t\bar{t}$ & 50\% & $\chi^0_{i>1} \chi^0_{i>1}$ & 27\% & $\chi^+_1 \chi^-_1$ & 12\% & $\chi^0_1, \chi^0_{i>1}$ & 9\% & $\chi^0_1 \chi^0_1$ & 2\% &&\\
$A_3$ & $H_2 Z$ & 99\% &&&&&&&&&&\\
$A_4$ & $H_2 Z$ & 36\% & $\chi^0_{i>1} \chi^0_{i>1}$ & 33\% & $\chi^+_1 \chi^-_1$ & 21\% & $\chi^+_1 \chi^-_2$ & 4\% & $\chi^0_1, \chi^0_{i>1}$ & 3\% & $\chi^+_2 \chi^-_2$ & 1\% \\
\end{tabular}
\end{flushleft}
\newpage

Eigenvectors/rotation matrices 
 \[R_H = 
 \left[ \begin {array}{cccccc}\scriptstyle  0.68&\scriptstyle 0.73&\scriptstyle 0.00054&\scriptstyle- 0.033&\scriptstyle- 0.025&\scriptstyle-
 0.0086\\\noalign{\medskip}\scriptstyle- 0.0080&\scriptstyle\scriptstyle- 0.0097&\scriptstyle\scriptstyle 0.018&\scriptstyle\scriptstyle- 0.81&\scriptstyle\scriptstyle 0.58&\scriptstyle\scriptstyle-
 0.060\\\noalign{\medskip}\scriptstyle 0.73&\scriptstyle\scriptstyle- 0.68&\scriptstyle\scriptstyle- 0.013&\scriptstyle\scriptstyle- 0.043&\scriptstyle\scriptstyle- 0.062&\scriptstyle\scriptstyle- 0.019
\\\noalign{\medskip}\scriptstyle 0.085&\scriptstyle- 0.024&\scriptstyle 0.17&\scriptstyle 0.55&\scriptstyle 0.78&\scriptstyle 0.24
\\\noalign{\medskip}\scriptstyle 0.0012&\scriptstyle 0.0011&\scriptstyle 0.084&\scriptstyle 0.18&\scriptstyle 0.15&\scriptstyle- 0.97
\\\noalign{\medskip}\scriptstyle- 0.0052&\scriptstyle- 0.0053&\scriptstyle 0.98&\scriptstyle- 0.096&\scriptstyle- 0.16&\scriptstyle 0.041
\end {array} \right] 
\]
\[ R_A = 
 \left[ \begin {array}{cccccc}\scriptstyle - 0.015&\scriptstyle- 0.015&\scriptstyle 0.031&\scriptstyle 0.77&\scriptstyle- 0.64&\scriptstyle-
 0.042\\\noalign{\medskip}\scriptstyle 0.70&\scriptstyle\scriptstyle 0.71&\scriptstyle\scriptstyle- 0.0054&\scriptstyle\scriptstyle- 0.015&\scriptstyle\scriptstyle- 0.053&\scriptstyle\scriptstyle 0.011
\\\noalign{\medskip}\scriptstyle- 0.0013&\scriptstyle- 0.0013&\scriptstyle- 0.063&\scriptstyle 0.16&\scriptstyle 0.13&\scriptstyle 0.98
\\\noalign{\medskip}\scriptstyle 0.0099&\scriptstyle 0.0099&\scriptstyle 0.99&\scriptstyle 0.083&\scriptstyle 0.14&\scriptstyle 0.031
\\\noalign{\medskip}\scriptstyle- 0.060&\scriptstyle 0.13&\scriptstyle- 0.15&\scriptstyle 0.61&\scriptstyle 0.74&\scriptstyle- 0.21
\\\noalign{\medskip}\scriptstyle 0.71&\scriptstyle- 0.69&\scriptstyle- 0.021&\scriptstyle 0.083&\scriptstyle 0.10&\scriptstyle- 0.028
\end {array} \right] 
\]
\[ R_{\chi^0} = 
 \left[ \begin {array}{ccccccccc}\scriptstyle  0.25&\scriptstyle 0.0017&\scriptstyle- 0.00074&\scriptstyle 0.21&\scriptstyle 0.21&\scriptstyle
- 0.57&\scriptstyle\scriptstyle 0.29&\scriptstyle\scriptstyle- 0.47&\scriptstyle\scriptstyle 0.48\\\noalign{\medskip}\scriptstyle- 0.0019&\scriptstyle\scriptstyle 0.85&\scriptstyle\scriptstyle- 0.16&\scriptstyle\scriptstyle-
 0.35&\scriptstyle\scriptstyle 0.35&\scriptstyle\scriptstyle- 0.0013&\scriptstyle\scriptstyle- 0.0013&\scriptstyle\scriptstyle 0.0016&\scriptstyle\scriptstyle- 0.0035\\\noalign{\medskip}\scriptstyle-
 0.24&\scriptstyle- 0.00017&\scriptstyle 0.00012&\scriptstyle- 0.072&\scriptstyle- 0.073&\scriptstyle 0.16&\scriptstyle 0.66&\scriptstyle- 0.48&\scriptstyle- 0.49
\\\noalign{\medskip}\scriptstyle- 0.30&\scriptstyle- 0.0044&\scriptstyle- 0.000085&\scriptstyle 0.22&\scriptstyle 0.22&\scriptstyle- 0.66&\scriptstyle-
 0.19&\scriptstyle\scriptstyle 0.16&\scriptstyle\scriptstyle- 0.56\\\noalign{\medskip}\scriptstyle- 0.00024&\scriptstyle\scriptstyle 0.51&\scriptstyle\scriptstyle 0.40&\scriptstyle\scriptstyle 0.54&\scriptstyle\scriptstyle-
 0.54&\scriptstyle\scriptstyle- 0.0019&\scriptstyle\scriptstyle- 0.00021&\scriptstyle\scriptstyle- 0.0000081&\scriptstyle\scriptstyle- 0.00051\\\noalign{\medskip}\scriptstyle-
 0.0046&\scriptstyle 0.00028&\scriptstyle- 0.00029&\scriptstyle 0.63&\scriptstyle 0.63&\scriptstyle 0.44&\scriptstyle 0.018&\scriptstyle 0.011&\scriptstyle- 0.026
\\\noalign{\medskip}\scriptstyle 0.0000069&\scriptstyle- 0.073&\scriptstyle 0.90&\scriptstyle- 0.30&\scriptstyle 0.30&\scriptstyle 0.00022&\scriptstyle-
 0.000038&\scriptstyle\scriptstyle- 0.000035&\scriptstyle\scriptstyle- 0.000042\\\noalign{\medskip}\scriptstyle 0.48&\scriptstyle\scriptstyle 0.000014&\scriptstyle\scriptstyle-
 0.000055&\scriptstyle- 0.025&\scriptstyle- 0.025&\scriptstyle 0.093&\scriptstyle- 0.54&\scriptstyle- 0.58&\scriptstyle- 0.36
\\\noalign{\medskip}\scriptstyle- 0.75&\scriptstyle- 0.0000057&\scriptstyle 0.0000083&\scriptstyle- 0.014&\scriptstyle- 0.014&\scriptstyle
 0.077&\scriptstyle- 0.38&\scriptstyle- 0.43&\scriptstyle 0.31\end {array} \right] 
\]
\[ R_{\chi^+} = 
 \left[ \begin {array}{cccc}\scriptstyle - 0.29&\scriptstyle- 0.64&\scriptstyle 0.29&\scriptstyle 0.64
\\\noalign{\medskip}\scriptstyle 0.29&\scriptstyle 0.64&\scriptstyle 0.29&\scriptstyle 0.64\\\noalign{\medskip}\scriptstyle 0.64&\scriptstyle-
 0.29&\scriptstyle\scriptstyle 0.64&\scriptstyle\scriptstyle- 0.29\\\noalign{\medskip}\scriptstyle 0.64&\scriptstyle\scriptstyle- 0.29&\scriptstyle\scriptstyle- 0.64&\scriptstyle\scriptstyle 0.29
\end {array} \right] 
\]

    % This is DATA[94]
\section{Typical light $H_1 \rightarrow A_1 A_1$ dominant, $A_1 \rightarrow {\rm SM}$}
\label{h1a1a1dominant}
\begin{flushleft}
\begin{tabular}[b]{l|c|c|c|c|c|c|c|c|c}
  \multicolumn{2}{c|}{$\tan{\beta} = 1.99$} &   \multicolumn{2}{c|}{$M_{Z^\prime} = 1374$ GeV} &   \multicolumn{2}{c|}{$M_{H^+} = 118$ GeV} &   \multicolumn{3}{c}{$\alpha_{ZZ^\prime} = $$8.9\!\cdot\!10^{-4}$} \\
\hline 
  $M_{H}$ & 137 & 147 & 183 & 1340 & 1924 & 2167 &&& \\
  $\xi_{\rm MSSM}$ & 0.19 & 0.81 & 1 & $4.7\!\cdot\!10^{-4}$ & $1.4\!\cdot\!10^{-5}$ & $2.0\!\cdot\!10^{-5}$ &&& \\
  $\sigma(H_i Z)$ & 4.1 &  13 &  31 & & & & & & \\
  $\sigma(H_i \nu\overline{\nu})$ & 5.4&  16&  30& & & & & & \\
  $\sigma(H_ie^+e^-)$ & 0.52& 1.6& 2.9& & & & & & \\
\hline 
  $M_{A}$ & 66 & 190 & 1904 & 2167 & 0 & 0 &&& \\
  $\xi_{\rm MSSM}$ & $7.6\!\cdot\!10^{-4}$ & 1 & 0 & $2.2\!\cdot\!10^{-5}$ & 1 & $2.8\!\cdot\!10^{-4}$ &&& \\
$\sigma(H_1 A)$  & 0.0034 & 2.5 & & & & & & & \\
$\sigma(H_2 A)$  & 0.016 &  12 & & & & & & & \\
$\sigma(H_3 A)$  & 0.0075 & 4.9 & & & & & & & \\
\hline 
  $M_{\chi^0}$               & 71 & 130 & 157 & 231 & 394 & 481 & 661 & 1095 & 1808 \\
  $\xi_{\rm MSSM}$          & 0.72 & 0.99 & 0.62 & 0.66 & $3.2\!\cdot\!10^{-5}$ & $2.3\!\cdot\!10^{-5}$ & 1 & $1.9\!\cdot\!10^{-4}$ & $1.8\!\cdot\!10^{-4}$ \\
  $\xi_{\tilde{s}}$         & 0.28 & $7.6\!\cdot\!10^{-3}$ & 0.38 & 0.34 & 0.99 & 1 & $1.5\!\cdot\!10^{-5}$ & 0.66 & 0.35 \\
  $\xi_{\tilde{Z^\prime}}$ & 0 & 0 & $1.0\!\cdot\!10^{-5}$ & 0 & $9.5\!\cdot\!10^{-3}$ & $4.2\!\cdot\!10^{-3}$ & 0 & 0.34 & 0.65 \\
\hline 
  $M_{\chi^+}$ & 104 & 661 &&&&&&& \\
\end{tabular}
\end{flushleft}
\vspace{-10pt}
Cross sections quoted are in fb for a linear $e^+ e^-$ collider at center-of-mass energy 500 GeV.\vspace{-10pt}
\begin{eqnarray*}
  v_2       = 156 \gev   & v_1       = 78 \gev   & v_s       = 116 \gev   \\
               v_{s1}    = 2059 \gev   & v_{s2}    = 351 \gev   & v_{s3}    = 1827 \gev   \\
               m_{H_u}^2 = (203 \gev)^2 & m_{H_d}^2 = (239 \gev)^2 & m_{S}^2   = (2125 \gev)^2 \\
               m_{S_1}^2 = (369 \gev)^2 & m_{S_2}^2 = (1824 \gev)^2 & m_{S_3}^2 = -(496 \gev)^2 \\
               h = 0.978                 & A_h       = 129 \gev   & \mu = h v_s = 113 \gev \\
               \lambda = 0.169          & A_\lambda = 1867 \gev  & \\
               M_1      = -122  \gev   & M_1^\prime = 799 \gev & M_2 = 650 \gev \\
               m_{S S_1}^2 = -(495 \gev)^2 & m_{S S_2}^2 = -(313 \gev)^2 &
\end{eqnarray*}
Branching Ratios for dominant decay modes (greater than 1\% excluding model-dependent squark, slepton, $Z^\prime$ and exotic decays; $\chi^0_{i>1}$ are summed): 
\begin{flushleft}\begin{tabular}{c|lr|lr|lr|lr|lr|lr}
$H_1$ & $A_1 A_1$ & 100\% &&&&&&&&&&\\
$H_2$ & $A_1 A_1$ & 87\% & $\chi^0_1 \chi^0_1$ & 11\% & $b\bar{b}$ & 2\% &&&&&&\\
$H_3$ & $W^+W^-$ & 64\% & $\chi^0_1 \chi^0_1$ & 27\% & $Z Z$ & 8\% & $b\bar{b}$ & 1\% &&&&\\
$H_4$ & $H_3 H_3$ & 20\% & $W^+W^-$ & 19\% & $H_2 H_2$ & 17\% & $H_1 H_1$ & 10\% & $H^+ W^-$ & 10\% & $Z Z$ & 10\% \\
$H_5$ & $\chi^0_{i>1} \chi^0_{i>1}$ & 41\% & $H_1 H_4$ & 23\% & $A_1 A_1$ & 9\% & $\chi^+_1 \chi^-_1$ & 8\% & $H_2 H_4$ & 6\% & $\chi^0_1 \chi^0_{i>1}$ & 5\% \\
$H_6$ & $\chi^+_1 \chi^-_1$ & 48\% & $\chi^0_1 \chi^0_{i>1}$ & 31\% & $\chi^0_{i>1} \chi^0_{i>1}$ & 18\% & $\chi^0_1 \chi^0_1$ & 2\% &&&&\\
\hline
$A_1$ & $b\bar{b}$ & 92\% & $\tau^+ \tau^-$ & 5\% & $c\bar{c}$ & 2\% &&&&&&\\
$A_2$ & $\chi^0_1 \chi^0_1$ & 99\% & $b\bar{b}$ & 1\% &&&&&&&&\\
$A_3$ & $H_2 Z$ & 100\% &&&&&&&&&&\\
$A_4$ & $\chi^+_1 \chi^-_1$ & 49\% & $\chi^0_1, \chi^0_{i>1}$ & 31\% & $\chi^0_{i>1} \chi^0_{i>1}$ & 18\% & $\chi^0_1 \chi^0_1$ & 2\% &&&&\\
\end{tabular}
\end{flushleft}
\newpage

Eigenvectors/rotation matrices 
 \[R_H = 
 \left[ \begin {array}{cccccc}\scriptstyle  0.40&\scriptstyle- 0.18&\scriptstyle- 0.046&\scriptstyle- 0.76&\scriptstyle- 0.20&\scriptstyle-
 0.44\\\noalign{\medskip}\scriptstyle 0.78&\scriptstyle\scriptstyle- 0.45&\scriptstyle\scriptstyle 0.017&\scriptstyle\scriptstyle 0.36&\scriptstyle\scriptstyle 0.097&\scriptstyle\scriptstyle 0.22
\\\noalign{\medskip}\scriptstyle- 0.48&\scriptstyle- 0.87&\scriptstyle 0.00072&\scriptstyle- 0.023&\scriptstyle- 0.0098&\scriptstyle- 0.034
\\\noalign{\medskip}\scriptstyle 0.019&\scriptstyle 0.0094&\scriptstyle 0.055&\scriptstyle 0.53&\scriptstyle- 0.22&\scriptstyle- 0.82
\\\noalign{\medskip}\scriptstyle 0.0035&\scriptstyle 0.0014&\scriptstyle 0.092&\scriptstyle- 0.079&\scriptstyle 0.95&\scriptstyle- 0.30
\\\noalign{\medskip}\scriptstyle 0.0044&\scriptstyle- 0.00088&\scriptstyle 0.99&\scriptstyle- 0.064&\scriptstyle- 0.086&\scriptstyle 0.049
\end {array} \right] 
\]
\[ R_A = 
 \left[ \begin {array}{cccccc}\scriptstyle  0.012&\scriptstyle 0.025&\scriptstyle- 0.040&\scriptstyle 0.86&\scriptstyle- 0.23&\scriptstyle
 0.46\\\noalign{\medskip}\scriptstyle 0.45&\scriptstyle\scriptstyle 0.89&\scriptstyle\scriptstyle- 0.0031&\scriptstyle\scriptstyle- 0.032&\scriptstyle\scriptstyle 0.0052&\scriptstyle\scriptstyle 0.0017
\\\noalign{\medskip}\scriptstyle 0.00035&\scriptstyle 0.00069&\scriptstyle 0.13&\scriptstyle- 0.16&\scriptstyle- 0.96&\scriptstyle- 0.18
\\\noalign{\medskip}\scriptstyle 0.0021&\scriptstyle 0.0042&\scriptstyle 0.99&\scriptstyle 0.068&\scriptstyle 0.12&\scriptstyle 0.018
\\\noalign{\medskip}\scriptstyle 0.89&\scriptstyle- 0.45&\scriptstyle- 0.000033&\scriptstyle 0.00059&\scriptstyle 0.00010&\scriptstyle- 0.0010
\\\noalign{\medskip}\scriptstyle 0.0063&\scriptstyle 0.015&\scriptstyle- 0.027&\scriptstyle 0.49&\scriptstyle 0.083&\scriptstyle- 0.87
\end {array} \right] 
\]
\[ R_{\chi^0} = 
 \left[ \begin {array}{ccccccccc}\scriptstyle  0.0028&\scriptstyle- 0.18&\scriptstyle- 0.11&\scriptstyle 0.16&\scriptstyle- 0.81&\scriptstyle
 0.52&\scriptstyle\scriptstyle 0.021&\scriptstyle\scriptstyle 0.0022&\scriptstyle\scriptstyle- 0.019\\\noalign{\medskip}\scriptstyle 0.00022&\scriptstyle\scriptstyle- 0.98&\scriptstyle\scriptstyle 0.019
&\scriptstyle\scriptstyle- 0.14&\scriptstyle\scriptstyle 0.13&\scriptstyle\scriptstyle- 0.087&\scriptstyle\scriptstyle- 0.0031&\scriptstyle\scriptstyle 0.00033&\scriptstyle\scriptstyle 0.0025\\\noalign{\medskip}\scriptstyle-
 0.0032&\scriptstyle- 0.084&\scriptstyle- 0.087&\scriptstyle 0.75&\scriptstyle- 0.22&\scriptstyle- 0.62&\scriptstyle- 0.014&\scriptstyle- 0.0028&\scriptstyle 0.013
\\\noalign{\medskip}\scriptstyle 0.00012&\scriptstyle 0.072&\scriptstyle- 0.016&\scriptstyle- 0.63&\scriptstyle- 0.52&\scriptstyle- 0.58&\scriptstyle-
 0.00074&\scriptstyle\scriptstyle 0.00021&\scriptstyle\scriptstyle 0.00048\\\noalign{\medskip}\scriptstyle- 0.097&\scriptstyle\scriptstyle- 0.00065&\scriptstyle\scriptstyle
 0.00031&\scriptstyle 0.0017&\scriptstyle 0.0054&\scriptstyle- 0.0076&\scriptstyle 0.68&\scriptstyle- 0.70&\scriptstyle 0.22
\\\noalign{\medskip}\scriptstyle 0.064&\scriptstyle 0.00014&\scriptstyle 0.00089&\scriptstyle 0.0030&\scriptstyle 0.0036&\scriptstyle- 0.0066&\scriptstyle
 0.59&\scriptstyle\scriptstyle 0.69&\scriptstyle\scriptstyle 0.41\\\noalign{\medskip}\scriptstyle 0.00067&\scriptstyle\scriptstyle- 0.0079&\scriptstyle\scriptstyle 0.99&\scriptstyle\scriptstyle 0.075&\scriptstyle\scriptstyle-
 0.12&\scriptstyle\scriptstyle- 0.0035&\scriptstyle\scriptstyle 0.00032&\scriptstyle\scriptstyle- 0.00049&\scriptstyle\scriptstyle- 0.0014\\\noalign{\medskip}\scriptstyle- 0.58&\scriptstyle\scriptstyle-
 0.00041&\scriptstyle 0.00041&\scriptstyle 0.0055&\scriptstyle 0.013&\scriptstyle- 0.018&\scriptstyle 0.34&\scriptstyle 0.19&\scriptstyle- 0.71
\\\noalign{\medskip}\scriptstyle- 0.80&\scriptstyle- 0.00017&\scriptstyle 0.00052&\scriptstyle- 0.0064&\scriptstyle- 0.012&\scriptstyle 0.018&\scriptstyle
- 0.28&\scriptstyle 0.0014&\scriptstyle 0.52\end {array} \right] 
\]
\[ R_{\chi^+} = 
 \left[ \begin {array}{cccc}\scriptstyle  0.12&\scriptstyle- 0.70&\scriptstyle- 0.074&\scriptstyle 0.70
\\\noalign{\medskip}\scriptstyle- 0.12&\scriptstyle 0.70&\scriptstyle- 0.074&\scriptstyle 0.70\\\noalign{\medskip}\scriptstyle
 0.70&\scriptstyle\scriptstyle 0.12&\scriptstyle\scriptstyle- 0.70&\scriptstyle\scriptstyle- 0.074\\\noalign{\medskip}\scriptstyle 0.70&\scriptstyle\scriptstyle 0.12&\scriptstyle\scriptstyle 0.70&\scriptstyle\scriptstyle 0.074
\end {array} \right] 
\]

    % This is DATA[59]
\section{Typical light $A_1 \rightarrow $ invisible dominant}
\label{a1invisible}
\begin{flushleft}
\begin{tabular}[b]{l|c|c|c|c|c|c|c|c|c}
  \multicolumn{2}{c|}{$\tan{\beta} = 1.08$} &   \multicolumn{2}{c|}{$M_{Z^\prime} = 2831$ GeV} &   \multicolumn{2}{c|}{$M_{H^+} = 622$ GeV} &   \multicolumn{3}{c}{$\alpha_{ZZ^\prime} = $$2.7\!\cdot\!10^{-5}$} \\
\hline 
  $M_{H}$ & 116 & 564 & 629 & 2739 & 3077 & 8917 &&& \\
  $\xi_{\rm MSSM}$ & 1 & $1.9\!\cdot\!10^{-4}$ & 1 & $4.3\!\cdot\!10^{-5}$ & $1.3\!\cdot\!10^{-4}$ & 0 &&& \\
  $\sigma(H_i Z)$ &  58 & & & & & & & & \\
  $\sigma(H_i \nu\overline{\nu})$ &  88& & & & & & & & \\
  $\sigma(H_ie^+e^-)$ & 8.5& & & & & & & & \\
\hline 
  $M_{A}$ & 78 & 621 & 3045 & 8916 & 0 & 0 &&& \\
  $\xi_{\rm MSSM}$ & $3.0\!\cdot\!10^{-4}$ & 1 & $1.5\!\cdot\!10^{-4}$ & 0 & 1 & $1.0\!\cdot\!10^{-4}$ &&& \\
$\sigma(H_1 A)$  & $5.0\!\cdot\!10^{-6}$ & & & & & & & & \\
\hline 
  $M_{\chi^0}$               & 36 & 159 & 176 & 191 & 335 & 666 & 696 & 2236 & 3630 \\
  $\xi_{\rm MSSM}$          & 0.17 & 1 & 0.83 & 1 & 1 & 0 & 0 & $5.2\!\cdot\!10^{-5}$ & $4.1\!\cdot\!10^{-5}$ \\
  $\xi_{\tilde{s}}$         & 0.83 & $4.1\!\cdot\!10^{-4}$ & 0.17 & $9.0\!\cdot\!10^{-4}$ & 0 & 1 & 1 & 0.63 & 0.38 \\
  $\xi_{\tilde{Z^\prime}}$ & 0 & 0 & 0 & 0 & 0 & $4.9\!\cdot\!10^{-3}$ & $3.0\!\cdot\!10^{-3}$ & 0.37 & 0.62 \\
\hline 
  $M_{\chi^+}$ & 154 & 335 &&&&&&& \\
\end{tabular}
\end{flushleft}
\vspace{-10pt}
Cross sections quoted are in fb for a linear $e^+ e^-$ collider at center-of-mass energy 500 GeV.\vspace{-10pt}
\begin{eqnarray*}
  v_2       = 128 \gev   & v_1       = 118 \gev   & v_s       = 308 \gev   \\
               v_{s1}    = 118 \gev   & v_{s2}    = 4171 \gev   & v_{s3}    = 3800 \gev   \\
               m_{H_u}^2 = (673 \gev)^2 & m_{H_d}^2 = (702 \gev)^2 & m_{S}^2   = (2934 \gev)^2 \\
               m_{S_1}^2 = (8914 \gev)^2 & m_{S_2}^2 = (699 \gev)^2 & m_{S_3}^2 = -(1201 \gev)^2 \\
               h = 0.456                 & A_h       = 1371 \gev   & \mu = h v_s = 140 \gev \\
               \lambda = 0.128          & A_\lambda = 4510 \gev  & \\
               M_1      = 185  \gev   & M_1^\prime = -1425 \gev & M_2 = 321 \gev \\
               m_{S S_1}^2 = -(770 \gev)^2 & m_{S S_2}^2 = -(814 \gev)^2 &
\end{eqnarray*}
Branching Ratios for dominant decay modes (greater than 1\% excluding model-dependent squark, slepton, $Z^\prime$ and exotic decays; $\chi^0_{i>1}$ are summed): 
\begin{flushleft}\begin{tabular}{c|lr|lr|lr|lr|lr|lr}
$H_1$ & $\chi^0_1 \chi^0_1$ & 97\% & $b\bar{b}$ & 3\% &&&&&&&&\\
$H_2$ & $H_1 H_1$ & 26\% & $W^+W^-$ & 21\% & $\chi^+_1 \chi^-_1$ & 15\% & $\chi^0_{i>1} \chi^0_{i>1}$ & 12\% & $Z Z$ & 10\% & $t\bar{t}$ & 8\% \\
$H_3$ & $t\bar{t}$ & 70\% & $\chi^0_1 \chi^0_{i>1}$ & 14\% & $\chi^0_{i>1} \chi^0_{i>1}$ & 10\% & $\chi^+_1 \chi^-_2$ & 4\% & $H_1 H_1$ & 1\% &&\\
$H_4$ & $A_1 A_2$ & 40\% & $H_3 H_3$ & 19\% & $\chi^+_1 \chi^-_1$ & 13\% & $\chi^0_{i>1} \chi^0_{i>1}$ & 13\% & $H_1 H_1$ & 4\% & $W^+W^-$ & 4\% \\
$H_5$ & $\chi^+_1 \chi^-_1$ & 37\% & $\chi^0_{i>1} \chi^0_{i>1}$ & 33\% & $\chi^0_1 \chi^0_{i>1}$ & 7\% & $A_1 A_2$ & 7\% & $H_1 H_1$ & 4\% & $W^+W^-$ & 4\% \\
$H_6$ & $\chi^0_{i>1} \chi^0_{i>1}$ & 96\% & $H_2 H_4$ & 2\% & $H_2 H_2$ & 1\% &&&&&&\\
\hline
$A_1$ & $\chi^0_1 \chi^0_1$ & 99\% & $b\bar{b}$ & 1\% &&&&&&&&\\
$A_2$ & $t\bar{t}$ & 73\% & $\chi^0_{i>1} \chi^0_{i>1}$ & 12\% & $\chi^0_1, \chi^0_{i>1}$ & 7\% & $\chi^0_1 \chi^0_1$ & 5\% & $\chi^+_1 \chi^-_1$ & 2\% &&\\
$A_3$ & $H_2 Z$ & 100\% &&&&&&&&&&\\
$A_4$ & $\chi^0_{i>1} \chi^0_{i>1}$ & 67\% & $H_5 Z$ & 27\% & $A_1 H_2$ & 4\% & $A_2 H_5$ & 1\% &&&&\\
\end{tabular}
\end{flushleft}
\newpage

Eigenvectors/rotation matrices 
 \[R_H = 
 \left[ \begin {array}{cccccc}\scriptstyle - 0.72&\scriptstyle- 0.69&\scriptstyle- 0.0098&\scriptstyle- 0.00066&\scriptstyle-
 0.0059&\scriptstyle\scriptstyle- 0.014\\\noalign{\medskip}\scriptstyle 0.013&\scriptstyle\scriptstyle 0.0038&\scriptstyle\scriptstyle- 0.069&\scriptstyle\scriptstyle- 0.039&\scriptstyle\scriptstyle-
 0.89&\scriptstyle\scriptstyle- 0.45\\\noalign{\medskip}\scriptstyle 0.69&\scriptstyle\scriptstyle- 0.72&\scriptstyle\scriptstyle- 0.00030&\scriptstyle\scriptstyle 0.00025&\scriptstyle\scriptstyle
 0.0057&\scriptstyle\scriptstyle 0.0032\\\noalign{\medskip}\scriptstyle 0.0050&\scriptstyle\scriptstyle 0.0042&\scriptstyle\scriptstyle 0.31&\scriptstyle\scriptstyle- 0.015&\scriptstyle\scriptstyle 0.41
&\scriptstyle\scriptstyle- 0.86\\\noalign{\medskip}\scriptstyle 0.0079&\scriptstyle\scriptstyle 0.0085&\scriptstyle\scriptstyle- 0.95&\scriptstyle\scriptstyle- 0.011&\scriptstyle\scriptstyle 0.20&\scriptstyle\scriptstyle- 0.24
\\\noalign{\medskip}\scriptstyle- 0.000028&\scriptstyle- 0.000027&\scriptstyle 0.0083&\scriptstyle- 1.0&\scriptstyle 0.026&\scriptstyle 0.033
\end {array} \right] 
\]
\[ R_A = 
 \left[ \begin {array}{cccccc}\scriptstyle  0.012&\scriptstyle 0.013&\scriptstyle- 0.060&\scriptstyle- 0.038&\scriptstyle 0.87&\scriptstyle
 0.48\\\noalign{\medskip}\scriptstyle- 0.68&\scriptstyle\scriptstyle- 0.73&\scriptstyle\scriptstyle 0.011&\scriptstyle\scriptstyle- 0.00064&\scriptstyle\scriptstyle 0.021&\scriptstyle\scriptstyle-
 0.00045\\\noalign{\medskip}\scriptstyle- 0.0083&\scriptstyle\scriptstyle- 0.0090&\scriptstyle\scriptstyle- 1.0&\scriptstyle\scriptstyle 0.011&\scriptstyle\scriptstyle- 0.070&\scriptstyle\scriptstyle
 0.0022\\\noalign{\medskip}\scriptstyle 0.0000081&\scriptstyle\scriptstyle 0.0000088&\scriptstyle\scriptstyle 0.0087&\scriptstyle\scriptstyle 1.0&\scriptstyle\scriptstyle 0.028&\scriptstyle\scriptstyle
 0.030\\\noalign{\medskip}\scriptstyle 0.73&\scriptstyle\scriptstyle- 0.68&\scriptstyle\scriptstyle 0.000069&\scriptstyle\scriptstyle- 0.000026&\scriptstyle\scriptstyle- 0.00094&\scriptstyle\scriptstyle
 0.0017\\\noalign{\medskip}\scriptstyle 0.0082&\scriptstyle\scriptstyle 0.0060&\scriptstyle\scriptstyle- 0.035&\scriptstyle\scriptstyle 0.014&\scriptstyle\scriptstyle 0.48&\scriptstyle\scriptstyle- 0.88
\end {array} \right] 
\]
\[ R_{\chi^0} = 
 \left[ \begin {array}{ccccccccc}\scriptstyle  0.00072&\scriptstyle- 0.0067&\scriptstyle 0.0075&\scriptstyle 0.28&\scriptstyle
 0.30&\scriptstyle\scriptstyle- 0.91&\scriptstyle\scriptstyle- 0.00084&\scriptstyle\scriptstyle- 0.028&\scriptstyle\scriptstyle 0.026\\\noalign{\medskip}\scriptstyle 0.000025&\scriptstyle\scriptstyle
 0.12&\scriptstyle- 0.16&\scriptstyle 0.69&\scriptstyle- 0.69&\scriptstyle- 0.020&\scriptstyle- 0.000032&\scriptstyle- 0.00028&\scriptstyle 0.00025
\\\noalign{\medskip}\scriptstyle- 0.00036&\scriptstyle 0.081&\scriptstyle- 0.0095&\scriptstyle- 0.65&\scriptstyle- 0.63&\scriptstyle- 0.41&\scriptstyle
 0.00051&\scriptstyle\scriptstyle- 0.0045&\scriptstyle\scriptstyle 0.0040\\\noalign{\medskip}\scriptstyle 0.000019&\scriptstyle\scriptstyle 0.99&\scriptstyle\scriptstyle 0.071&\scriptstyle\scriptstyle-
 0.026&\scriptstyle\scriptstyle 0.13&\scriptstyle\scriptstyle 0.030&\scriptstyle\scriptstyle- 0.000026&\scriptstyle\scriptstyle 0.00021&\scriptstyle\scriptstyle- 0.00019\\\noalign{\medskip}\scriptstyle-
 0.000021&\scriptstyle 0.050&\scriptstyle- 0.98&\scriptstyle- 0.11&\scriptstyle 0.13&\scriptstyle 0.0025&\scriptstyle 0.000036&\scriptstyle- 0.00014&\scriptstyle
 0.00011\\\noalign{\medskip}\scriptstyle 0.070&\scriptstyle\scriptstyle 0.000021&\scriptstyle\scriptstyle- 0.000052&\scriptstyle\scriptstyle 0.0015&\scriptstyle\scriptstyle
 0.0017&\scriptstyle\scriptstyle- 0.010&\scriptstyle\scriptstyle- 0.70&\scriptstyle\scriptstyle 0.65&\scriptstyle\scriptstyle 0.29\\\noalign{\medskip}\scriptstyle- 0.054&\scriptstyle\scriptstyle-
 0.0000082&\scriptstyle 0.000013&\scriptstyle 0.0018&\scriptstyle 0.0019&\scriptstyle- 0.0082&\scriptstyle 0.71&\scriptstyle 0.61&\scriptstyle 0.36
\\\noalign{\medskip}\scriptstyle 0.61&\scriptstyle 0.000013&\scriptstyle- 0.000025&\scriptstyle 0.0049&\scriptstyle 0.0053&\scriptstyle- 0.027
&\scriptstyle\scriptstyle 0.10&\scriptstyle\scriptstyle 0.35&\scriptstyle\scriptstyle- 0.70\\\noalign{\medskip}\scriptstyle 0.79&\scriptstyle\scriptstyle 0.0000054&\scriptstyle\scriptstyle- 0.0000096&\scriptstyle\scriptstyle-
 0.0044&\scriptstyle- 0.0047&\scriptstyle 0.022&\scriptstyle 0.033&\scriptstyle- 0.29&\scriptstyle 0.54\end {array} \right] 
\]
\[ R_{\chi^+} = 
 \left[ \begin {array}{cccc}\scriptstyle  0.13&\scriptstyle- 0.70&\scriptstyle- 0.11&\scriptstyle 0.70
\\\noalign{\medskip}\scriptstyle 0.13&\scriptstyle- 0.70&\scriptstyle 0.11&\scriptstyle- 0.70\\\noalign{\medskip}\scriptstyle 0.70
&\scriptstyle\scriptstyle 0.13&\scriptstyle\scriptstyle- 0.70&\scriptstyle\scriptstyle- 0.11\\\noalign{\medskip}\scriptstyle 0.70&\scriptstyle\scriptstyle 0.13&\scriptstyle\scriptstyle 0.70&\scriptstyle\scriptstyle 0.11
\end {array} \right] 
\]

    % This is DATA[4]
\section{Typical heavy $H_1 \rightarrow W^+ W^-$ dominant}
\label{h1wwdominant}
\begin{flushleft}
\begin{tabular}[b]{l|c|c|c|c|c|c|c|c|c}
  \multicolumn{2}{c|}{$\tan{\beta} = 0.909$} &   \multicolumn{2}{c|}{$M_{Z^\prime} = 1200$ GeV} &   \multicolumn{2}{c|}{$M_{H^+} = 326$ GeV} &   \multicolumn{3}{c}{$\alpha_{ZZ^\prime} = $-$1.8\!\cdot\!10^{-4}$} \\
\hline 
  $M_{H}$ & 178 & 346 & 367 & 1051 & 1224 & 3379 &&& \\
  $\xi_{\rm MSSM}$ & 1 & 1 & $1.2\!\cdot\!10^{-3}$ & $1.3\!\cdot\!10^{-4}$ & $4.1\!\cdot\!10^{-4}$ & $1.4\!\cdot\!10^{-5}$ &&& \\
  $\sigma(H_i Z)$ &  47 & 0.18 & 0.0025 & & & & & & \\
  $\sigma(H_i \nu\overline{\nu})$ &  47& 0.051& 0.00061& & & & & & \\
  $\sigma(H_ie^+e^-)$ & 4.5& 0.0049& $5.9\!\cdot\!10^{-5}$& & & & & & \\
\hline 
  $M_{A}$ & 351 & 367 & 1082 & 3378 & 0 & 0 &&& \\
  $\xi_{\rm MSSM}$ & 0.91 & 0.091 & 0 & $4.9\!\cdot\!10^{-5}$ & 1 & $5.7\!\cdot\!10^{-4}$ &&& \\
\hline 
  $M_{\chi^0}$               & 97 & 111 & 203 & 209 & 232 & 339 & 546 & 847 & 1711 \\
  $\xi_{\rm MSSM}$          & 0.81 & 0.5 & $3.1\!\cdot\!10^{-5}$ & $4.7\!\cdot\!10^{-5}$ & 0.68 & 1 & 1 & $2.7\!\cdot\!10^{-4}$ & $2.1\!\cdot\!10^{-4}$ \\
  $\xi_{\tilde{s}}$         & 0.19 & 0.5 & 1 & 1 & 0.32 & 0 & 0 & 0.67 & 0.33 \\
  $\xi_{\tilde{Z^\prime}}$ & 0 & $1.2\!\cdot\!10^{-5}$ & $1.5\!\cdot\!10^{-3}$ & $5.6\!\cdot\!10^{-4}$ & 0 & 0 & 0 & 0.33 & 0.67 \\
\hline 
  $M_{\chi^+}$ & 109 & 546 &&&&&&& \\
\end{tabular}
\end{flushleft}
\vspace{-10pt}
Cross sections quoted are in fb for a linear $e^+ e^-$ collider at center-of-mass energy 500 GeV.\vspace{-10pt}
\begin{eqnarray*}
  v_2       = 117 \gev   & v_1       = 129 \gev   & v_s       = 137 \gev   \\
               v_{s1}    = 3643 \gev   & v_{s2}    = 436 \gev   & v_{s3}    = 114 \gev   \\
               m_{H_u}^2 = -(574 \gev)^2 & m_{H_d}^2 = -(573 \gev)^2 & m_{S}^2   = (3473 \gev)^2 \\
               m_{S_1}^2 = -(837 \gev)^2 & m_{S_2}^2 = -(720 \gev)^2 & m_{S_3}^2 = (1572 \gev)^2 \\
               h = 0.907                 & A_h       = 502 \gev   & \mu = h v_s = 124 \gev \\
               \lambda = 0.058          & A_\lambda = 1343 \gev  & \\
               M_1      = 332  \gev   & M_1^\prime = 871 \gev & M_2 = 531 \gev \\
               m_{S S_1}^2 = -(601 \gev)^2 & m_{S S_2}^2 = -(729 \gev)^2 &
\end{eqnarray*}
Branching Ratios for dominant decay modes (greater than 1\% excluding model-dependent squark, slepton, $Z^\prime$ and exotic decays; $\chi^0_{i>1}$ are summed): 
\begin{flushleft}\begin{tabular}{c|lr|lr|lr|lr|lr|lr}
$H_1$ & $W^+W^-$ & 99\% & $b\bar{b}$ & 1\% &&&&&&&&\\
$H_2$ & $\chi^0_1 \chi^0_1$ & 38\% & $\chi^0_1 \chi^0_{i>1}$ & 30\% & $\chi^0_{i>1} \chi^0_{i>1}$ & 24\% & $W^+W^-$ & 5\% & $Z Z$ & 2\% &&\\
$H_3$ & $\chi^0_1 \chi^0_{i>1}$ & 30\% & $\chi^+_1 \chi^-_1$ & 27\% & $W^+W^-$ & 15\% & $H_1 H_1$ & 11\% & $\chi^0_{i>1} \chi^0_{i>1}$ & 8\% & $Z Z$ & 7\% \\
$H_4$ & $H_3 H_3$ & 24\% & $H_1 H_1$ & 16\% & $W^+W^-$ & 14\% & $H_2 H_2$ & 13\% & $\chi^0_{i>1} \chi^0_{i>1}$ & 12\% & $A_1 A_1$ & 11\% \\
$H_5$ & $H_3 H_3$ & 21\% & $H_2 H_2$ & 16\% & $A_1 A_1$ & 14\% & $H_1 H_1$ & 13\% & $\chi^0_{i>1} \chi^0_{i>1}$ & 13\% & $W^+W^-$ & 12\% \\
$H_6$ & $\chi^+_1 \chi^-_1$ & 45\% & $\chi^0_{i>1} \chi^0_{i>1}$ & 22\% & $\chi^0_1 \chi^0_{i>1}$ & 22\% & $\chi^0_1 \chi^0_1$ & 8\% & $H_2 H_2$ & 1\% & $A_1 A_1$ & 1\% \\
\hline
$A_1$ & $\chi^0_1, \chi^0_{i>1}$ & 63\% & $\chi^0_{i>1} \chi^0_{i>1}$ & 28\% & $\chi^+_1 \chi^-_1$ & 5\% & $H_1 Z$ & 2\% & $t\bar{t}$ & 1\% & $\chi^0_1 \chi^0_1$ & 1\% \\
$A_2$ & $H_1 Z$ & 94\% & $\chi^0_1, \chi^0_{i>1}$ & 2\% & $\chi^0_{i>1} \chi^0_{i>1}$ & 2\% & $\chi^+_1 \chi^-_1$ & 1\% & $t\bar{t}$ & 1\% &&\\
$A_3$ & $H_1 Z$ & 53\% & $\chi^0_{i>1} \chi^0_{i>1}$ & 39\% & $H_2 Z$ & 4\% & $A_1 H_1$ & 2\% & $\chi^+_1 \chi^-_1$ & 1\% &&\\
$A_4$ & $\chi^+_1 \chi^-_1$ & 46\% & $\chi^0_{i>1} \chi^0_{i>1}$ & 23\% & $\chi^0_1, \chi^0_{i>1}$ & 22\% & $\chi^0_1 \chi^0_1$ & 8\% & $H_5 Z$ & 1\% &&\\
\end{tabular}
\end{flushleft}
\newpage

Eigenvectors/rotation matrices 
 \[R_H = 
 \left[ \begin {array}{cccccc}\scriptstyle - 0.58&\scriptstyle- 0.81&\scriptstyle- 0.0031&\scriptstyle 0.024&\scriptstyle- 0.011&\scriptstyle-
 0.0022\\\noalign{\medskip}\scriptstyle- 0.81&\scriptstyle\scriptstyle 0.58&\scriptstyle\scriptstyle- 0.0023&\scriptstyle\scriptstyle 0.0055&\scriptstyle\scriptstyle- 0.030&\scriptstyle\scriptstyle-
 0.0087\\\noalign{\medskip}\scriptstyle- 0.034&\scriptstyle\scriptstyle 0.0079&\scriptstyle\scriptstyle 0.042&\scriptstyle\scriptstyle- 0.086&\scriptstyle\scriptstyle 0.95&\scriptstyle\scriptstyle 0.28
\\\noalign{\medskip}\scriptstyle 0.0076&\scriptstyle 0.0082&\scriptstyle 0.0062&\scriptstyle 0.44&\scriptstyle- 0.22&\scriptstyle 0.87
\\\noalign{\medskip}\scriptstyle 0.014&\scriptstyle 0.015&\scriptstyle 0.048&\scriptstyle 0.89&\scriptstyle 0.20&\scriptstyle- 0.41
\\\noalign{\medskip}\scriptstyle- 0.0030&\scriptstyle- 0.0023&\scriptstyle 1.0&\scriptstyle- 0.042&\scriptstyle- 0.049&\scriptstyle 0.0022
\end {array} \right] 
\]
\[ R_A = 
 \left[ \begin {array}{cccccc}\scriptstyle - 0.71&\scriptstyle- 0.64&\scriptstyle 0.018&\scriptstyle 0.063&\scriptstyle- 0.28&\scriptstyle
 0.082\\\noalign{\medskip}\scriptstyle 0.22&\scriptstyle\scriptstyle 0.20&\scriptstyle\scriptstyle 0.037&\scriptstyle\scriptstyle 0.12&\scriptstyle\scriptstyle- 0.91&\scriptstyle\scriptstyle 0.27
\\\noalign{\medskip}\scriptstyle- 0.00088&\scriptstyle- 0.00080&\scriptstyle- 0.016&\scriptstyle 0.025&\scriptstyle 0.28&\scriptstyle 0.96
\\\noalign{\medskip}\scriptstyle 0.0052&\scriptstyle 0.0047&\scriptstyle 1.0&\scriptstyle 0.032&\scriptstyle 0.048&\scriptstyle 0.0014
\\\noalign{\medskip}\scriptstyle 0.67&\scriptstyle- 0.74&\scriptstyle 0.00017&\scriptstyle- 0.0045&\scriptstyle- 0.00054&\scriptstyle 0.00028
\\\noalign{\medskip}\scriptstyle 0.020&\scriptstyle 0.012&\scriptstyle- 0.037&\scriptstyle 0.99&\scriptstyle 0.12&\scriptstyle- 0.062
\end {array} \right] 
\]
\[ R_{\chi^0} = 
 \left[ \begin {array}{ccccccccc}\scriptstyle - 0.0020&\scriptstyle 0.16&\scriptstyle- 0.15&\scriptstyle 0.79&\scriptstyle- 0.36&\scriptstyle-
 0.43&\scriptstyle\scriptstyle- 0.023&\scriptstyle\scriptstyle 0.0038&\scriptstyle\scriptstyle 0.0038\\\noalign{\medskip}\scriptstyle- 0.0035&\scriptstyle\scriptstyle- 0.098&\scriptstyle\scriptstyle
 0.093&\scriptstyle- 0.027&\scriptstyle 0.70&\scriptstyle- 0.70&\scriptstyle- 0.035&\scriptstyle 0.0067&\scriptstyle 0.0070
\\\noalign{\medskip}\scriptstyle 0.039&\scriptstyle- 0.000044&\scriptstyle 0.000059&\scriptstyle- 0.0041&\scriptstyle- 0.0038&\scriptstyle
 0.0041&\scriptstyle\scriptstyle- 0.16&\scriptstyle\scriptstyle 0.70&\scriptstyle\scriptstyle- 0.69\\\noalign{\medskip}\scriptstyle- 0.024&\scriptstyle\scriptstyle 0.000050&\scriptstyle\scriptstyle-
 0.000035&\scriptstyle- 0.0045&\scriptstyle- 0.0051&\scriptstyle 0.010&\scriptstyle- 0.026&\scriptstyle 0.70&\scriptstyle 0.71
\\\noalign{\medskip}\scriptstyle- 0.00023&\scriptstyle 0.0022&\scriptstyle- 0.0030&\scriptstyle 0.58&\scriptstyle 0.59&\scriptstyle 0.56&\scriptstyle
 0.0013&\scriptstyle\scriptstyle 0.0018&\scriptstyle\scriptstyle- 0.0018\\\noalign{\medskip}\scriptstyle 0.000043&\scriptstyle\scriptstyle 0.98&\scriptstyle\scriptstyle 0.073&\scriptstyle\scriptstyle-
 0.12&\scriptstyle\scriptstyle 0.12&\scriptstyle\scriptstyle- 0.0024&\scriptstyle\scriptstyle 0.00015&\scriptstyle\scriptstyle 0.000037&\scriptstyle\scriptstyle 0.000025\\\noalign{\medskip}\scriptstyle-
 0.00012&\scriptstyle- 0.039&\scriptstyle 0.98&\scriptstyle 0.14&\scriptstyle- 0.13&\scriptstyle 0.0010&\scriptstyle- 0.00027&\scriptstyle- 0.000040&\scriptstyle-
 0.000011\\\noalign{\medskip}\scriptstyle- 0.57&\scriptstyle\scriptstyle 0.000067&\scriptstyle\scriptstyle- 0.00010&\scriptstyle\scriptstyle 0.012&\scriptstyle\scriptstyle 0.011&\scriptstyle\scriptstyle
- 0.027&\scriptstyle\scriptstyle 0.80&\scriptstyle\scriptstyle 0.12&\scriptstyle\scriptstyle- 0.10\\\noalign{\medskip}\scriptstyle- 0.82&\scriptstyle\scriptstyle 0.000041&\scriptstyle\scriptstyle-
 0.000087&\scriptstyle- 0.011&\scriptstyle- 0.0099&\scriptstyle 0.023&\scriptstyle- 0.57&\scriptstyle- 0.068&\scriptstyle 0.019\end {array}
 \right] 
\]
\[ R_{\chi^+} = 
 \left[ \begin {array}{cccc}\scriptstyle - 0.13&\scriptstyle 0.70&\scriptstyle- 0.14&\scriptstyle 0.69
\\\noalign{\medskip}\scriptstyle- 0.13&\scriptstyle 0.70&\scriptstyle 0.14&\scriptstyle- 0.69\\\noalign{\medskip}\scriptstyle 0.70
&\scriptstyle\scriptstyle 0.13&\scriptstyle\scriptstyle 0.69&\scriptstyle\scriptstyle 0.14\\\noalign{\medskip}\scriptstyle- 0.70&\scriptstyle\scriptstyle- 0.13&\scriptstyle\scriptstyle 0.69&\scriptstyle\scriptstyle 0.14
\end {array} \right] 
\]

    % This is DATA[435]
\section{Light charged higgs}
\label{lightcharged}
\begin{flushleft}
\begin{tabular}[b]{l|c|c|c|c|c|c|c|c|c}
  \multicolumn{2}{c|}{$\tan{\beta} = 1.38$} &   \multicolumn{2}{c|}{$M_{Z^\prime} = 595$ GeV} &   \multicolumn{2}{c|}{$M_{H^+} = 76$ GeV} &   \multicolumn{3}{c}{$\alpha_{ZZ^\prime} = $$2.5\!\cdot\!10^{-3}$} \\
\hline 
  $M_{H}$ & 118 & 168 & 199 & 550 & 1767 & 1932 &&& \\
  $\xi_{\rm MSSM}$ & 1 & 0.42 & 0.58 & $3.6\!\cdot\!10^{-3}$ & 0 & $1.6\!\cdot\!10^{-5}$ &&& \\
  $\sigma(H_i Z)$ & 0.017 &  21 &  25 & & & & & & \\
  $\sigma(H_i \nu\overline{\nu})$ & 0.025&  22&  22& & & & & & \\
  $\sigma(H_ie^+e^-)$ & 0.0024& 2.1& 2.1& & & & & & \\
\hline 
  $M_{A}$ & 117 & 168 & 1760 & 1930 & 0 & 0 &&& \\
  $\xi_{\rm MSSM}$ & 0 & 1 & 0 & $3.3\!\cdot\!10^{-5}$ & 1 & $2.1\!\cdot\!10^{-3}$ &&& \\
$\sigma(H_1 A)$  & $8.2\!\cdot\!10^{-5}$ &  28 & & & & & & & \\
$\sigma(H_2 A)$  & $3.0\!\cdot\!10^{-8}$ & 0.0099 & & & & & & & \\
$\sigma(H_3 A)$  & $9.6\!\cdot\!10^{-10}$ & 0.00029 & & & & & & & \\
\hline 
  $M_{\chi^0}$               & 54 & 118 & 150 & 156 & 182 & 230 & 552 & 600 & 676 \\
  $\xi_{\rm MSSM}$          & 0.97 & 0.49 & 0.18 & 0.71 & $9.5\!\cdot\!10^{-5}$ & 0.65 & $1.6\!\cdot\!10^{-3}$ & 1 & $9.0\!\cdot\!10^{-4}$ \\
  $\xi_{\tilde{s}}$         & 0.031 & 0.51 & 0.8 & 0.29 & 0.98 & 0.35 & 0.6 & $7.4\!\cdot\!10^{-5}$ & 0.44 \\
  $\xi_{\tilde{Z^\prime}}$ & 0 & $1.3\!\cdot\!10^{-3}$ & 0.019 & $2.8\!\cdot\!10^{-3}$ & 0.017 & 0 & 0.4 & $6.6\!\cdot\!10^{-5}$ & 0.56 \\
\hline 
  $M_{\chi^+}$ & 115 & 600 &&&&&&& \\
\end{tabular}
\end{flushleft}
\vspace{-10pt}
Cross sections quoted are in fb for a linear $e^+ e^-$ collider at center-of-mass energy 500 GeV.\vspace{-10pt}
\begin{eqnarray*}
  v_2       = 141 \gev   & v_1       = 102 \gev   & v_s       = 109 \gev   \\
               v_{s1}    = 538 \gev   & v_{s2}    = 1714 \gev   & v_{s3}    = 141 \gev   \\
               m_{H_u}^2 = -(325 \gev)^2 & m_{H_d}^2 = -(309 \gev)^2 & m_{S}^2   = (1963 \gev)^2 \\
               m_{S_1}^2 = (57 \gev)^2 & m_{S_2}^2 = -(375 \gev)^2 & m_{S_3}^2 = (1783 \gev)^2 \\
               h = -0.976                 & A_h       = 126 \gev   & \mu = h v_s = -107 \gev \\
               \lambda = 0.114          & A_\lambda = 3860 \gev  & \\
               M_1      = -90  \gev   & M_1^\prime = 157 \gev & M_2 = 591 \gev \\
               m_{S S_1}^2 = -(266 \gev)^2 & m_{S S_2}^2 = -(462 \gev)^2 &
\end{eqnarray*}
Branching Ratios for dominant decay modes (greater than 1\% excluding model-dependent squark, slepton, $Z^\prime$ and exotic decays; $\chi^0_{i>1}$ are summed): 
\begin{flushleft}\begin{tabular}{c|lr|lr|lr|lr|lr|lr}
$H_1$ & $\chi^0_1 \chi^0_1$ & 56\% & $b\bar{b}$ & 39\% & $\tau^+\tau^-$ & 2\% & $c\bar{c}$ & 2\% &&&&\\
$H_2$ & $W^+W^-$ & 77\% & $H^+ W^-$ & 11\% & $H^+ H^-$ & 6\% & $\chi^0_1 \chi^0_1$ & 6\% & $b\bar{b}$ & 1\% &&\\
$H_3$ & $W^+W^-$ & 56\% & $H^+ W^-$ & 20\% & $Z Z$ & 20\% & $\chi^0_1 \chi^0_1$ & 2\% & $H^+ H^-$ & 2\% & $\chi^0_1 \chi^0_{i>1}$ & 1\% \\
$H_4$ & $H_2 H_3$ & 18\% & $H_1 H_1$ & 17\% & $W^+W^-$ & 15\% & $\chi^0_{i>1} \chi^0_{i>1}$ & 11\% & $H^+ W^-$ & 9\% & $Z Z$ & 7\% \\
$H_5$ & $\chi^0_{i>1} \chi^0_{i>1}$ & 79\% & $H_2 H_4$ & 10\% & $H_3 H_4$ & 8\% & $\chi^+_1 \chi^-_1$ & 1\% & $H_2 H_2$ & 1\% &&\\
$H_6$ & $\chi^+_1 \chi^-_1$ & 47\% & $\chi^0_{i>1} \chi^0_{i>1}$ & 36\% & $\chi^0_1 \chi^0_{i>1}$ & 13\% & $\chi^0_1 \chi^0_1$ & 3\% & $H_1 H_1$ & 1\% &&\\
\hline
$A_1$ & $\chi^0_1 \chi^0_1$ & 91\% & $b\bar{b}$ & 8\% &&&&&&&&\\
$A_2$ & $H^+ W^-$ & 60\% & $H^+ H^-$ & 24\% & $\chi^0_1 \chi^0_1$ & 11\% & $b\bar{b}$ & 4\% &&&&\\
$A_3$ & $\chi^0_{i>1} \chi^0_{i>1}$ & 97\% & $A_2 H_3$ & 1\% & $\chi^+_1 \chi^-_1$ & 1\% & $A_1 H_2$ & 1\% &&&&\\
$A_4$ & $\chi^+_1 \chi^-_1$ & 46\% & $\chi^0_{i>1} \chi^0_{i>1}$ & 35\% & $\chi^0_1, \chi^0_{i>1}$ & 13\% & $\chi^0_1 \chi^0_1$ & 3\% & $A_2 H_4$ & 1\% & $\chi^+_1 \chi^-_2$ & 1\% \\
\end{tabular}
\end{flushleft}
\newpage

Eigenvectors/rotation matrices 
 \[R_H = 
 \left[ \begin {array}{cccccc}\scriptstyle  0.60&\scriptstyle- 0.80&\scriptstyle- 0.0023&\scriptstyle- 0.013&\scriptstyle- 0.0033&\scriptstyle
- 0.0037\\\noalign{\medskip}\scriptstyle- 0.51&\scriptstyle\scriptstyle- 0.40&\scriptstyle\scriptstyle 0.024&\scriptstyle\scriptstyle 0.73&\scriptstyle\scriptstyle 0.092&\scriptstyle\scriptstyle 0.20
\\\noalign{\medskip}\scriptstyle 0.61&\scriptstyle 0.45&\scriptstyle 0.010&\scriptstyle 0.63&\scriptstyle 0.00080&\scriptstyle 0.17
\\\noalign{\medskip}\scriptstyle- 0.049&\scriptstyle- 0.034&\scriptstyle- 0.065&\scriptstyle 0.096&\scriptstyle- 0.99&\scriptstyle- 0.079
\\\noalign{\medskip}\scriptstyle 0.0018&\scriptstyle 0.0011&\scriptstyle 0.077&\scriptstyle 0.26&\scriptstyle 0.096&\scriptstyle- 0.96
\\\noalign{\medskip}\scriptstyle- 0.0040&\scriptstyle- 0.00067&\scriptstyle- 0.99&\scriptstyle 0.037&\scriptstyle 0.074&\scriptstyle- 0.062
\end {array} \right] 
\]
\[ R_A = 
 \left[ \begin {array}{cccccc}\scriptstyle - 0.00089&\scriptstyle- 0.0012&\scriptstyle 0.0010&\scriptstyle 0.92&\scriptstyle- 0.32
&\scriptstyle\scriptstyle- 0.22\\\noalign{\medskip}\scriptstyle- 0.59&\scriptstyle\scriptstyle- 0.81&\scriptstyle\scriptstyle 0.0030&\scriptstyle\scriptstyle 0.012&\scriptstyle\scriptstyle 0.043&\scriptstyle\scriptstyle-
 0.0068\\\noalign{\medskip}\scriptstyle 0.00023&\scriptstyle\scriptstyle 0.00031&\scriptstyle\scriptstyle 0.055&\scriptstyle\scriptstyle- 0.25&\scriptstyle\scriptstyle- 0.076&\scriptstyle\scriptstyle-
 0.96\\\noalign{\medskip}\scriptstyle 0.0034&\scriptstyle\scriptstyle 0.0047&\scriptstyle\scriptstyle 1.0&\scriptstyle\scriptstyle 0.031&\scriptstyle\scriptstyle 0.061&\scriptstyle\scriptstyle 0.044
\\\noalign{\medskip}\scriptstyle 0.81&\scriptstyle- 0.59&\scriptstyle 0.000044&\scriptstyle- 0.00021&\scriptstyle- 0.00068&\scriptstyle
 0.00011\\\noalign{\medskip}\scriptstyle 0.027&\scriptstyle\scriptstyle 0.036&\scriptstyle\scriptstyle- 0.060&\scriptstyle\scriptstyle 0.29&\scriptstyle\scriptstyle 0.94&\scriptstyle\scriptstyle- 0.15
\end {array} \right] 
\]
\[ R_{\chi^0} = 
 \left[ \begin {array}{ccccccccc}\scriptstyle  0.0028&\scriptstyle 0.76&\scriptstyle 0.077&\scriptstyle- 0.32&\scriptstyle 0.53&\scriptstyle-
 0.17&\scriptstyle\scriptstyle 0.011&\scriptstyle\scriptstyle- 0.029&\scriptstyle\scriptstyle- 0.0033\\\noalign{\medskip}\scriptstyle- 0.037&\scriptstyle\scriptstyle 0.42&\scriptstyle\scriptstyle- 0.030
&\scriptstyle\scriptstyle- 0.24&\scriptstyle\scriptstyle- 0.50&\scriptstyle\scriptstyle 0.69&\scriptstyle\scriptstyle- 0.12&\scriptstyle\scriptstyle 0.14&\scriptstyle\scriptstyle 0.093\\\noalign{\medskip}\scriptstyle 0.14&\scriptstyle\scriptstyle
 0.25&\scriptstyle- 0.037&\scriptstyle 0.23&\scriptstyle- 0.25&\scriptstyle- 0.0062&\scriptstyle 0.60&\scriptstyle- 0.35&\scriptstyle- 0.56
\\\noalign{\medskip}\scriptstyle- 0.053&\scriptstyle 0.43&\scriptstyle- 0.069&\scriptstyle 0.65&\scriptstyle- 0.32&\scriptstyle- 0.38&\scriptstyle- 0.26&\scriptstyle
 0.12&\scriptstyle\scriptstyle 0.24\\\noalign{\medskip}\scriptstyle 0.13&\scriptstyle\scriptstyle 0.00078&\scriptstyle\scriptstyle 0.00095&\scriptstyle\scriptstyle 0.0039&\scriptstyle\scriptstyle 0.0088
&\scriptstyle\scriptstyle- 0.017&\scriptstyle\scriptstyle- 0.66&\scriptstyle\scriptstyle 0.089&\scriptstyle\scriptstyle- 0.74\\\noalign{\medskip}\scriptstyle 0.00040&\scriptstyle\scriptstyle- 0.011&\scriptstyle\scriptstyle-
 0.017&\scriptstyle- 0.60&\scriptstyle- 0.53&\scriptstyle- 0.59&\scriptstyle 0.0025&\scriptstyle 0.0018&\scriptstyle 0.0024
\\\noalign{\medskip}\scriptstyle- 0.63&\scriptstyle- 0.00093&\scriptstyle 0.00068&\scriptstyle 0.026&\scriptstyle 0.031&\scriptstyle- 0.053&\scriptstyle
 0.28&\scriptstyle\scriptstyle 0.66&\scriptstyle\scriptstyle- 0.28\\\noalign{\medskip}\scriptstyle 0.0081&\scriptstyle\scriptstyle- 0.0073&\scriptstyle\scriptstyle 0.99&\scriptstyle\scriptstyle 0.060&\scriptstyle\scriptstyle-
 0.097&\scriptstyle\scriptstyle- 0.0028&\scriptstyle\scriptstyle 0.0027&\scriptstyle\scriptstyle 0.0077&\scriptstyle\scriptstyle 0.00043\\\noalign{\medskip}\scriptstyle 0.75&\scriptstyle\scriptstyle
 0.00062&\scriptstyle- 0.010&\scriptstyle 0.015&\scriptstyle 0.024&\scriptstyle- 0.033&\scriptstyle 0.21&\scriptstyle 0.63&\scriptstyle 0.017\end {array}
 \right] 
\]
\[ R_{\chi^+} = 
 \left[ \begin {array}{cccc}\scriptstyle  0.097&\scriptstyle- 0.70&\scriptstyle 0.060&\scriptstyle- 0.70
\\\noalign{\medskip}\scriptstyle- 0.097&\scriptstyle 0.70&\scriptstyle 0.060&\scriptstyle- 0.70\\\noalign{\medskip}\scriptstyle-
 0.70&\scriptstyle\scriptstyle- 0.097&\scriptstyle\scriptstyle 0.70&\scriptstyle\scriptstyle 0.060\\\noalign{\medskip}\scriptstyle 0.70&\scriptstyle\scriptstyle 0.097&\scriptstyle\scriptstyle 0.70&\scriptstyle\scriptstyle
 0.060\end {array} \right] 
\]

    % This is DATA[808]
\section{Heaviest $H_1$}
\label{h1heaviest}
\begin{flushleft}
\begin{tabular}[b]{l|c|c|c|c|c|c|c|c|c}
  \multicolumn{2}{c|}{$\tan{\beta} = 0.813$} &   \multicolumn{2}{c|}{$M_{Z^\prime} = 1085$ GeV} &   \multicolumn{2}{c|}{$M_{H^+} = 1142$ GeV} &   \multicolumn{3}{c}{$\alpha_{ZZ^\prime} = $-$4.9\!\cdot\!10^{-4}$} \\
\hline 
  $M_{H}$ & 192 & 713 & 1048 & 1149 & 2240 & 2608 &&& \\
  $\xi_{\rm MSSM}$ & 1 & $2.4\!\cdot\!10^{-4}$ & $2.0\!\cdot\!10^{-3}$ & 1 & $2.0\!\cdot\!10^{-5}$ & $5.3\!\cdot\!10^{-5}$ &&& \\
  $\sigma(H_i Z)$ &  44 & & & & & & & & \\
  $\sigma(H_i \nu\overline{\nu})$ &  41& & & & & & & & \\
  $\sigma(H_ie^+e^-)$ & 3.9& & & & & & & & \\
\hline 
  $M_{A}$ & 764 & 1149 & 2234 & 2585 & 0 & 0 &&& \\
  $\xi_{\rm MSSM}$ & $2.4\!\cdot\!10^{-5}$ & 1 & $5.3\!\cdot\!10^{-4}$ & $1.1\!\cdot\!10^{-3}$ & 0.055 & 0.95 &&& \\
\hline 
  $M_{\chi^0}$               & 47 & 265 & 338 & 424 & 430 & 570 & 602 & 1052 & 1122 \\
  $\xi_{\rm MSSM}$          & 0.075 & 1 & 1 & $9.5\!\cdot\!10^{-5}$ & 0 & 1 & 0.93 & $6.6\!\cdot\!10^{-4}$ & 0 \\
  $\xi_{\tilde{s}}$         & 0.93 & $3.2\!\cdot\!10^{-4}$ & $1.8\!\cdot\!10^{-4}$ & 1 & 1 & $9.6\!\cdot\!10^{-4}$ & 0.074 & 0.51 & 0.49 \\
  $\xi_{\tilde{Z^\prime}}$ & $6.1\!\cdot\!10^{-5}$ & 0 & 0 & $1.8\!\cdot\!10^{-3}$ & $8.1\!\cdot\!10^{-5}$ & 0 & $2.4\!\cdot\!10^{-4}$ & 0.49 & 0.51 \\
\hline 
  $M_{\chi^+}$ & 264 & 561 &&&&&&& \\
\end{tabular}
\end{flushleft}
\vspace{-10pt}
Cross sections quoted are in fb for a linear $e^+ e^-$ collider at center-of-mass energy 500 GeV.\vspace{-10pt}
\begin{eqnarray*}
  v_2       = 110 \gev   & v_1       = 135 \gev   & v_s       = 558 \gev   \\
               v_{s1}    = 3256 \gev   & v_{s2}    = 133 \gev   & v_{s3}    = 178 \gev   \\
               m_{H_u}^2 = (436 \gev)^2 & m_{H_d}^2 = -(252 \gev)^2 & m_{S}^2   = (2590 \gev)^2 \\
               m_{S_1}^2 = -(606 \gev)^2 & m_{S_2}^2 = (2016 \gev)^2 & m_{S_3}^2 = (1411 \gev)^2 \\
               h = 0.992                 & A_h       = 1174 \gev   & \mu = h v_s = 553 \gev \\
               \lambda = -0.132          & A_\lambda = 3379 \gev  & \\
               M_1      = -346  \gev   & M_1^\prime = 65 \gev & M_2 = 257 \gev \\
               m_{S S_1}^2 = -(1014 \gev)^2 & m_{S S_2}^2 = -(826 \gev)^2 &
\end{eqnarray*}
Branching Ratios for dominant decay modes (greater than 1\% excluding model-dependent squark, slepton, $Z^\prime$ and exotic decays; $\chi^0_{i>1}$ are summed): 
\begin{flushleft}\begin{tabular}{c|lr|lr|lr|lr|lr|lr}
$H_1$ & $W^+W^-$ & 59\% & $\chi^0_1 \chi^0_1$ & 22\% & $Z Z$ & 18\% &&&&&&\\
$H_2$ & $H_1 H_1$ & 40\% & $W^+W^-$ & 33\% & $Z Z$ & 16\% & $t\bar{t}$ & 7\% & $\chi^0_1 \chi^0_{i>1}$ & 4\% &&\\
$H_3$ & $H_1 H_1$ & 33\% & $W^+W^-$ & 28\% & $Z Z$ & 14\% & $t\bar{t}$ & 13\% & $\chi^0_{i>1} \chi^0_{i>1}$ & 8\% & $\chi^0_1 \chi^0_{i>1}$ & 3\% \\
$H_4$ & $t\bar{t}$ & 70\% & $\chi^0_1 \chi^0_{i>1}$ & 23\% & $\chi^+_1 \chi^-_2$ & 3\% & $\chi^0_{i>1} \chi^0_{i>1}$ & 3\% & $H_1 H_1$ & 1\% &&\\
$H_5$ & $\chi^0_{i>1} \chi^0_{i>1}$ & 44\% & $\chi^+_2 \chi^-_2$ & 27\% & $A_1 A_2$ & 21\% & $\chi^0_1 \chi^0_{i>1}$ & 6\% &&&&\\
$H_6$ & $\chi^+_2 \chi^-_2$ & 32\% & $\chi^0_{i>1} \chi^0_{i>1}$ & 31\% & $A_1 A_2$ & 21\% & $H_4 H_4$ & 7\% & $\chi^0_1 \chi^0_{i>1}$ & 6\% & $H^+ H^-$ & 1\% \\
\hline
$A_1$ & $\chi^0_1, \chi^0_{i>1}$ & 86\% & $t\bar{t}$ & 11\% & $\chi^0_1 \chi^0_1$ & 3\% &&&&&&\\
$A_2$ & $t\bar{t}$ & 72\% & $\chi^0_1, \chi^0_{i>1}$ & 16\% & $\chi^0_1 \chi^0_1$ & 6\% & $\chi^0_{i>1} \chi^0_{i>1}$ & 5\% &&&&\\
$A_3$ & $\chi^0_{i>1} \chi^0_{i>1}$ & 39\% & $\chi^+_2 \chi^-_2$ & 26\% & $H_2 Z$ & 15\% & $A_1 H_2$ & 10\% & $A_2 H_2$ & 6\% & $\chi^0_1, \chi^0_{i>1}$ & 3\% \\
$A_4$ & $H_2 Z$ & 98\% & $\chi^+_2 \chi^-_2$ & 1\% & $\chi^0_{i>1} \chi^0_{i>1}$ & 1\% &&&&&&\\
\end{tabular}
\end{flushleft}
\newpage

Eigenvectors/rotation matrices 
 \[R_H = 
 \left[ \begin {array}{cccccc}\scriptstyle - 0.62&\scriptstyle- 0.78&\scriptstyle 0.0038&\scriptstyle 0.028&\scriptstyle 0.0021&\scriptstyle
 0.0040\\\noalign{\medskip}\scriptstyle- 0.013&\scriptstyle\scriptstyle- 0.0079&\scriptstyle\scriptstyle- 0.11&\scriptstyle\scriptstyle- 0.35&\scriptstyle\scriptstyle- 0.32&\scriptstyle\scriptstyle-
 0.87\\\noalign{\medskip}\scriptstyle 0.044&\scriptstyle\scriptstyle- 0.0040&\scriptstyle\scriptstyle 0.20&\scriptstyle\scriptstyle 0.91&\scriptstyle\scriptstyle- 0.098&\scriptstyle\scriptstyle- 0.36
\\\noalign{\medskip}\scriptstyle 0.78&\scriptstyle- 0.62&\scriptstyle- 0.00047&\scriptstyle- 0.037&\scriptstyle 0.0030&\scriptstyle 0.0078
\\\noalign{\medskip}\scriptstyle- 0.0041&\scriptstyle 0.0017&\scriptstyle 0.38&\scriptstyle- 0.12&\scriptstyle 0.86&\scriptstyle- 0.31
\\\noalign{\medskip}\scriptstyle- 0.0070&\scriptstyle 0.0022&\scriptstyle 0.89&\scriptstyle- 0.20&\scriptstyle- 0.38&\scriptstyle 0.11
\end {array} \right] 
\]
\[ R_A = 
 \left[ \begin {array}{cccccc}\scriptstyle - 0.0038&\scriptstyle- 0.0031&\scriptstyle 0.018&\scriptstyle 0.12&\scriptstyle- 0.33&\scriptstyle
 0.94\\\noalign{\medskip}\scriptstyle- 0.78&\scriptstyle\scriptstyle- 0.63&\scriptstyle\scriptstyle 0.035&\scriptstyle\scriptstyle 0.031&\scriptstyle\scriptstyle- 0.0031&\scriptstyle\scriptstyle- 0.011
\\\noalign{\medskip}\scriptstyle- 0.018&\scriptstyle- 0.015&\scriptstyle- 0.42&\scriptstyle- 0.071&\scriptstyle 0.85&\scriptstyle 0.31
\\\noalign{\medskip}\scriptstyle 0.026&\scriptstyle 0.021&\scriptstyle 0.89&\scriptstyle 0.15&\scriptstyle 0.41&\scriptstyle 0.11
\\\noalign{\medskip}\scriptstyle- 0.13&\scriptstyle 0.20&\scriptstyle- 0.16&\scriptstyle 0.95&\scriptstyle 0.039&\scriptstyle- 0.10
\\\noalign{\medskip}\scriptstyle 0.62&\scriptstyle- 0.75&\scriptstyle- 0.039&\scriptstyle 0.23&\scriptstyle 0.0093&\scriptstyle- 0.025
\end {array} \right] 
\]
\[ R_{\chi^0} = 
 \left[ \begin {array}{ccccccccc}\scriptstyle  0.0078&\scriptstyle- 0.0087&\scriptstyle- 0.016&\scriptstyle 0.21&\scriptstyle 0.17
&\scriptstyle\scriptstyle- 0.95&\scriptstyle\scriptstyle- 0.17&\scriptstyle\scriptstyle 0.0060&\scriptstyle\scriptstyle 0.011\\\noalign{\medskip}\scriptstyle 0.00074&\scriptstyle\scriptstyle 0.0069&\scriptstyle\scriptstyle-
 0.99&\scriptstyle- 0.056&\scriptstyle 0.082&\scriptstyle 0.018&\scriptstyle 0.0030&\scriptstyle- 0.00043&\scriptstyle 0.00018
\\\noalign{\medskip}\scriptstyle- 0.00070&\scriptstyle- 0.98&\scriptstyle- 0.025&\scriptstyle 0.13&\scriptstyle- 0.14&\scriptstyle 0.013&\scriptstyle
 0.0022&\scriptstyle\scriptstyle- 0.00013&\scriptstyle\scriptstyle- 0.00029\\\noalign{\medskip}\scriptstyle 0.043&\scriptstyle\scriptstyle- 0.000082&\scriptstyle\scriptstyle
 0.00069&\scriptstyle 0.0072&\scriptstyle 0.0066&\scriptstyle- 0.014&\scriptstyle 0.12&\scriptstyle- 0.71&\scriptstyle 0.70
\\\noalign{\medskip}\scriptstyle 0.0090&\scriptstyle- 0.00020&\scriptstyle 0.000045&\scriptstyle- 0.00090&\scriptstyle- 0.00050&\scriptstyle
 0.0042&\scriptstyle\scriptstyle 0.045&\scriptstyle\scriptstyle 0.71&\scriptstyle\scriptstyle 0.71\\\noalign{\medskip}\scriptstyle- 0.0021&\scriptstyle\scriptstyle 0.19&\scriptstyle\scriptstyle- 0.094&\scriptstyle\scriptstyle
 0.69&\scriptstyle\scriptstyle- 0.69&\scriptstyle\scriptstyle 0.031&\scriptstyle\scriptstyle 0.0039&\scriptstyle\scriptstyle 0.00020&\scriptstyle\scriptstyle- 0.00015\\\noalign{\medskip}\scriptstyle
 0.016&\scriptstyle- 0.0045&\scriptstyle 0.022&\scriptstyle 0.68&\scriptstyle 0.68&\scriptstyle 0.27&\scriptstyle 0.027&\scriptstyle 0.0056&\scriptstyle- 0.0078
\\\noalign{\medskip}\scriptstyle- 0.70&\scriptstyle 0.00014&\scriptstyle- 0.00014&\scriptstyle 0.018&\scriptstyle 0.018&\scriptstyle- 0.12&\scriptstyle
 0.70&\scriptstyle\scriptstyle 0.021&\scriptstyle\scriptstyle- 0.057\\\noalign{\medskip}\scriptstyle- 0.72&\scriptstyle\scriptstyle 0.000076&\scriptstyle\scriptstyle- 0.00023&\scriptstyle\scriptstyle-
 0.0021&\scriptstyle 0.0014&\scriptstyle 0.12&\scriptstyle- 0.68&\scriptstyle- 0.054&\scriptstyle 0.11\end {array} \right] 
\]
\[ R_{\chi^+} = 
 \left[ \begin {array}{cccc}\scriptstyle  0.70&\scriptstyle 0.088&\scriptstyle 0.71&\scriptstyle 0.049
\\\noalign{\medskip}\scriptstyle- 0.70&\scriptstyle- 0.088&\scriptstyle 0.71&\scriptstyle 0.049\\\noalign{\medskip}\scriptstyle-
 0.088&\scriptstyle\scriptstyle 0.70&\scriptstyle\scriptstyle- 0.049&\scriptstyle\scriptstyle 0.71\\\noalign{\medskip}\scriptstyle 0.088&\scriptstyle\scriptstyle- 0.70&\scriptstyle\scriptstyle- 0.049&\scriptstyle\scriptstyle
 0.71\end {array} \right] 
\]

    % This is DATA[401]
\section{Heaviest $A_1$}
\label{a1heaviest}
\begin{flushleft}
\begin{tabular}[b]{l|c|c|c|c|c|c|c|c|c}
  \multicolumn{2}{c|}{$\tan{\beta} = 0.71$} &   \multicolumn{2}{c|}{$M_{Z^\prime} = 1263$ GeV} &   \multicolumn{2}{c|}{$M_{H^+} = 1275$ GeV} &   \multicolumn{3}{c}{$\alpha_{ZZ^\prime} = $-$5.8\!\cdot\!10^{-4}$} \\
\hline 
  $M_{H}$ & 130 & 973 & 1209 & 1283 & 2003 & 2369 &&& \\
  $\xi_{\rm MSSM}$ & 1 & $5.2\!\cdot\!10^{-4}$ & 0.012 & 0.99 & $2.3\!\cdot\!10^{-3}$ & $1.0\!\cdot\!10^{-4}$ &&& \\
  $\sigma(H_i Z)$ &  56 & & & & & & & & \\
  $\sigma(H_i \nu\overline{\nu})$ &  77& & & & & & & & \\
  $\sigma(H_ie^+e^-)$ & 7.4& & & & & & & & \\
\hline 
  $M_{A}$ & 998 & 1269 & 1966 & 2364 & 0 & 0 &&& \\
  $\xi_{\rm MSSM}$ & $8.8\!\cdot\!10^{-4}$ & 0.99 & 0.011 & $2.6\!\cdot\!10^{-4}$ & $8.7\!\cdot\!10^{-3}$ & 0.99 &&& \\
\hline 
  $M_{\chi^0}$               & 33 & 198 & 198 & 398 & 412 & 507 & 509 & 809 & 1973 \\
  $\xi_{\rm MSSM}$          & 0.08 & 1 & 1 & 1 & 0.92 & 0 & 0 & $6.2\!\cdot\!10^{-4}$ & $9.0\!\cdot\!10^{-5}$ \\
  $\xi_{\tilde{s}}$         & 0.92 & $1.7\!\cdot\!10^{-3}$ & $1.0\!\cdot\!10^{-3}$ & $2.4\!\cdot\!10^{-3}$ & 0.076 & 1 & 1 & 0.71 & 0.29 \\
  $\xi_{\tilde{Z^\prime}}$ & $1.5\!\cdot\!10^{-5}$ & 0 & 0 & 0 & $2.4\!\cdot\!10^{-5}$ & $3.3\!\cdot\!10^{-4}$ & $3.1\!\cdot\!10^{-4}$ & 0.29 & 0.71 \\
\hline 
  $M_{\chi^+}$ & 197 & 389 &&&&&&& \\
\end{tabular}
\end{flushleft}
\vspace{-10pt}
Cross sections quoted are in fb for a linear $e^+ e^-$ collider at center-of-mass energy 500 GeV.\vspace{-10pt}
\begin{eqnarray*}
  v_2       = 101 \gev   & v_1       = 142 \gev   & v_s       = 544 \gev   \\
               v_{s1}    = 102 \gev   & v_{s2}    = 3826 \gev   & v_{s3}    = 93 \gev   \\
               m_{H_u}^2 = (734 \gev)^2 & m_{H_d}^2 = (133 \gev)^2 & m_{S}^2   = (2149 \gev)^2 \\
               m_{S_1}^2 = (1836 \gev)^2 & m_{S_2}^2 = -(829 \gev)^2 & m_{S_3}^2 = (1833 \gev)^2 \\
               h = -0.693                 & A_h       = 2048 \gev   & \mu = h v_s = -377 \gev \\
               \lambda = 0.133          & A_\lambda = 3752 \gev  & \\
               M_1      = -208  \gev   & M_1^\prime = 1164 \gev & M_2 = 187 \gev \\
               m_{S S_1}^2 = -(707 \gev)^2 & m_{S S_2}^2 = -(729 \gev)^2 &
\end{eqnarray*}
Branching Ratios for dominant decay modes (greater than 1\% excluding model-dependent squark, slepton, $Z^\prime$ and exotic decays; $\chi^0_{i>1}$ are summed): 
\begin{flushleft}\begin{tabular}{c|lr|lr|lr|lr|lr|lr}
$H_1$ & $\chi^0_1 \chi^0_1$ & 97\% & $b\bar{b}$ & 3\% &&&&&&&&\\
$H_2$ & $W^+W^-$ & 22\% & $H_1 H_1$ & 21\% & $\chi^+_2 \chi^-_2$ & 17\% & $\chi^0_{i>1} \chi^0_{i>1}$ & 17\% & $Z Z$ & 11\% & $\chi^0_1 \chi^0_{i>1}$ & 10\% \\
$H_3$ & $t\bar{t}$ & 35\% & $\chi^0_{i>1} \chi^0_{i>1}$ & 28\% & $\chi^+_2 \chi^-_2$ & 21\% & $\chi^0_1 \chi^0_{i>1}$ & 12\% & $\chi^+_1 \chi^-_2$ & 4\% &&\\
$H_4$ & $t\bar{t}$ & 71\% & $\chi^0_1 \chi^0_{i>1}$ & 14\% & $\chi^0_{i>1} \chi^0_{i>1}$ & 9\% & $\chi^+_1 \chi^-_2$ & 6\% &&&&\\
$H_5$ & $\chi^+_2 \chi^-_2$ & 33\% & $\chi^0_{i>1} \chi^0_{i>1}$ & 32\% & $W^+W^-$ & 9\% & $H_1 H_1$ & 9\% & $\chi^0_1 \chi^0_{i>1}$ & 6\% & $Z Z$ & 5\% \\
$H_6$ & $\chi^0_{i>1} \chi^0_{i>1}$ & 50\% & $\chi^+_2 \chi^-_2$ & 20\% & $A_1 A_2$ & 9\% & $W^+W^-$ & 4\% & $H_1 H_1$ & 4\% & $\chi^0_1 \chi^0_{i>1}$ & 4\% \\
\hline
$A_1$ & $\chi^+_2 \chi^-_2$ & 33\% & $t\bar{t}$ & 25\% & $\chi^0_{i>1} \chi^0_{i>1}$ & 25\% & $\chi^0_1, \chi^0_{i>1}$ & 15\% & $\chi^0_1 \chi^0_1$ & 2\% &&\\
$A_2$ & $t\bar{t}$ & 73\% & $\chi^0_1, \chi^0_{i>1}$ & 12\% & $\chi^0_{i>1} \chi^0_{i>1}$ & 10\% & $\chi^0_1 \chi^0_1$ & 3\% & $\chi^+_2 \chi^-_2$ & 1\% & $\chi^+_1 \chi^-_1$ & 1\% \\
$A_3$ & $\chi^0_{i>1} \chi^0_{i>1}$ & 33\% & $\chi^+_2 \chi^-_2$ & 31\% & $H_2 Z$ & 28\% & $A_2 H_2$ & 4\% & $\chi^0_1, \chi^0_{i>1}$ & 3\% & $t\bar{t}$ & 2\% \\
$A_4$ & $H_2 Z$ & 99\% & $H_3 Z$ & 1\% &&&&&&&&\\
\end{tabular}
\end{flushleft}
\newpage

Eigenvectors/rotation matrices 
 \[R_H = 
 \left[ \begin {array}{cccccc}\scriptstyle  0.57&\scriptstyle 0.82&\scriptstyle 0.038&\scriptstyle 0.0073&\scriptstyle- 0.0038&\scriptstyle
 0.0071\\\noalign{\medskip}\scriptstyle- 0.0068&\scriptstyle\scriptstyle 0.022&\scriptstyle\scriptstyle- 0.15&\scriptstyle\scriptstyle- 0.47&\scriptstyle\scriptstyle- 0.30&\scriptstyle\scriptstyle- 0.81
\\\noalign{\medskip}\scriptstyle- 0.090&\scriptstyle 0.066&\scriptstyle- 0.25&\scriptstyle 0.15&\scriptstyle- 0.90&\scriptstyle 0.29
\\\noalign{\medskip}\scriptstyle- 0.81&\scriptstyle 0.57&\scriptstyle 0.00028&\scriptstyle- 0.010&\scriptstyle 0.12&\scriptstyle- 0.014
\\\noalign{\medskip}\scriptstyle- 0.047&\scriptstyle- 0.011&\scriptstyle 0.91&\scriptstyle 0.22&\scriptstyle- 0.28&\scriptstyle- 0.20
\\\noalign{\medskip}\scriptstyle- 0.0095&\scriptstyle- 0.0033&\scriptstyle 0.28&\scriptstyle- 0.84&\scriptstyle- 0.069&\scriptstyle 0.46
\end {array} \right] 
\]
\[ R_A = 
 \left[ \begin {array}{cccccc}\scriptstyle - 0.024&\scriptstyle- 0.017&\scriptstyle 0.077&\scriptstyle- 0.50&\scriptstyle 0.067&\scriptstyle
 0.86\\\noalign{\medskip}\scriptstyle 0.81&\scriptstyle\scriptstyle 0.58&\scriptstyle\scriptstyle- 0.099&\scriptstyle\scriptstyle- 0.0092&\scriptstyle\scriptstyle- 0.033&\scriptstyle\scriptstyle 0.040
\\\noalign{\medskip}\scriptstyle- 0.085&\scriptstyle- 0.061&\scriptstyle- 0.95&\scriptstyle 0.19&\scriptstyle- 0.13&\scriptstyle 0.20
\\\noalign{\medskip}\scriptstyle 0.013&\scriptstyle 0.0094&\scriptstyle 0.26&\scriptstyle 0.85&\scriptstyle 0.036&\scriptstyle 0.46
\\\noalign{\medskip}\scriptstyle- 0.035&\scriptstyle 0.086&\scriptstyle- 0.14&\scriptstyle 0.026&\scriptstyle 0.98&\scriptstyle- 0.048
\\\noalign{\medskip}\scriptstyle 0.58&\scriptstyle- 0.81&\scriptstyle- 0.013&\scriptstyle 0.0024&\scriptstyle 0.090&\scriptstyle- 0.0044
\end {array} \right] 
\]
\[ R_{\chi^0} = 
 \left[ \begin {array}{ccccccccc}\scriptstyle  0.0039&\scriptstyle- 0.024&\scriptstyle- 0.035&\scriptstyle 0.23&\scriptstyle 0.16&\scriptstyle
- 0.95&\scriptstyle\scriptstyle 0.0037&\scriptstyle\scriptstyle- 0.14&\scriptstyle\scriptstyle 0.0040\\\noalign{\medskip}\scriptstyle 0.00076&\scriptstyle\scriptstyle 0.014&\scriptstyle\scriptstyle-
 0.99&\scriptstyle- 0.065&\scriptstyle 0.13&\scriptstyle 0.042&\scriptstyle- 0.00023&\scriptstyle 0.0048&\scriptstyle 0.000021
\\\noalign{\medskip}\scriptstyle- 0.00079&\scriptstyle- 0.98&\scriptstyle- 0.043&\scriptstyle 0.14&\scriptstyle- 0.16&\scriptstyle 0.032&\scriptstyle-
 0.00012&\scriptstyle\scriptstyle 0.0050&\scriptstyle\scriptstyle- 0.00022\\\noalign{\medskip}\scriptstyle- 0.0018&\scriptstyle\scriptstyle 0.22&\scriptstyle\scriptstyle- 0.13&\scriptstyle\scriptstyle
 0.68&\scriptstyle\scriptstyle- 0.68&\scriptstyle\scriptstyle 0.049&\scriptstyle\scriptstyle- 0.00035&\scriptstyle\scriptstyle 0.0056&\scriptstyle\scriptstyle- 0.00053\\\noalign{\medskip}\scriptstyle-
 0.0049&\scriptstyle 0.011&\scriptstyle- 0.057&\scriptstyle- 0.68&\scriptstyle- 0.68&\scriptstyle- 0.27&\scriptstyle 0.0028&\scriptstyle- 0.015&\scriptstyle- 0.0022
\\\noalign{\medskip}\scriptstyle- 0.018&\scriptstyle- 0.0000097&\scriptstyle 0.000040&\scriptstyle 0.0015&\scriptstyle 0.0019&\scriptstyle
 0.0069&\scriptstyle\scriptstyle 0.71&\scriptstyle\scriptstyle- 0.043&\scriptstyle\scriptstyle- 0.71\\\noalign{\medskip}\scriptstyle 0.018&\scriptstyle\scriptstyle 0.000019&\scriptstyle\scriptstyle-
 0.000015&\scriptstyle- 0.00074&\scriptstyle- 0.0013&\scriptstyle 0.0065&\scriptstyle 0.71&\scriptstyle- 0.0069&\scriptstyle 0.71
\\\noalign{\medskip}\scriptstyle 0.54&\scriptstyle- 0.00033&\scriptstyle 0.00036&\scriptstyle- 0.018&\scriptstyle- 0.017&\scriptstyle 0.12&\scriptstyle-
 0.030&\scriptstyle\scriptstyle- 0.83&\scriptstyle\scriptstyle 0.0076\\\noalign{\medskip}\scriptstyle- 0.84&\scriptstyle\scriptstyle 0.000078&\scriptstyle\scriptstyle- 0.00017&\scriptstyle\scriptstyle-
 0.0081&\scriptstyle- 0.0048&\scriptstyle 0.075&\scriptstyle- 0.020&\scriptstyle- 0.53&\scriptstyle 0.035\end {array} \right] 
\]
\[ R_{\chi^+} = 
 \left[ \begin {array}{cccc}\scriptstyle - 0.69&\scriptstyle- 0.15&\scriptstyle 0.71&\scriptstyle 0.047
\\\noalign{\medskip}\scriptstyle 0.69&\scriptstyle 0.15&\scriptstyle 0.71&\scriptstyle 0.047\\\noalign{\medskip}\scriptstyle 0.15&\scriptstyle
- 0.69&\scriptstyle\scriptstyle- 0.047&\scriptstyle\scriptstyle 0.71\\\noalign{\medskip}\scriptstyle- 0.15&\scriptstyle\scriptstyle 0.69&\scriptstyle\scriptstyle- 0.047&\scriptstyle\scriptstyle 0.71
\end {array} \right] 
\]

\pagebreak
 
\end{document}